# Equipartition of energy defines the size-thickness relationship in liquid-exfoliated nanosheets

Claudia Backes,[1*] Davide Campi,[2] Beata M. Szydlowska,[1,3] Kevin Synnatschke,[1] Ezgi Ojala,[1] Farnia Rashvand,[1] Andrew Harvey,[3] Aideen Griffin,[3] Zdenek Sofer,[4] Nicola Marzari,[2] Jonathan N. Coleman,[3*] David D. O'Regan[3*]

[1]*Chair of Applied Physical Chemistry, University of Heidelberg, Im Neuenheimer Feld 253, 69120 Heidelberg, Germany*

[2]*Theory and Simulation of Materials (THEOS) and National Centre for Computational Design and Discovery of Novel Materials (MARVEL), École Polytechnique Fédérale de Lausanne, CH-1015 Lausanne, Switzerland*

[3]*School of Physics and CRANN & AMBER Research Centers, Trinity College Dublin, The University of Dublin, Dublin 2, Ireland*

[4]*Department of Inorganic Chemistry, University of Chemistry and Technology Prague, Technická 5, 166 28 Prague 6, Czech Republic*

* backes@uni-heidelberg.de ; colemaj@tcd.ie ; david.o.regan@tcd.ie

ABSTRACT: Liquid phase exfoliation (LPE) is a commonly-used method to produce 2D nanosheets from a range of layered crystals. However, such nanosheets display broad size and thickness distributions and correlations between area and thickness, issues which limit nanosheet application-potential. To understand the factors controlling the exfoliation process, we have liquid-exfoliated 12 different layered materials, size-selecting each into fractions before using AFM to measure the nanosheet length, width and thickness distributions for each fraction. The resultant data shows a clear power-law scaling of nanosheet area with thickness for each material. We have developed a simple non-equilibrium thermodynamics-based model predicting that the power-law pre-factor is proportional to both the ratios of in-plane-tearing/out-of-plane-peeling energies and in-plane/out-of-plane moduli. By comparing the experimental data with the modulus ratio calculated from first principles, we find close agreement between experiment and theory. This supports our hypothesis that energy equipartition holds between nanosheet tearing and peeling during sonication-assisted exfoliation.

Two-dimensional (2D) materials are a diverse family of nanostructures consisting of chemically bonded 2D monolayers that are often arranged in van der Waals-bonded few-layer stacks.[1-4] They are interesting in their own right for fundamental studies and useful in a broad range of applications.[4-6] Importantly, almost all 2D materials have physical and chemical properties that depend on the number of layers in a stack.[7] This makes control of stack (i.e. nanosheet) thickness important.

2D materials can be fabricated by various top-down and bottom-up techniques.[7] One important top-down technique is liquid phase exfoliation (LPE).[8, 9] This method converts 3D layered crystals into large numbers of few-layer 2D nanosheets by using an energy input to remove nanosheets from their parent crystal in a liquid environment. Due to its simplicity, scalability and compatibility with solution processing, LPE has recently gained increasing attention. Importantly, this technique is applicable to a whole host of van der Waals crystals with dozens of 2D materials having been produced in this way. A range of methods[10, 11] for inputting energy have been described, including ultrasonication[12] or shear exfoliation in rotor stator mixers,[13] kitchen blenders,[14, 15] microfluidizers[16, 17], via compressive flow[18] etc. Overall, sonication is still the most widely used technique at the laboratory scale.

However, a significant disadvantage of LPE is that it always yields broad nanosheet size and thickness distributions,[19] rendering a precise characterisation of exfoliated nanosheets challenging and making it hard to assess exfoliation quality. This also limits the suitability of LPE nanosheets for applications.

In order to further develop LPE and to identify its intrinsic limitations as well as future opportunities, it will be essential to develop a general understanding of the fundamental physics of the exfoliation procedure. Such an understanding must be based on a broad combination of experimental data and theoretical modelling.

Here we perform a comparative study using sonication-assisted LPE to exfoliate a range of van der Waals crystals under comparable conditions. By performing extensive AFM analysis, we demonstrate a fundamental relationship between nanosheet size and thickness, which is not affected by solvent choice or sonication conditions. The existence of such a relationship allows us to propose an experimental metric for exfoliation quality. These experimental observations are underpinned by a minimal theoretical model based on the thermodynamic principle of equipartition, which results in a simple analytical relationship between nanosheet area and thickness. This model very closely describes the experimental data and links the exfoliation-quality metric to energetic parameters associated with the breaking of both chemical and van der Waals bonds during the exfoliation process.

## RESULTS AND DISCUSSION:

*Size-selection and AFM analysis of nanosheets*

LPE yields stock dispersions with broad lateral size and thickness distributions (see Figure 1A and SI, Fig. S1). For technical reasons, this polydispersity makes statistical atomic force microscopy (AFM) characterisation challenging. As a result, we generally performed size selection of all stock dispersions by liquid cascade centrifugation (LCC), as introduced elsewhere.[19] In this iterative centrifugation process, size-selected fractions are produced in a two-step process with a low-centrifugation-rate step removing large nanosheets and a higher-centrifugation-rate step removing small nanosheets. A number of such two-step processes can be connected in a cascade to result in a set of fractions, each labelled via the midpoint of the centrifugation rates. This procedure yields fractions with reasonably narrow size and thickness distributions. The final supernatant obtained after the end of the cascade contains very small nanosheets (<20 nm long) and was discarded due to difficulties in accurately measuring nanosheet sizes at small lengthscales.

The size-selected fractions were deposited onto $Si/SiO_2$ and then subjected to size/thickness quantification by atomic force microscopy (AFM). The contrast of nanosheets deposited onto opaque layered substrates was exploited to identify regions promising for AFM as discussed elsewhere.[20] To minimise aggregation, the substrate was heated on a hotplate to above the boiling point of the solvent. This leads to a flash evaporation as illustrated in our previous video publication.[21] From the AFM images (examples Fig. 1, for all data see SI), it is clear that the dispersed objects are 2-dimensional with lateral sizes and thicknesses that vary across fractions. For example, figure 1 shows representative images of an unselected stock dispersion of graphite exfoliated in aqueous sodium cholate by tip sonication (figure 1A) and images of the fractions after LCC (figure 1B-F).

From such images, we measured the longest lateral dimension ($L$), the dimension perpendicular (width, $W$) and the nanosheet thickness. This was done by manually cropping widefield images into smaller regions with only a few objects in each image and manually drawing line profiles across the nanosheets to extract the dimensions. For a visualisation see ref [21]. Only those objects that had the distinct appearance of 2-dimensional sheets lying flat on the substrate were counted.

Other deposits such as aggregates or impurities from residual solvent or surfactant were not taken into account (examples SI figure S3-4). By using previously established step height analysis, the apparent AFM thickness was converted to nanosheet layer number, *N*.[13, 19, 22-27] The resultant data can be used to calculate a number of statistical parameters such as arithmetic means: <*L*>, <*W*> and <*N*>. Example histograms are presented in figure 2 for the fractions containing the largest (L-Gra) and the smallest (S-Gra) nanosheets produced in this particular cascade.

*Comparison of graphite exfoliated under different conditions*

Before investigating nanosheet exfoliation for various materials, it is first important to first understand the impact of exfoliation conditions on the resultant nanosheets. To this end, graphite was exfoliated by tip sonication in aqueous sodium cholate (SC) and *N*-methyl-2-pyrrolidone (NMP) and in SC by bath sonication. One would expect the various conditions to produce nanosheet dispersions that are distinct, i.e. at different yield and different nanosheet dimensions. However, a detailed comparison of statistically measured size and thickness distributions under such circumstances has not yet been reported. In all three cases, size selection by LCC was performed as explained above and the fractions subjected to statistical AFM analysis (distribution histograms see SI figure S6-S8). The aqueous dispersions were centrifuged for 2 h in each step, while for NMP-based dispersions the centrifugation time was 3.5 h to balance the effect of higher viscosity.

While size selection is required to make the statistical analysis of the nanosheet dimensions more feasible, it is not clear whether the resultant lateral sizes and thicknesses in the fractions are representative of the nanosheet population in the stock dispersion. In liquid cascade centrifugation, to a first approximation, the sample is fractionated by hydrodynamic volume. One may argue that the correlation between nanosheet lateral dimension and layer number typically observed[19] is thus a result of the centrifugation. However, this is not necessarily the case. As we will show below, such a correlation is present before size selection and is a consequence of the details of the exfoliation process. The fact that larger flakes tend to be thicker reflects that fact that it costs more energy to exfoliate larger-area nanosheets of a given thickness compared to smaller ones. Due to the difficulties of directly measuring nanosheet area by AFM for relative small nanosheets such as these, we use $L \times W$ as a proxy for nanosheet face area. Figure 3A shows a scatter plot of *LW* as function of layer number, *N*, of the nanosheets in the stock dispersion produced from graphite. Here, each data point represents a single nanosheet. It is clear from this plot that larger area nanosheets tend to be thicker and smaller nanosheets tend to be thinner. Exactly the same picture is obtained when combining the data from the analysis of the fractions of the same sample after LCC (Fig. 3B). This demonstrates that the distribution of nanosheet sizes in the fractions after size selection is a consequence of that in the stock. For a more detailed discussion see SI figure S2.

Broadly similar data point clouds were obtained for the sample exfoliated in SC using a sonic bath and in NMP exfoliated with a sonic tip (figure 3C-D). Representative widefield AFM images are shown in figure S3-4. However, some differences can be observed: the correlation between area and layer number is softened in the case of the graphite exfoliated using SC in the bath (figure 3C) and more large monolayers are produced at relatively low centrifugal acceleration compared to graphite exfoliated in SC with the sonic tip. In contrast, for the tip exfoliation in NMP, the area-layer number correlation is even more well defined (figure 3D). In addition, the data cloud is shifted toward larger/thicker nanosheets with hardly any monolayers observed. Unlike the SC-based samples, no few-layer graphene could be isolated at centrifugal acceleration higher than 6,000 *g* which suggests that fewer small/thin nanosheets are produced in NMP compared to exfoliation in aqueous surfactant. The reason for this is currently unclear, but it might be related to differences in interfacial stress transfer at solvent-nanosheet versus surfactant-nanosheet interfaces. However, in spite of these differences, the area-layer number data clouds seem to be centred around a similar mean area-layer number relationship as will be discussed further below.

The observation that nanosheets were isolated above 6,000 *g* in SC, but not in NMP suggests that the sonication conditions have an impact on the relative population of nanosheets in a certain size windows as one would expect. This can be illustrated clearly when determining the yield (i.e. fraction of graphite mass converted to graphene) of nanosheets in each fraction (see methods). In all cases, the yield decreases with increasing centrifugal acceleration (figure 3E). However, for tip-sonication the yield decreases much more steeply for NMP compared to SC, illustrating that relatively few small/thin nanosheets are produced in NMP. The overall yield summed over the fractions isolated above 100 *g* is 19.5% for the exfoliation by tip sonication in SC, only 0.3% for bath sonication in SC and 5% for tip sonication in NMP. This emphasizes that bath sonication is not suitable to produce large masses and that exfoliation in aqueous surfactant gives the best yield of few-layered material.

The overall larger population of larger/thicker nanosheets in NMP also has an impact on the mean lateral size and layer number isolated in the cascade. This is shown by the plots of mean area (expressed as $\langle LW \rangle$) and $\langle N \rangle$ as function of the midpoint of the pair of centrifugal accelerations used in the cascade in figure 3F-G. The NMP data sits consistently above the SC data for both area and layer number and is also characterised by a different power law exponent relating the dimensions to RCF. Interestingly, the nanosheets produced and isolated in SC in tip and bath sonication are very similar on average even though larger monolayers and bilayers can be produced in the bath as shown by the data cloud in figure 3C. However, since the nanosheets in NMP are larger, but also thicker, it is important to analyse whether the relationship between <*LW*> and <*N*> varies across the samples. As mentioned above, the data clouds in figure 3B-D seem to be centred around similar averages. To test this, we plot <*LW*> as function of <*N*> in figure 3H. As expected from figure 3F-G, the samples obtained by bath and tip sonication in SC fall on the same curve. However, the data for graphite exfoliated in NMP by tip sonication is shifted to higher values of <*LW*> and <*N*> and slightly offset. In all cases, <*LW*> is related to <*N*> as a powerlaw. Empirical fitting shows that the powerlaw exponent of the NMP samples is slightly different to the exponent of the samples exfoliated in SC. Interestingly, all curves project to the same value of <*LW*> at <*N*>=1. This is intriguing, as it would suggest that the (extrapolated) average size of the monolayer is identical in all cases and thus independent on the exfoliation conditions. The same behaviour is observed for $WS_2$ (figure S5). One could therefore consider this value to be an interesting descriptor to evaluate the exfoliation quality across different materials. This concept will be discussed in detail below. Thus to summarise this section, while exfoliation yield, i.e. quantity depends strongly on sonication conditions (e.g. environment, power), nanosheet aspect ratios depend much more weakly on exfoliation conditions.

*Materials comparison*

In order to compare liquid exfoliation among various materials, we selected 12 layered van der Waals crystals with a range of structures, chemical compositions, crystallite shape and inter- and intra-layer bonding strengths: graphite, four transition metal dichalcogenides ($WS_2$, $MoS_2$, $MoSe_2$, $PtSe_2$), hexagonal boron nitride, a post-transition metal chalcogenide with metal-metal bonds (GaS, which is interpreted here as layered $Ga_2S_2$), a complex layered silicate (talc) and four layered hydroxides ($Mg(OH)_2$, $Ni(OH)_2$, $Co(OH)_2$, and $Zn(OH)_2$). In most cases, liquid phase exfoliation by sonication has previously been demonstrated and yields dispersions of nanosheets with unaltered chemical composition.[13, 19, 22-28] Here, in all cases but GaS, the crystals were exfoliated in aqueous sodium cholate by tip sonication according to established procedures (see methods). Since GaS is prone to oxidation, it was exfoliated in the solvent NMP using bath sonication.[23]

For each material, we isolated 4-6 size-selected fractions by LCC. In a few cases ($WS_2$, $MoS_2$), the size selection and centrifugation procedure was repeated under slightly varying conditions to confirm the robustness of the approach. Nanosheet dimensions were determined by AFM statistics in all cases as described above (SI, figures S6-22). The scaling of mean nanosheet dimensions ($<L>$, $<W>$, $<N>$) with central acceleration (the midpoint of the pair of centrifugal accelerations used during preparation of each fraction) is shown in figure 4 for representative materials (all data see Fig. S23-29). It is clear that nanosheet length ($<L>$, Fig. 4A), width ($<W>$, Fig. 4B) and layer number ($<N>$, Fig. 4C) decrease as power laws with increasing centrifugal acceleration in all cases. This also implies that nanosheet dimensions (including monolayer content) scale with each other, as shown in the SI (Fig. S30). Figure 4 demonstrates that, depending on the material, different lateral dimensions and layer numbers are accessible. Since the mean dimensions of the nanosheets in the fractions reflect the population in the stock dispersions, this means that sonication produces nanosheets of different length-scales and layer numbers, depending on the material. For example, the lateral dimensions of GaS are only slightly smaller than those of graphene (i.e., the data points of graphene and GaS in figure 4A and 4B sit relatively close together), but the nanosheets are significantly thicker (Fig. 4C). In contrast, the $WS_2$ nanosheets isolated in the fractions are significantly smaller than graphene (Fig. 4A and 4B), but have a comparable thickness (Fig. 4C).

The data shown in Fig. 4 (see also SI Fig. S30) implies that the correlation between the thickness and lateral size observed for liquid-exfoliated graphene applies to a wide range of nanosheet types. To show this, we calculate $\langle LW \rangle$ and $\langle N \rangle$ for each fraction as plotted for all 12 materials in Fig. 5A (and in the SI, Fig. S31). In all cases, $\langle LW \rangle$ clearly scales with $\langle N \rangle$, with data from within the same class of materials (e.g. TMDs, hydroxides) sitting close beside each other (see also SI, Fig. S32). Interestingly, there is a different offset in the data for the different material classes. This means that, depending on the material in question, the lateral dimensions achievable by sonication-based LPE for a given thickness vary significantly. For example, graphene nanosheets with a mean area <LW> of 0.01 $\mu m^2$ are 2-3 layers thick on average, while the hydroxides with similar areas have a mean layer number of 20-25. Thus, the $<LW>$ vs $<N>$ data allows us to quantify the exfoliation quality for a given material.

To do this, we note that, in all materials, we can nicely fit an empirical power-law scaling of $\langle LW \rangle$ with $\langle N \rangle$ (SI, figure S31), which we write as:

$$\langle LW \rangle = D_{ML}^{2} \langle N \rangle^{\beta} \tag{1}$$

Here, $D_{ML}$ represents the characteristic lateral nanosheet size associated with monolayers (i.e., when $\langle N \rangle = 1$) and is a measure of exfoliation quality (high $D_{ML}$ is consistent with larger, thinner nanosheets). From fitting the data in figure 5A, we find $\beta$ to lie in the range 2-3 while $D_{ML}$ tends to fall between ~0.5 and ~40 nm (Fig. 5B).

One would naturally expect the experimentally-observed exfoliation quality (i.e. represented by $D_{ML}$) to reflect the strength of the interlayer binding energy. To test this, we have computed from quantum-mechanical first principles the inter-monolayer binding energy [J/m$^2$], $E_S$, from the difference between the ground-state total energy of the optimized 3D bulk structure and that of its isolated monolayers.[29] In figure 5B, we plot $D_{ML}$ versus $E_S$, finding a reasonable correlation, albeit with considerable scatter. This suggests that, while the interlayer binding energy clearly plays an important role in defining the exfoliation quality, it may not be the only contributing parameter. This is a key insight that will prove important in explaining the observed nanosheet sizes, as we will discuss.

Ultrasonication is a relatively high-energy process that is known to result in sonication-induced scission during nanosheet exfoliation.[30-32] While scission is usually considered as an inconvenience that reduces flake size, we will demonstrate that it is a critical component in the determination of nanosheet dimensions. Indeed, in many cases, nanosheets with sizes of hundreds of nanometres are exfoliated from micron-sized layered crystallites, showing that scission must occur during the exfoliation process. This implies that the intra-layer bonding strength should also play a role in determining nanosheet size. One would expect high intra-layer bonding strength (i.e., high tensile strength) to result in larger nanosheets as more energy would be required to cut them to smaller sizes. Qualitatively, this could explain the scatter in figure 5C. For example the conjugated carbon-carbon bonds in the graphene lattice are the reason for its extraordinary material strength, which would explain why larger graphene nanosheets are obtained in spite of its inter-layer binding energies being similar to those of GaS which is known to be considerably weaker then graphene.[33]

*Model development*

We will next build a minimal model that will help us to understand the experimentally-observed behaviour. In what follows, we will make a series of reasonable assumptions and approximations in an effort to reduce the complexity of what is, in its full extent, a formidable problem in non-equilibrium, quantum-mechanical statistical mechanics.

We may picture liquid-phase exfoliation as a violent process occurring at the nanoscale, whereby incident shock-waves of sufficiently high energy lead to the removal of small nanosheets from large, layered crystallites.[34] The process is thus characterized by rare, irreversible events that result in the breakage of both inter-layer van der Waals bonds and intra-layer chemical bonds.[34, 35] In general, the sample undergoing liquid phase exfoliation by means of external excitation (such as sonication) may be considered to be in a quasi-steady state over a suitable period of time (insofar as significant quantities of parent crystallites remain), but it is out of thermodynamic equilibrium. This lack of equilibrium, not to mention the fact that bond-breaking at nanosheet edges is an inherently quantum-mechanical but nonetheless high-energy (compared to the average thermal energy) process, implies that we ordinarily cannot appeal to basic thermodynamic principles such as the equipartition of energy between separable degrees of freedom and, indeed, even the concept of a global temperature breaks down out of equilibrium.[34, 35]

One way forward is to note that the hypothesis of ergodicity (simplistically, that averages over large sample numbers or long times give the same results[34, 35]) can reasonably be applied to the aggregated degrees of freedom involved in edge tearing (for nanosheets with a given edge geometric area) and, separately, surface delamination (with a given surface geometric area). Firstly, we assume that the tearing energy, $E_{Tearing}$, required to break enough intra-layer chemical bonds to remove a nanosheet from its parent crystallite is independent of the sheet shape, and is the product of the created edge geometric area, $A_E$, and the energy per unit area required to create edges, $E_E$. Similarly, we assume that the pealing or surface delamination (exfoliation) energy, $E_{Pealing}$, is the product of the new surface area, $A_S$, and the energy per unit area, $E_S$, required to peel a sheet from its parent crystal (i.e., the destruction of a weakly-bonded interface, and the creation of two exposed surfaces). Secondly, and bringing in ergodicity, we may suppose that if a sufficiently large sample is taken (over time or volume, equivalently, assuming steady-state conditions macroscopically), then all nanosheet face shapes would be represented with an equal probability.

As an aside, but one relevant to our first-principles results presented later, we argue that $E_E$ and $E_S$ comprise only the energies required to break in-plane chemical and out-of-plane van der Waals bonds, respectively, including local charge and ionic reorganisation but *not* solvation effects. To see this, we note that, were the system allowed to come to equilibrium (i.e. sonication being switched off), transition-state theory predicts that the relevant energies are the net energy changes in going from an initial state (unexfoliated crystal) to a final state (remaining crystallite plus solvated, exfoliated nanosheets), with an energy barrier controlling the kinetics. However, in the out-of-equilibrium conditions relevant to sonication, the rare and violent events that bring the system from the initial state to the transition state are reaction-limiting. As a result, we understand $E_E$ and $E_S$ to be properties associated only with nanosheet mechanics. While exfoliation yields are linked to solvent-nanosheet interactions as previously reported[36] and shown above, such interactions govern the stabilisation of exfoliated nanosheets rather than the actual exfoliation process.[37] In other words, as clearly demonstrated by the direct comparison of graphite and $WS_2$ exfoliated in aqueous surfactant solution and NMP, the relationship between lateral size and layer number is largely independent of the medium chosen for LPE while nanosheet yield is not.

Simplifying matters further is the observation that neither tearing nor delamination may occur without the other, if the net result is to be a free flake in solution. Both processes occur simultaneously during events that deliver sufficient energy to remove a nanosheet from the parent crystal. Still assuming that the relevant energies do not explicitly depend on sheet shape, we may further suppose that the tearing and delamination energies are, on average, disbursed in a fixed ratio, *a*, a factor which reflects the microscopic details of the exfoliation process. This gives rise to an assumed quasi-equipartition of energies for the aggregated degrees of freedom responsible for edge tearing and surface delamination. We can express this hypothesis as

$$\left\langle E_{Pealing} \right\rangle = a \times \left\langle E_{Tearing} \right\rangle \tag{2a}$$

where $\left\langle E_{Pealing} \right\rangle$ is the average contribution to the exfoliation energy associated with pealing (or delamination) of a nanosheet from its parent crystal. Similarly, $\left\langle E_{Tearing} \right\rangle$ is the average contribution associated with the breakage of in-plane bonds during nanosheet removal. In the case of perfect equipartition, we would expect $a$=1.

We can express equation 2a in terms of the pealing and tearing energies, $E_S$ and $E_E$, as well as nanosheet dimensions:

$$2\left\langle A_S \right\rangle E_S = 2aE_E \left\langle A_E \right\rangle / 2 \tag{2b}$$

The factor of 2 on the left hand side comes from the fact that there are two surfaces created when removing a nanosheet from the layered crystal, i.e. on the nanosheet and on the parent crystallite. Furthermore, we note that while some nanosheets will need to have their entire perimeter created in an exfoliation event (the case of sheets being removed from the centre of the surface of the parent crystal), others will require only a very small amount of edge to be created (for example, sections connected to the main crystal by only a thin neck). Thus, on average, approximately half of a nanosheet perimeter will be newly formed during an exfoliation event, resulting in the factor of 2 in the denominator on the right hand side. However, the equivalent amount of edge must also be formed on the parent crystal, resulting in the factor of 2 in the numerator on the right hand side.

The edge area is just the product of nanosheet perimeter, $P$, and thickness, $h_0N$, where $h_0$ is the monolayer thickness (the parent crystallite out-of-plane unit-cell height, or the fraction of that unit-cell height associated with a single monolayer), allowing us to write

$$\langle A_S \rangle = \frac{ah_0}{2}\frac{E_E}{E_S}\langle PN \rangle \qquad (3)$$

In principle, AFM images can be analysed automatically using appropriate software[34] on a sheet-by-sheet basis to obtain $\langle A_S \rangle$ and $\langle PN \rangle$ in order to test this equation. In practice, however, for the small nanosheet sizes associated with LPE nanosheets, residual solvent and aggregated sheets make automated analysis challenging. This means that manual analysis, such as that employed in this work, results in a rather limited data set consisting of length, width and thickness data for each fraction. Within a given fraction, the $L$, $W$, and $N$ distributions are reasonably narrow (at least compared to the stock). In order to obtain averaged nanosheet area ($A_S$) and perimeter ($P$) from length ($L$) and width ($W$) data, we approximate the nanosheets as having a single fixed shape. This approximation is necessary even though nanosheet morphology is quite diverse with a large variety of nanosheet shapes present.

Although it appears more obvious to approximate the characteristic nanosheet shape as rectangular or diamond-like, here we approximate the nanosheets as elliptical. This choice is not based on an attempt to best match the observed shapes. Rather, it is an approximation designed to simplify the mathematics of the model described below (the simplification is based on the fact that within the approximation for perimeter given below, the area of a low-aspect-ratio ellipse is proportional to the square of its perimeter. This factor allows the model to be developed without any additional assumptions. See SI section 1.9). The area of an ellipse is given by the product of the semi-major and semi-minor axes. Using our notation for nanosheets, this yields

$$\langle A_S \rangle = \frac{\pi}{4}\langle LW \rangle . \tag{4}$$

Next, we must address the nanosheet perimeter *P*. For ellipses with aspect ratios sufficiently close to unity, and given the approximate nature of the ellipse assumption to begin with, it is sufficient to use a low-order approximation for the ellipse perimeter, involving the geometric mean of *L* and *W* in place of the diameter of a circle. As long as the aspect ratio of each flake is not too high, we may write

$$\langle PN \rangle = \left\langle \pi\sqrt{LW}N \right\rangle = \pi\left\langle N\sqrt{LW} \right\rangle \tag{5}$$

Combining equations 3, 4 and 5:

$$\langle LW \rangle = 2ah_0\frac{E_E}{E_S}\left\langle N\sqrt{LW} \right\rangle \tag{6}$$

Any uniform deviation from the low-aspect ratio ellipse shape assumption for the newly-created edge, such as due to roughness, is expected to primarily be hidden in the factor *a*. In order to manipulate these multi-parameter means, we make the assumption that our *L*, *W*, and *N* distribution data are consistent with:

$$\left\langle N\sqrt{LW} \right\rangle = \sqrt{\langle LW \rangle}\langle N \rangle^b \tag{7}$$

Here, the exponent *b* is expected to be close to 1 but will depend on the details of the *LW* and *N* distributions. As shown in figure S33-34 (SI), this expression applies well to the materials studied, with values of *b* found to be mostly close to 1 and always <1.5. It is worth emphasising here that the factor <N> in this work is that of size selected fractions (see Fig. S2). In principle, we can envisage no reason why our expressions should not remain valid if the nanosheets are fractionated in a different way.

Equation 7 allows us to simplify and re-cast equation 6 into the convenient form

$$\langle LW \rangle = \left(2ah_0\frac{E_E}{E_S}\right)^2\langle N \rangle^{2b} \tag{8}$$

This equation gives a theoretical expression for the relationship between nanosheet area (represented by $\langle LW \rangle$) and thickness (represented by $\langle N \rangle$). Importantly, it has the same form as the empirical expression used to fit the experimental data in figure 5A (equation 1). The fit quality shows equation 8 to be consistent with experiments across a diverse spectrum of chemical species and bonding types. We note that, while the observed power-law dependence of $\langle LW \rangle$ on $\langle N \rangle$ cannot conclusively prove the validity of our quasi-equipartition hypothesis, it strongly supports it. Indeed, it seems difficult to envisage how such a power law could come about given an absence of systematic exfoliation-energy partitioning.

*Comparing model and experiment*

Assuming this model does indeed describe the data, then equation 8 should be equivalent to the empirical relationship represented by equation 1. Comparing equations 1 and 8 shows that $\beta=2b$ and:

$$\frac{D_{ML}}{h_0} = 2a\frac{E_E}{E_S} \tag{9}$$

We first address the relationship between exponents. The $\beta$-exponents can be extracted by fitting the data in figure 5A using equation 1 and are given in figure 5B. The *b*-exponents can be found by fitting the statistical nanosheet size data using equation 7 as shown in figure S33. Any systematic non-constancy of the equipartition factor *a* in the form of a power-law dependence on <N> will inevitably be hidden in the exponent *b*. As shown in figure S34 however, the data is reasonably close to $\beta=2b$, in line with our model.

Equation 9 is very interesting, as it allows us to test the validity of our model by plotting $D_{ML}/h_0$ versus $E_E/E_S$. Here, $D_{ML}$ is obtained from fitting experimental data (such as Fig. 5A) while $h_0$ is the interlayer distance obtained from published crystallographic lattice constants (see SI, table S2) and checked for consistency against first-principles density-functional theory calculations (SI section 3), which agree very well. In contrast, although neither $E_E$ or $E_S$ are readily available for all materials, both values (or their proxies) can be computed in various ways. Assuming that data reflecting $E_E/E_S$ were available for a range of materials, a straight-line relationship between $D_{ML}/h_0$ and $E_E/E_S$ would strongly support our model and allow for the estimation of *a*.

To test this model, the $E_S$ values described above (the interlayer binding energy) should be combined with an estimated in-plane bonding energy ($E_E$) for each material to obtain a proxy of $E_E/E_S$. One of the most computationally inexpensive ways to estimate $E_E$ is by using the integral crystal orbital Hamiltonian population (ICOHP) based on a Kohn-Sham Hamiltonian of approximate density-functional theory.[38] Shown in figure 5D is a plot of $D_{ML} / h_0$ versus $E_{E,ICOHP} / E_S$. We find reasonable linearity, albeit with some scatter. Fitting the data to equation 9 gives a value of a=0.8±0.1.

However, $E_{E,ICOHP}$ is expected to be a crude approximation for $E_E$. It would be useful to have a proxy for $E_E / E_S$ which can be calculated to a reasonable degree of accuracy. Very recently, Ji et al. proposed that exfoliation quality scales with the ratio of the in-plane to out-of-plane Young's moduli of the layered material: $Y_{In-plane} / Y_{Out-of-plane}$.[39] Although this proposal was originally made without the any rigorous theoretical support, here we use a simple model to demonstrate that $E_E / E_S$ and $Y_{In-plane} / Y_{Out-of-plane}$ are roughly equal.

To achieve this (see SI section 4), we model the dependence of the out-of-plane interlayer interaction energy ($E_{OOP}$) on the inter-layer distance, $r_{IL}$, using a Lennard-Jones-like potential with exponents $a$ and $b$:

$$E_{OOP}(r_{IL}) = \varepsilon_{vdW}\left[\frac{b}{a-b}\left(\frac{r_{vdW}}{r_{IL}}\right)^a - \frac{a}{a-b}\left(\frac{r_{vdW}}{r_{IL}}\right)^b\right] \tag{10a}$$

where $r_{vdW}$ is the equilibrium inter-layer separation and $\varepsilon_{vdW}$ is the binding curve well depth. Furthermore, we model the in-plane bond energy, $E_{IP}$, versus inter-atomic separation, $r_{IA}$, using a similar function

$$E_{IP}(r_{IA}) = \varepsilon_{bond}\left[\frac{\beta}{\alpha-\beta}\left(\frac{r_{bond}}{r_{IA}}\right)^\alpha - \frac{\alpha}{\alpha-\beta}\left(\frac{r_{bond}}{r_{IA}}\right)^\beta\right] \tag{10b}$$

Here $r_{bond}$ is the equilibrium inter-atomic separation and $\varepsilon_{bond}$ is the binding curve well depth. We justify using this function by noting that it is very similar in shape to the well-known Morse potential which is widely used to model bond potentials.[40] By calculating the associated spring constants from the second derivatives of the equations 10a and 10b, it is possible to show that (see SI section 4):

$$\frac{Y_{In-plane}}{Y_{Out-of-plane}} \approx \frac{\alpha\beta r_{vdW}}{abl_{bond}}\frac{E_E}{E_S} \tag{11}$$

As shown in the SI (section 4), by fitting DFT data, we argue that, at least for graphene, $\alpha\beta r_{vdW} \approx abl_{bond}$, making the modulus ratio a very good proxy for $E_E / E_S$.

The combination of equations 9 and 11 show that we would expect $D_{ML} / h_0$ to scale linearly with $Y_{In-plane} / Y_{Out-of-plane}$. To test this, the in-plane and out-of-plane moduli were calculated from first principles, as detailed in the methods section. Briefly, the in-plane Young's modulus for each material was estimated from its calculated elastic coefficients $c_{ij}$ using the formula

$$Y_{In\text{-}plane} = (c_{11}c_{22} - c_{12}c_{12}) / \sqrt{c_{11}c_{22}} \tag{12}$$

that results from applying symmetry considerations to the two-dimensional Reuss average as discussed in detail in Ref. [41] and its supplemental information. This modulus is more generally computed as the inverse of the average of the compliance over in-plane angles. It assumes uniform stress rather than uniform strain conditions, as is more appropriate to the simulation of exfoliation. The coupling between the in-plane and out-of-plane degrees of freedom is neglected, i.e., elements such as $c_{13}$. For all but the case of talc, where the above formula is a numerically small approximation, we have $c_{11} = c_{22}$ and $c_{66} = (c_{11} - c_{22}) / 2$ and the in-plane compliance becomes isotropic. This simple formula then recovers the two-dimensional Reuss Young's modulus exactly. The corresponding out-of-plane modulus is simply the out-of-plane elastic coefficient, $Y_{Out\text{-}of\text{-}plane} = c_{33}$. So, effectively we apply the approximation that

$$\frac{E_E}{E_S} \approx \frac{Y_{In-plane}}{Y_{Out-of-plane}} = \frac{(c_{11}c_{22} - c_{12}c_{12})}{c_{33}\sqrt{c_{11}c_{22}}} \tag{13}$$

We note that $Y_{In-plane} / Y_{Out-of-plane}$ falls below 1 for the transition-metal hydroxides due to their anomalously high two-dimensional Poisson's ratio (~85% surface area conservation is predicted, assuming uniform in-plane stress).

In figure 4D, we plot $D_{ML} / h_0$ versus $Y_{In-plane} / Y_{Out-of-plane}$, finding clear linearity and much less scatter than in figure 5C. Assuming that $\alpha\beta r_{vdW} \approx abl_{bond}$ holds generally such that $Y_{In-plane} / Y_{Out-of-plane} \approx E_E / E_S$ can be applied to all data, fitting yields a value of a=1.0±0.1. As outlined above, *a*=1 is the hallmark of equipartition of energy between pealing and tearing during the exfoliation process.

The observation that our combined experimental data and theoretical modelling gives *a*-values that are reasonably close to 1 is an important result may begin to shed light on the details of the exfoliation mechanised. However, it should be noted that there is some scope for uncertainty here due to possible cancellation of errors associated with factors such as: the assumption that the nanosheets resemble low-aspect ellipses on average; the numerous assumptions invoked in using ICOHP or the in-and-out-of-plane Young's moduli; and the assumption of negligible <*N*>-dependence in *a* and indeed in the energy densities (which we have calculated always in the bulk limit). Nevertheless, taken together our results provide strong evidence supporting our hypothesis that quasi-equipartition holds, at least, particularly as it explains the observed power law of equation 1.

Finally, it is interesting to note that equation 8 can be rearranged to reflect the lateral size/thickness aspect ratio, *k*, of arbitrarily thick nanosheets:

$$k = \frac{\sqrt{\langle LW \rangle}}{\langle N \rangle h_0} = 2a\frac{E_E}{E_S}\langle N \rangle^{b-1} \tag{14}$$

Assuming that $a \approx 1$ and in the ideal case $b=1$, we find that *k* loses its <*N*>-dependence and becomes $k \approx 2E_E / E_S$, so that this is a quantity set by fundamental material parameters. This is an important result that suggests fundamental limitations of LPE nanosheets produced by means of sonication for applications such as mechanics, where high aspect ratios are important.

## CONCLUSIONS

In summary, we have measured the length, width and thickness distributions of 12 different liquid-exfoliated 2D materials. We have found clear correlations between nanosheet area and thickness, allowing us to propose a metric for exfoliation quality. By developing a model for nanosheet size based on non-equilibrium thermodynamics, we give a theoretical explanation for the observed scaling and link the exfoliation quality metric to the ratio of basal-plane to edge formation energies. Using a simple model we show that this energy ratio is close to the ratio of in-plane to out-of-plane nanosheet moduli. By computing these moduli, we find the model to be completely consistent with the data. Comparing data with theory strongly suggests that a generalized energy equipartition holds, on average, between nanosheet tearing and peeling during sonication, providing valuable insight into the basis physics of exfoliation.

## METHODS

### *Materials*

β-Nickel Hydroxide powder (>95% 283662), Magnesium Hydroxide (95% 310093), Cobalt hydroxide (342440), Zinc hydroxide (96466), Talc (243604), Boron Nitride (255475),

Graphite (332461-2.5 kg), $WS_2$ (C1254), $MoS_2$ (69860) and Sodium Cholate were purchased from Sigma Aldrich. Gallium Sulphide was purchased from American Elements (99.999% GaS-05-P), $MoSe_2$ (13112.14) from VWR. Selenium (99.999% and ammonium hexachloroplatinate (99.99%) were obtained from STREM. Platinum sponge was prepared by thermal decomposition of ammonium hexachloroplatinate in hydrogen at 500 °C for 1 hour. De-ionized water was prepared in house, and solvents (*N*-methyl-2-pyrrolidone and 2-propanol) used were purchased with the highest available purity.

Synthesis $PtSe_2$: Platinum diselenide was prepared by direct reaction of the elements. Stoichiometric amounts of selenium and platinum sponge corresponding to 5 g of $PtSe_2$ was placed in a quartz ampoule and melt-sealed under high vacuum. The reaction mixture was heated to 1,260 °C for one hour with a heating and cooling rate of 5 °C/min.

*Exfoliation*

Graphite, group VI-TMD and BN dispersions were prepared by probe sonicating the powder with an initial concentration 20 $gL^{-1}$ in an aqueous sodium cholate (SC) solution. The powder was immersed in 80 mL of aqueous surfactant solution ($C_{SC}$= 6g/L). The mixture was sonicated under ice-cooling in a 100 mL metal beaker by probe sonication using a solid flathead tip (Sonics VXC-500, *i.e. 500* W) for 1 h at 60 % amplitude with a pulse of 6 s on and 2 s off. During the sonication, the sonic probe was placed 1.5 cm from the bottom of the beaker. The dispersion was centrifuged in 20 mL aliquots using 50 mL vials in a Hettich Mikro 220R centrifuge equipped with a fixed-angle rotor 1016 at 2,660 *g* for 1.5 h. The supernatant was discarded and the sediment collected in 80 mL of fresh surfactant ($C_{SC}$= 2 $gL^{-1}$) and subjected to a second sonication using a solid flathead tip (Sonics VX-500) for 5 h at 60 % amplitude with a pulse of 6 s on and 2 s off. From our experience, this two-step sonication procedure yields a higher concentration of exfoliated material and removes impurities. Graphite exfoliation in *N*-methyl-2-pyrrolidone was performed under identical conditions except for the solvent.

For the hydroxides, 1.6 g of powder was pre-treated by sonicating using a sonic tip in 80 mL deionised water for 1 h. The dispersion was then centrifuged at 2,150 *g* for 1 h and decanted with the sediment being retained and dried. The pre-treated material (20 $gL^{-1}$) was then sonicated in 9 g/L of sodium cholate and de-ionized water solution using a flat head tip (Sonics VCX-750) with 60% amplitude and 6s on/ 2s off for 4 h under ice cooling.

The $PtSe_2$ crystal (0.5 $gL^{-1}$) was immersed in 35 mL of aqueous sodium cholate (SC) solution ($C_{surf}$ = 1.7 $gL^{-1}$). The mixture was sonicated under cooling in a metal beaker by probe sonication using a solid horn probe tip (Sonics VX-750) for 7.5 h at 30% amplitude with a pulse of 6 s on and 4 s off.

Gallium sulfide powder (45 $gL^{-1}$) was sonicated in *N*-methyl-2-pyrrolidone using an ultrasonic bath (P30 H Ultrasonic from Fischer scientific). The sonication was performed for 6 h with an amplitude of 100% and a frequency of 37 kHz in 50 mL plastic centrifuge tubes. The water in the sonic bath was cooled by a water cooling system to maintain a temperature below 30°C enabled by cold water being pumped through piping which was wrapped around the interior of the bath.

For exfoliation of graphite in aqueous sodium cholate by bath sonication, two vials each containing 40 mL of dispersed bulk graphite material in SC solution (8 $gL^{-1}$) were positioned in hot spots of a Branson CPX3800 sonication bath. After a sonication time of 1 h for the purification step, the dispersion was centrifuged at 2,660 *g* for 1.5 h, the impurity rich supernatant discarded. The sediment was collected in fresh SC solution (2 $gL^{-1}$) for the second exfoliation step with a sonication time of 5 h. During the bath sonication, the bath water was replaced every 30 min with new water to avoid overheating.

*Size selection*

To select nanosheets by size, we used liquid cascade centrifugation with sequentially increasing rotation speeds. Centrifugation conditions are expressed as relative centrifugal field (RCF) in units of $10^3$ *x g* (or k *g*) with *g* being the gravitational force. Two different centrifuges were used: For centrifugal accelerations < 3,000 *g*, a Hettich Mikro 220R centrifuge equipped with a fixed-angle rotor 1195-A was used; above 3,000 *g*, a Beckman Coulter Avanti XP centrifuge with a JA25.50 fixed angle rotor was used. Graphite, BN, TMDs: All centrifugation runs were performed for 2 h (10°C). Unexfoliated material was removed by centrifugation at 100 *g*. The supernatant was subjected to further centrifugation at 400 *g*. The sediment was collected in fresh surfactant ($C_{SC}$= 0.1 $gL^{-1}$) at reduced volume (3-8 mL), while the supernatant was centrifuged at 1,000 *g*. Again, the sediment was collected and the supernatant subjected to centrifugation at higher speeds. This procedure was repeated with the following RCF: 5k *g*, 10k *g*, 22k *g*, 74k *g*. As sample nomenclature, the lower and upper boundary of the centrifugation are indicated. For graphite exfoliated in NMP, centrifugation was performed for 3.5 h to balance the slower sedimentation rate in the higher viscosity solvent. Since fewer small/thin nanosheets are produced, steps with lower centrifugal acceleration were included with the following overall cascade: 30 *g*, 100 *g*, 400 *g*, 1k *g*, 3k *g*, 6k *g*.

For $PtSe_2$, centrifugation was performed for 2 h in each step. Unexfoliated material was first removed by centrifugation at 100 *g* (50 mL vial). The supernatant was than subjected to further centrifugation at 400 *g* (50 mL vial). The sediment was collected in fresh SC-$H_2O$ (0.1 $gL^{-1}$), while the supernatant was centrifuged at 1,000 *g* (50 ml vial). Again, the sediment was collected, and the supernatant subjected to centrifugation at higher speeds. This procedure was repeated with the following speeds: 3k *g* (50 ml vial), 5k *g* (50 ml vial), 10k *g* (50 ml vial), 30k *g* (2x12 ml vial).

In the case of the hydroxide and talc dispersions, centrifugation parameters were adjusted: the stock obtained after sonication was centrifuged at 25 *g* for 60 min. The sediment was discarded and the supernatant was centrifuged at 100 *g* for 60 min. The sediment after this centrifugation step was redispersed in fresh surfactant solution (1 h bath sonication, $c_{SC}$=9 $gL^{-1}$) producing the largest size. The supernatant after the 100 *g* centrifugation step was centrifuged at 250 *g* for 60 min, producing the second largest size in the redispersed sediment. These steps were repeated in further increments of 400 *g*, 1k *g*, and 3k *g*, thus producing five sizes. For GaS, centrifugation was performed for 2 h in each step using 25 *g* (unexfoliated removed), 100 *g*, 400 *g*, 1k *g*, 3k *g*, 10k *g*. In this case, the sediment was redispersed in 2-propanol to facilitate deposition for AFM. The final supernatants were discarded in all cases.

*Characterisation*

Atomic force microscopy (AFM) was carried out on a Dimension ICON3 scanning probe microscope (Bruker AXS S.A.S.) in ScanAsyst in air under ambient conditions using aluminium coated silicon cantilevers (OLTESPA-R3). The concentrated dispersions were diluted with water (or 2-propanol in the case of GaS) to optical densities <0.1 across the resonant spectral region. A drop of the dilute dispersions (20 μL) was deposited on pre-heated (180 °C) Si/$SiO_2$ wafers (0.5×0.5 $cm^2$) with an oxide layer of 300 nm. After deposition, the wafers were rinsed with ~5 mL of water and ~5 mL of isopropanol. Typical image sizes ranged from 15×15 for larger nanosheets to 5×5 $\mu m^2$ at scan rates of 0.5-0.8 Hz with 1024 lines per image. Published values for step heights were used to convert apparent AFM thickness to layer number.[13, 19, 22-27] The step height analysis of $PtSe_2$ is shown in the SI (Fig. S14). Previously published length corrections were used to correct lateral dimensions from cantilever broadening.[42] A detailed description of the statistical analysis is provided in the SI (section 5).

The yield of the exfoliation in graphite was determined as follows: Extinction spectra were measured with a known dilution factor for each fraction on a Agilent Carry 6000i (quartz cuvettes, 0.4 cm pathlength). With the size-independent extinction coefficient of 5,450 $Lg^{-1}m^{-1}$ at 750 nm, the concentration of dispersed graphite was calculated.[43] The volume of each fraction was measured to calculate the mass and thus the yield through dividing by the initial mass of graphite (2.4 g).

*Calculation of binding energies*

All first-principles calculations were carried out using Kohn-Sham density-functional theory as implemented the Quantum Espresso package.[44] The binding energies for non-magnetic materials has been computed in [29] using SSSP efficiency v.07 pseudopotentials[45] and suggested cut-offs, a k-point density of 0.2 A^[27] and a Marzari-Vanderbilt cold smearing[46] of 0.02 Ry. The vdW-DF2-C09[47-49] van der Waals functional was used. The binding energy of magnetic materials was validated using the RVV10 functional[50-52] and by performing collinear spin-polarized calculations considering the magnetic ground state of the bulk structure and each isolated substructure. Despite this, the binding energies of magnetic materials $Ni(OH)_2$ and $Co(OH)_2$ are likely subject to larger errors due the difficulties of dealing with d electrons in standard DFT. The calculation of the integral crystal orbital Hamiltonian population (ICOHP) has been performed with the Lobster code[53] post-processing calculations carried out with PAW pseudopotentials[54] form the PSlibrary set.[46] An indication of the energy per unit area required to create edges, $E_E$ , has been obtained summing up the ICOHP energies of the minumum number of bonds per unit cell that have to be broken to create the edge divided by the layer thickness $h_0$.

The calculation of the integral crystal orbital Hamiltonian population (ICOHP) was performed using the Lobster code[55] post-processing calculations carried out with PAW pseudopotentials[54] from the PSlibrary set.[56] The data is summarised in SI section 3.

*Calculation of elastic constants*

Elastic constants were computed by finite deformations and exploiting the stress-strain relation as implemented in the ElaStic module[57] using 9 deformed structures for each symmetry-independent strain and a maximum strain of 0.004. Both the vdW-DF2-C09 and RVV10 functionals were used for non-magnetic structures, while only RVV10 has been used for magnetic materials due to the current limitations in the Quantum Espresso implementation. A refined k-point density of 0.05 A^{-1} was employed. Forces were relaxed down to 5.10^-5

Ry/a.u., while stress has been optimized up to a threshold of 0.05 kbar within structural optimizations. The data is summarised in SI section 3.

## Acknowledgements

We acknowledge the European Union under grant agreements n°785219 Graphene Flagship-core 2. C.B. acknowledges support from the German research foundation (DFG) under grant agreement Emmy-Noether, BA4856/2-1 and Jana Zaumseil for the access to the infrastructure at the Chair of Applied Physical Chemistry. Additional support was provided by Science Foundation Ireland (SFI/12/RC/2278) and the European Research Council Advanced Grant (FUTURE-PRINT). D.C. and N.M. acknowledge support from the H2020 MaX Centre of Excellence and the MARVEL NCCR. D.C. acknowledge support from the EPFL Fellows fellowship pro-gramme co-funded by Marie Sklodowska-Curie, Horizon 2020 grant agreement no. 665667. Simulation time was awarded PRACE on Marconi at Cineca, Italy (project id. 2016163963). Z.S. was supported by the Czech Science Foundation (GACR No. 17-11456S) and by the project Advanced Functional Nanorobots (reg. No. CZ.02.1.01/0.0/0.0/15_003/0000444 financed by the EFRR).

## Author contributions

C.B. designed the experiments, performed AFM and analyzed all data; D.C. and N.M. performed first-principles simulations; D.D.O'R calculated elastic moduli; B.M.S., K.S., E.O., F.R., A.H., A.G. prepared samples; B.M.S., K.S., performed AFM; Z.S. synthesized bulk $PtSe_2$; J.N.C. and D.D.O'R. developed the model, and C.B., J.N.C. and D.D.O'R wrote the manuscript.

## Additional information

Supplementary information is available in the online version of the paper. Correspondence and requests for materials should be addressed to C. B., D.D.O'R, or J.N.C.

## Competing financial interests

The authors declare no competing financial interests.

FIGURES

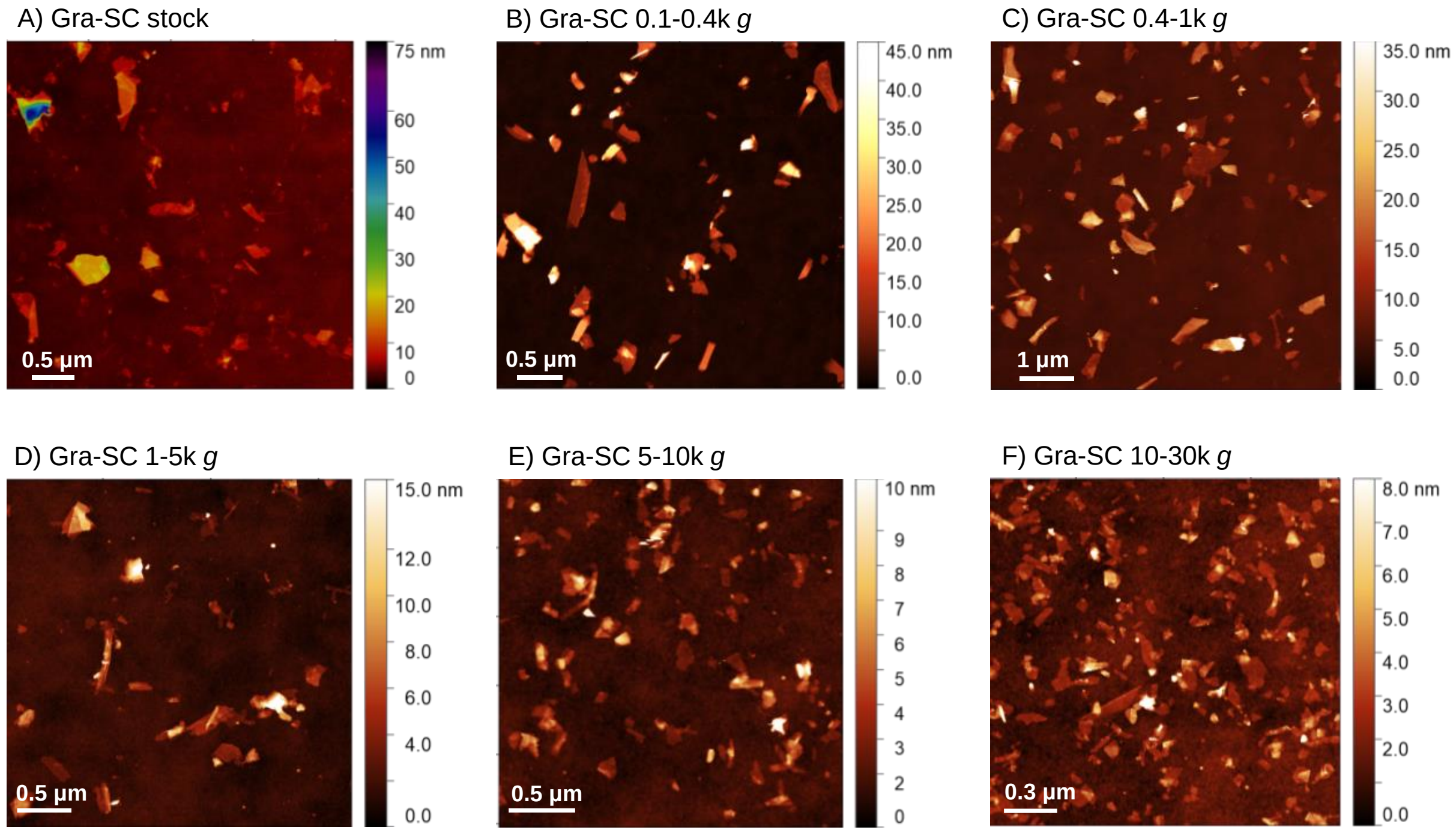

**Figure 1: AFM images of graphite exfoliated in aqueous SC solution by tip sonication both before and after size selection.** A) Unselected stock dispersion, B-F) fractions after liquid cascade centrifugation isolated from the stock dispersion shown in A. B) Fraction of largest nanosheets isolated by centrifugation between 100-400 *g*. C) 400-1,000 *g* fraction, D) 1-5k *g* fraction, E) 5-10k *g* fraction, F) Fraction of smallest nanosheets isolated by centrifugation between 10-30k *g*. Material remaining in the supernatant after centrifugation at 30k *g* was discarded.

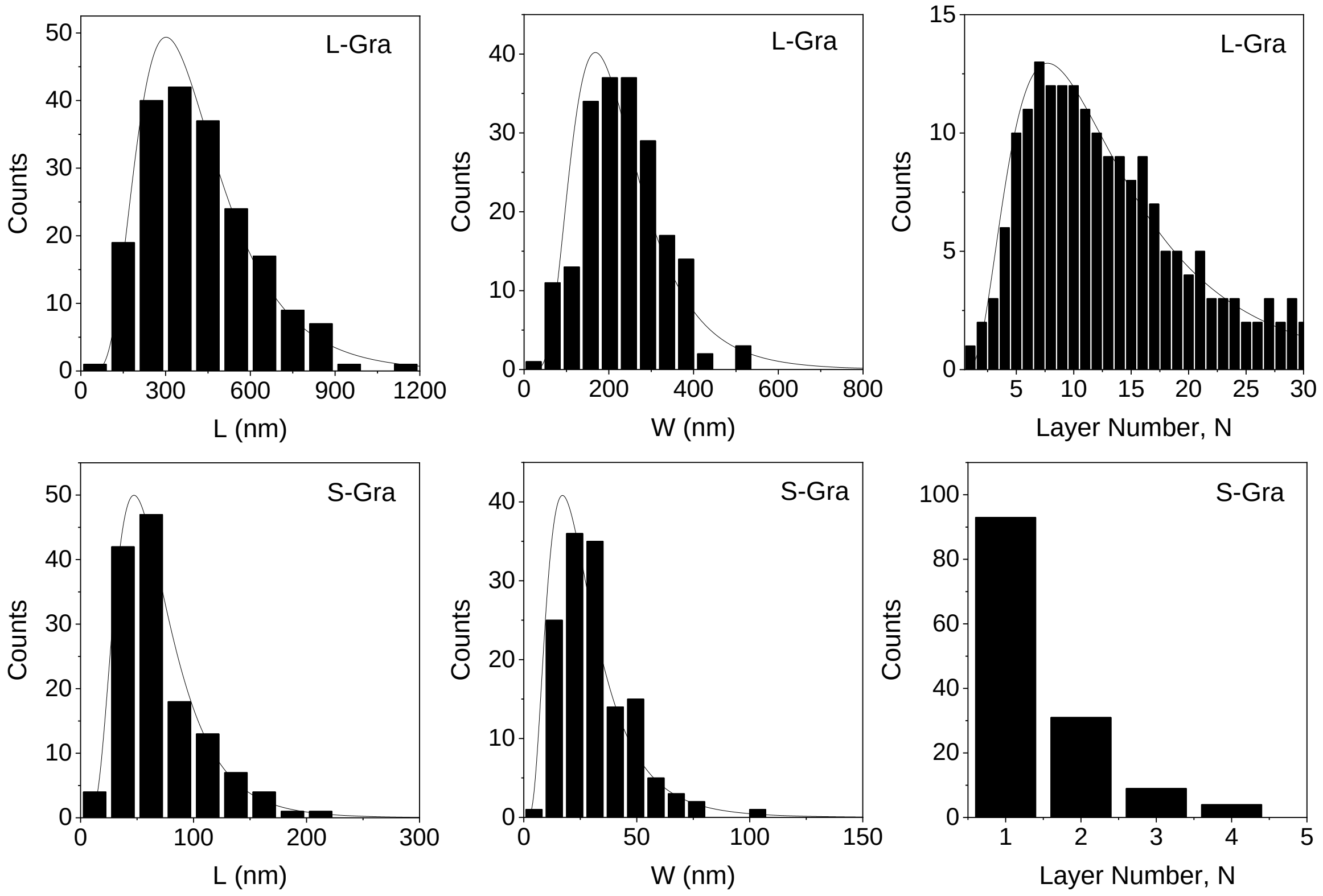

**Figure 2: Size distribution histograms determined from AFM statistics**. A-B) Longest lateral dimension, length L, for A) large nanosheets, B) small nanosheets. C-D) Dimension perpendicular to longest dimension, width W, for C) large nanosheets, D) small nanosheets. E-F) Nanosheet layer number, N, for E) large nanosheets, F) small nanosheets. For additional data, see SI. These particular distributions are for graphene nanosheets exfoliated in SC using a sonic tip. Large and small nanosheets were prepared using the centrifugation parameters 0.1-0.4krpm and 10-30krpm respectively. Solid lines are lognormal fits.

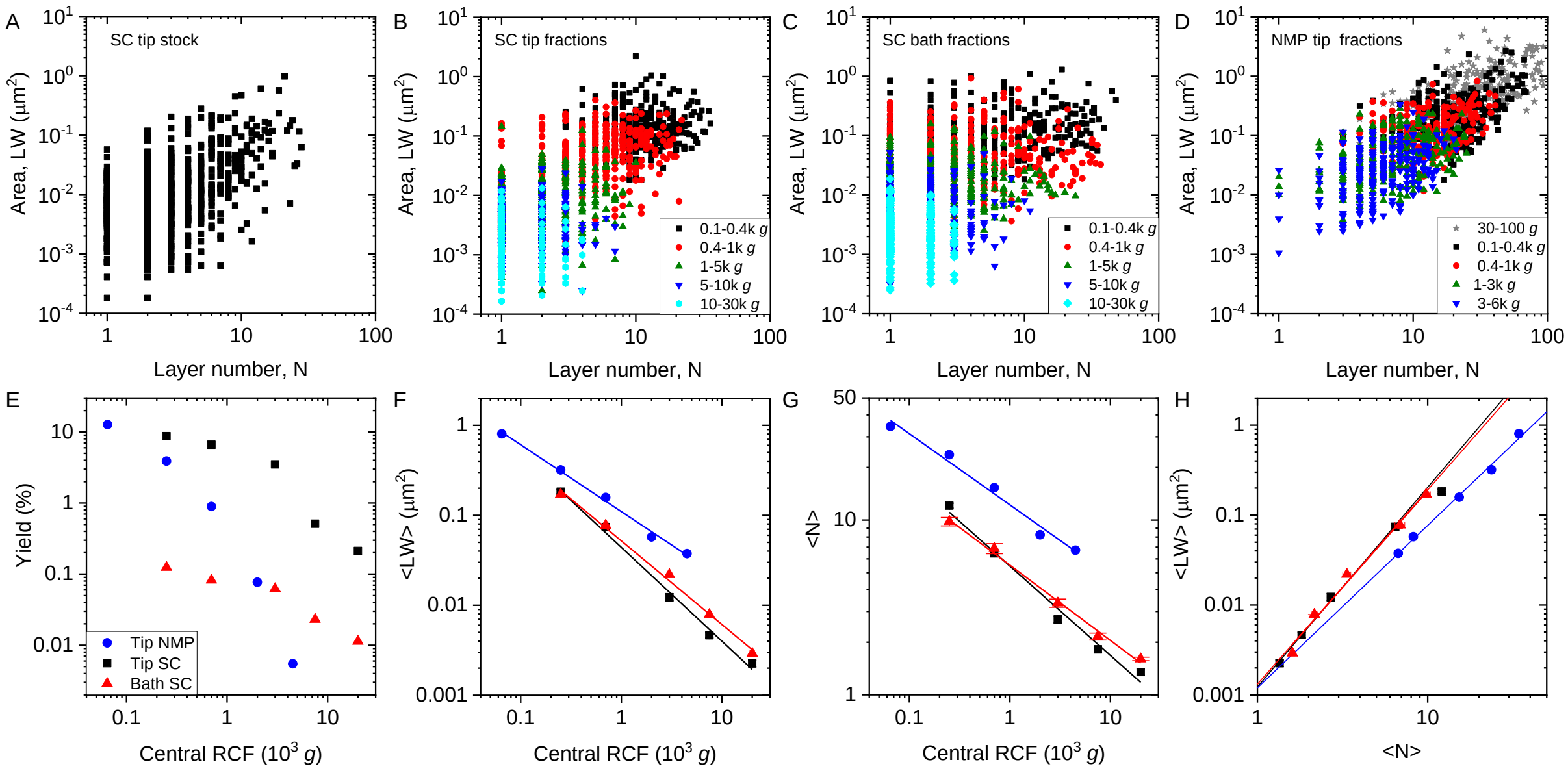

**Figure 3: Size analysis of graphite exfoliated using different conditions.** A-D) Scatter plots of nanosheet area (L×W) as function of layer number (N). Each data point represents an individual nanosheet. A) Stock dispersion after tip sonication in aqueous sodium cholate. B) Same sample as in (A) after size selection by liquid cascade centrifugation. The different fractions are color-coded. C) Nanosheet dimension data cloud for graphite exfoliated in aqueous sodium cholate by bath sonication. D) Nanosheet dimension data cloud for graphite exfoliated in NMP by tip sonication. E) Plot of nanosheet yield as function of midpoint of the pair of centrifugal accelerations used in the centrifugation cascade (central RCF). F) Plot of mean nanosheet area (<LW>) of the fractions as function of the central RCF. G) Plot of mean nanosheet layer number (<N>) as function of the central RCF. H) Plot of mean nanosheet area as function of layer number.

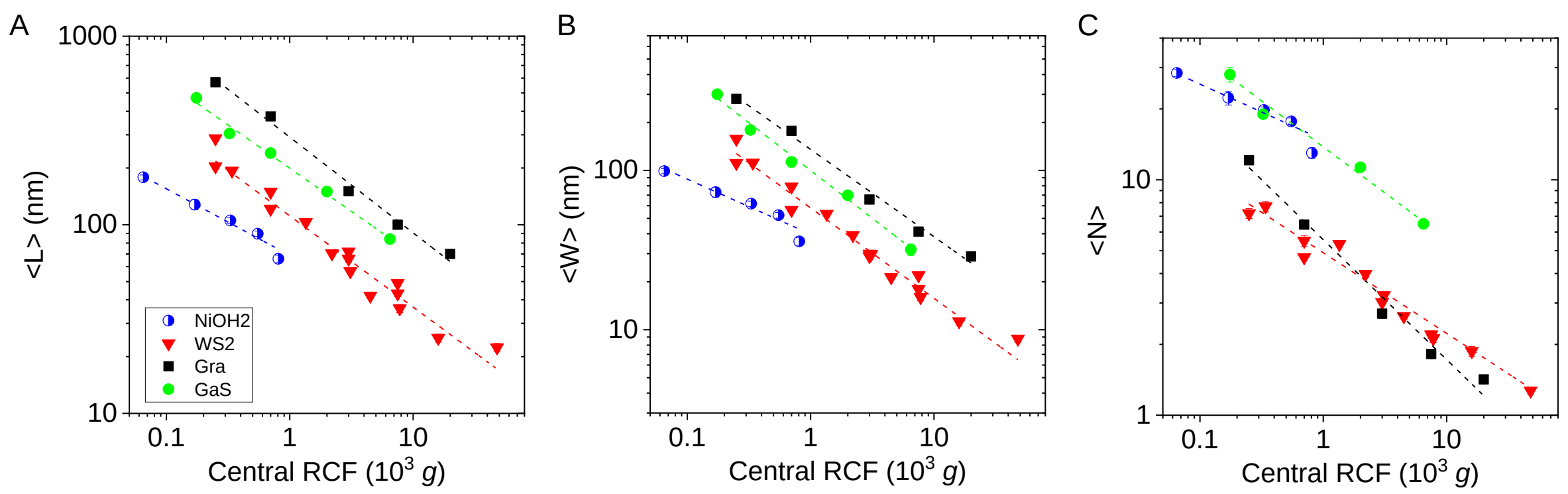

**Figure 4: Scaling of nanosheet dimension with centrifugal acceleration.** Plots of A) Mean nanosheet length, <*L*>, B) mean width, <*W*> and C) Mean layer number, <*N*> as function of the midpoint of the pair of centrifugal acceleration used for the size selection. Data for four representative materials under study are shown. For additional data, see SI. In all cases, the reduction in lateral dimensions and layer number with increasing centrifugal acceleration is evident. Lines are power law fits provided as guide for the eye.

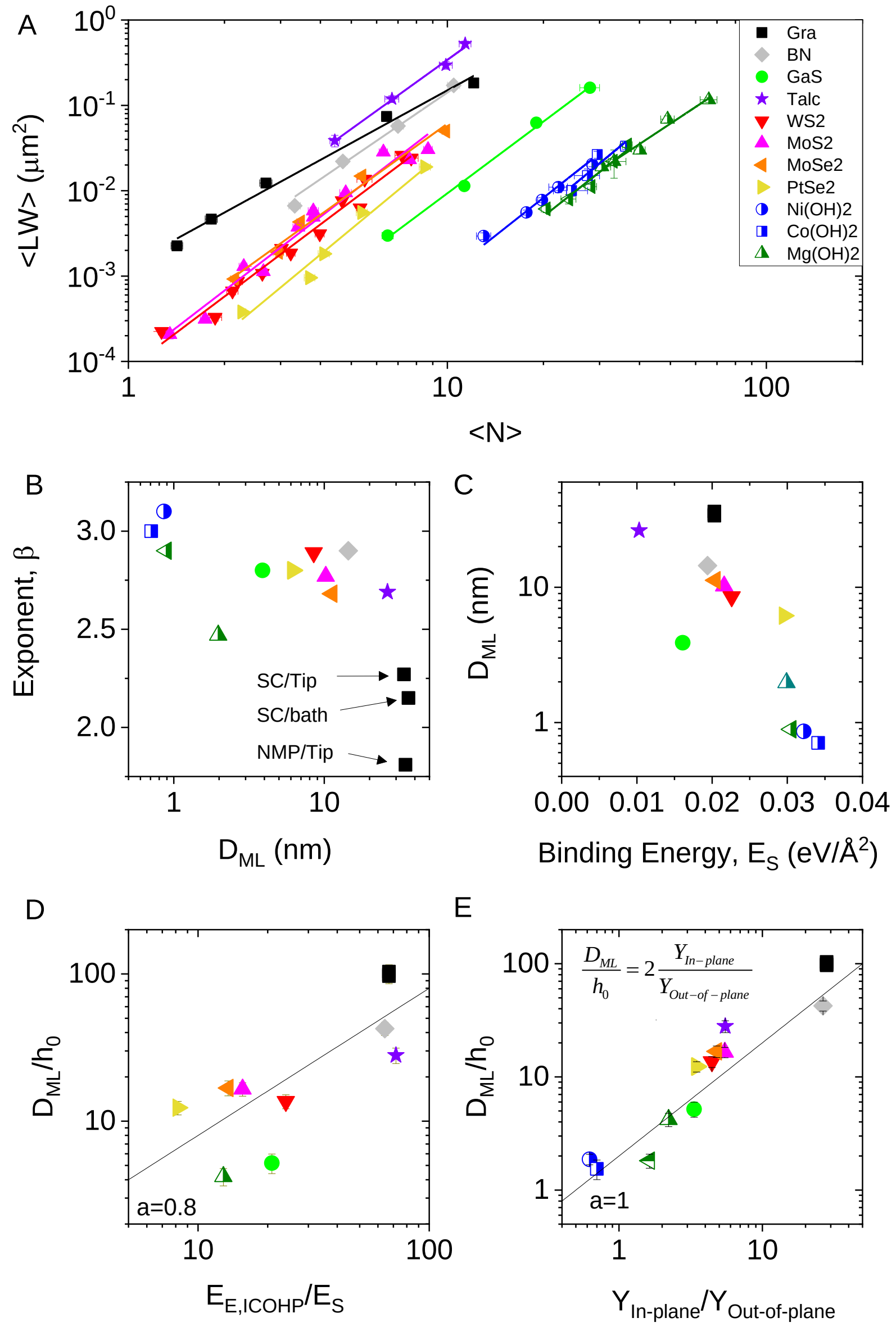

**Figure 5: Evaluation of the exfoliation efficiency compared to theoretical data.** A) Plot of the (experimental) mean nanosheet area approximated as the product of length and width, <*LW*> as function of mean layer number, <*N*> for all materials under study. The nanosheet area increases as a power-law with thickness as illustrated by solid lines. Extrapolation of the fit lines to <*N*>*=1* gives a quantitative description of the exfoliation efficiency, $D_{ML}^2$, where $D_{ML}$ is a characteristic lateral size associated with monolayers. Materials within the same class (transition metal dichalcogenides or hydroxides) sit very close beside each other. B) Map showing power-law fit parameters, $\beta$ and $D_{ML}$. C) Plots of $D_{ML}$ against calculated interlayer binding energies, $E_S$. D-E) Plots of $D_{ML}/h_0$ against parameters designed to approximate the ratio of in-plane to out-of-plane bonding strengths. In D, this ratio is represented by the ratio of the integral crystal orbital Hamiltonian population (ICOHP) surface density over the interlayer

binding energy density while in E) we use the calculated ratio of the in-plane to out-of-plane Young's moduli. The solid lines are fits to equation 9. In B-E, the three graphene data points reflect the samples prepared by sonication with both tip and bath in NMP and sodium cholate (SC).

# SI for

# Equipartition of energy describes the size-thickness relationship in liquid-exfoliated nanosheets

Claudia Backes,[1*] Davide Campi,[2] Beata M. Szydlowska,[1,3] Kevin Synnatschke,[1] Ezgi Ojala,[1] Farnia Rashvand,[1] Andrew Harvey,[3] Aideen Griffin,[3] Zdenek Sofer,[4] Nicola Marzari,[2] Jonathan N. Coleman,[3*] David D. O'Regan[3*]

[1]*Chair of Applied Physical Chemistry, University of Heidelberg, Im Neuenheimer Feld 253, 69120 Heidelberg, Germany*

[2]*Theory and Simulation of Materials (THEOS) and National Centre for Computational Design and Discovery of Novel Materials (MARVEL), École Polytechnique Fédérale de Lausanne, CH-1015 Lausanne, Switzerland*

[3]*School of Physics and CRANN & AMBER Research Centers, Trinity College Dublin, The University of Dublin, Dublin 2, Ireland*

[4]*Department of Inorganic Chemistry, University of Chemistry and Technology Prague, Technická 5, 166 28 Prague 6, Czech Republic*

* backes@uni-heidelberg.de ; colemaj@tcd.ie ; david.o.regan@tcd.ie

**Content**

# 1 Experimental Results

## 1.1 AFM characterisation of the graphene stock dispersion (SC tip)

Liquid-phase exfoliation (LPE) produces dispersions with broad distributions of lateral size and layer number. This is illustrated by the representative AFM image of stock dispersion obtained from graphite (Fig. S1A). Lateral sizes in this case range from 10 nm to ~ 2 µm with layer numbers ranging from 1-30. Statistical analysis of such a sample is extremely challenging: There are a few practical problems associated with AFM. In scanning microscopy, for instance AFM, the resolution is limited by the number of lines in each image, which defines the number of pixels. Hence, when imaging smaller objects, it is practical to record smaller images (~4x4 µm$^2$ in this case) to achieve sufficient resolution. This limits the field of view so that larger objects are often overlooked or only partially imaged, which does not allow for a statistical analysis of their length and thickness. If the field of view is too large, small nanosheets cannot be resolved. This means that many images of various magnifications are required to obtain a representative population of the stock dispersions and it is necessary to also count the smallest nanosheets that are often not resolved in wide-view images.

In spite of this restriction, an AFM analysis was exemplarily performed for the graphene stock dispersion where in total > 550 nanosheets were counted. The length and layer number histograms are shown in figure S1B-C. Even though characteristic log-normal distribution histograms with a long tail are obtained, overall, the analysis of the stock dispersion was not found to be feasible due to the uncertainty associated with "correctly" choosing imaging conditions (high resolution in combination with wide field of view).

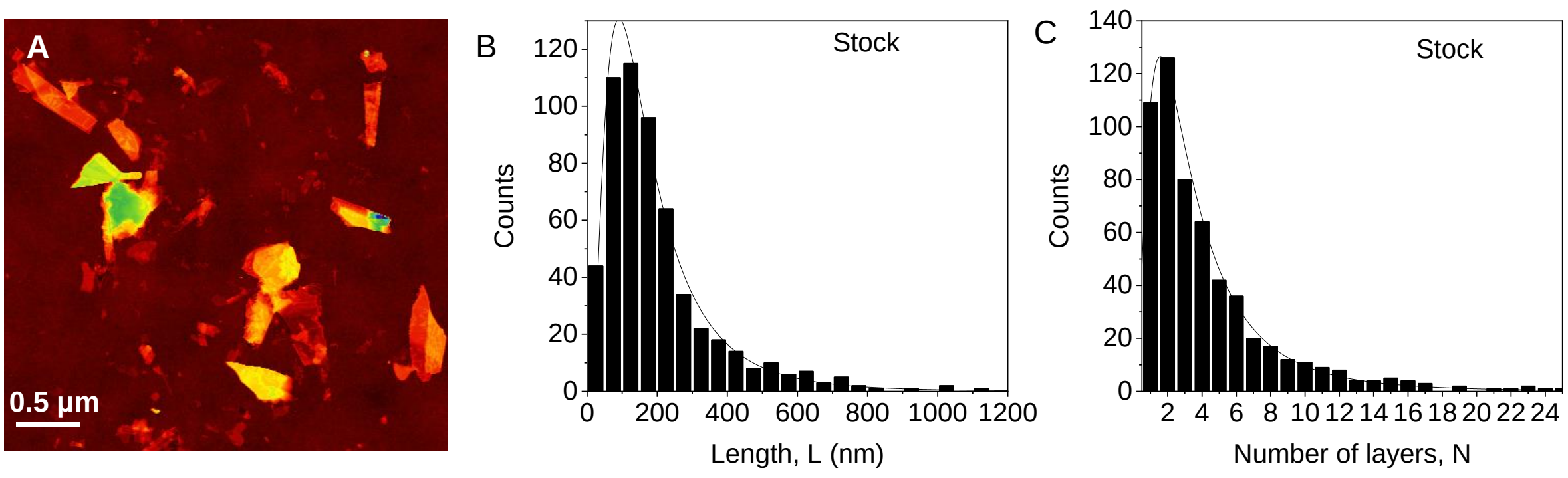

**Figure S1: AFM statistics of a graphene stock dispersion** after removal of unexfoliated nanosheets, but prior to size selection. A) Representative atomic force micrograph showing the polydispersity. B) Histogram of the longest nanosheet dimension (L). C) Layer number histogram. A total of > 550 nanosheets were counted.

To test the way in which the fractionation by centrifugation quantitatively impacts the area-layer number relationship, the data of the stock dispersion was analysed by grouping nanosheets of the same layer number into one bin. In other words, the population shown in figure S2A was sliced by vertical lines (rather than at an angle) and <*L*> was determined for each *N*-mer. This data is plotted in figure S2A. Clearly, a power law scaling is observed, albeit with a significantly greater scatter than in the case of the analysis of the size-selected fractions (compare with Fig. 2 of the main manuscript). In addition, we observe that the length/width aspect ratio <*L/W*> increases with decreasing layer number (Fig. S2B), i.e. nanosheets become more belt-like. This indicates that nanosheet scission/tearing occurs, as the shape of the nanosheets should not change if only delamination occured. It is surprising that such a change in nanosheet shape with layer number is apparent for a material such as graphite/graphene with an isotropic bonding situation in the layer. This points to a complex mechanism of nanosheet tearing in an unzippering-type fashion. Further studies will be required to understand this.

Finally, in figure S2C, we compare <*LW*> as function of <*N*> from *N*-mer binning of the stock to the arithmetic means of the fractions after LCC. Importantly, a power law scaling is observed in both cases, with different exponents. The fractions after LCC are characterised by an exponent of ~2, while it is ~1 in the case of the *N*-mer binned stock. This difference is a manifestation of the centrifugation, which essentially bins the data by volume rather than by layer number.

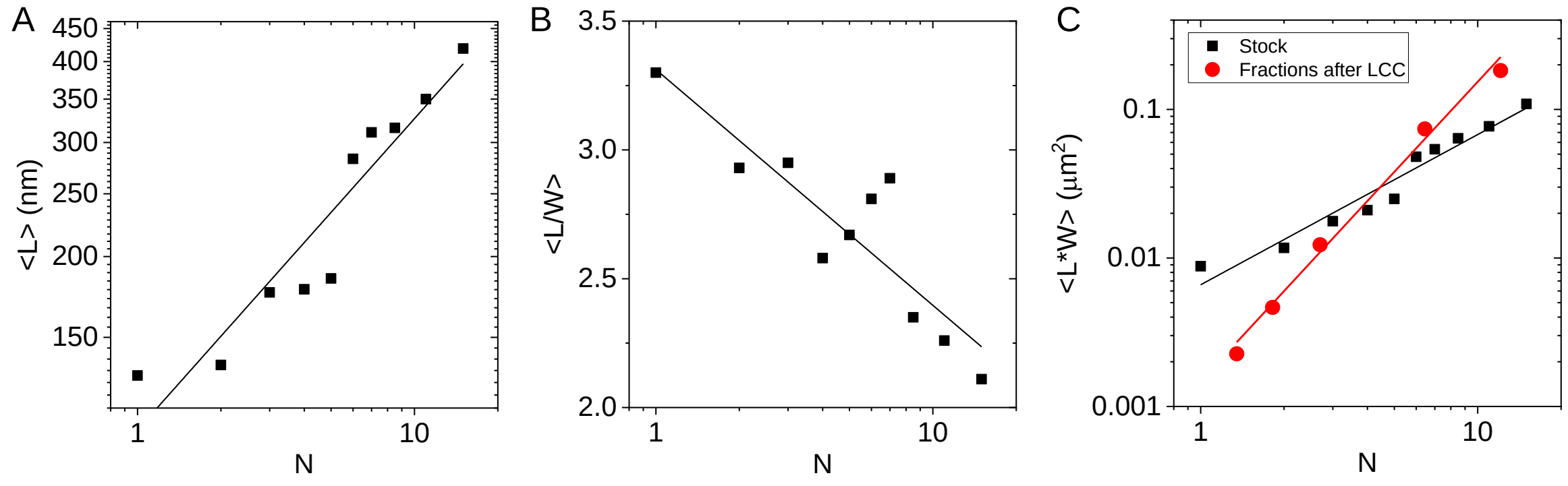

**Figure S2: Scaling of lateral nanosheet dimensions as a function of layer number in the stock.** The nanosheet population was grouped according to its layer number and the mean lateral size for each *N*-mer (all nanosheets in the group of a certain layer number) calculated from the AFM statistics. Note that in spite of >550 counts of the nanosheets in the stock dispersion, there is significant scatter in the data. A) Plot of longest dimension <*L*> as function of *N*-mer, B) Plot of mean length/width aspect ratio as function of *N*-mer, C) Plot of nanosheet area expressed as <*LW*> as function *N*-mer. In this plot, the result from the statistical analysis of the fractions after cascade centrifugation is included. In this case, the mean layer number <*LW*> is plotted as function of mean nanosheet layer number. Because centrifugation slices the initial population in an angle (Fig. S2), the exponent relating the two quantities is different, but statistical analysis is more feasible due to the reduced polydispersity.

We have consistently found that it is the exponent ~2 associated with fractions defined in terms of <*N*>, and not the exponent ~1 for *N*-mer binned fractions, that matches with the predictions of our energy quasi-equipartition model. This is unsurprising, perhaps, since the model makes statements concerning only steady-state averaged quantities such as <*N*> that characterise nanosheets within aggregates that were generated under similar conditions. What is subtle here, however, is the question of whether all nanosheets of a given species may be considered as having been generated under similar conditions, irrespective of *N*, or whether instead the quasi-equipartition only holds within aggregates of a given <*N*> (with a not-too-broad distribution) but then this aggregation is preserved by fractionation by volume via LCC. For the purposes of our model and experimental setup this distinction makes no practical difference (as long as the deviation-from equipartion factor *a*, if needed at all, is <*N*>-independent), however this and indeed the development of a more detailed non-equilibrium statistical mechanics account of LPE nanosheet size in terms of *N*-mer binning, would be worthy of future investigation.

## 1.2 Direct comparison of graphite exfoliated by tip and bath sonication in aqueous sodium cholate and by tip sonication in NMP

The data in figure 2 of the main manuscript shows that graphite exfoliation by tip sonication in NMP produces larger and thicker nanosheets compared to exfoliation in aqueous surfactant, whereas bath sonication produces a larger population of large and thin nanosheets compared to tip sonication (albeit at much lower yield). This can be clearly seen when comparing the images themselves without statistical analysis. Examples for a fraction isolated at low centrifugal acceleration (Fig. S3) and intermediate centrifugal acceleration (Fig. S4) are shown below. At centrifugal accelerations above 6k *g*, no nanosheets could be isolated in the NMP sample making a direct comparison of the fraction with the smallest/thinnest nanosheets impossible. Image dimensions (including height) are identical in the images below to facilitate comparison. The images are shown in two different colour scales. While the black and white colour (bottom rows) is rather unusual, it can be useful to clearly identify thin nanosheets that lie flat on the surface (or steps and terraces of incompletely exfoliated nanosheets). This is evident in both images of the graphite exfoliated in the sonic bath (middle panel).

Prior to statistical analysis, it is important to “choose” which objects should be counted. In this work, they were manually picked after cropping the images into smaller frames. In general, all aggregates were avoided. Examples are highlighted by the blue circles in figures S3-4. In addition, all (small) objects that did not have characteristic shapes of nanosheets were excluded, as they are mostly solvent/surfactant residues or other contaminations. Examples are highlighted by the orange circles in figure S3-4. Again, it can be useful to switch between different colour scales and play with the image contrast to facilitate the decision whether a deposit has a characteristic shape of a nanosheet lying flat on the surface. This is sometimes difficult to judge. For new materials of unknown morfology, it can be beneficial to have transmission electron micrographs(TEM), where the nanosheet shapes can be observed more distinctly. For all materials under study here, TEM images of LPE samples can be found throughout the literature. In general, the nanosheets should appear with a similar shape in AFM and TEM. Characteristics and typical shapes include belts, triangles, hexagons or sharp edges. Often nanosheets are incompletely exfoliated and have unmissable terraces, sometimes they appear folded (especially when they are large and thin) etc. All these features can be an indicators whether deposition and AFM imaging are reliable. A very detailed protocol of the procedure we use is provided in section 4.

In general, it is possible that some of the “contaminations” that were excluded from the size measurement are nanosheets, or nanosheets buried in impurities. However, since all images were treated in the same way and because the majority of deposits had a distinct 2D shape often with sharp edges, we do not expect this selection to have an impact on the statistical analysis. In fact, if aggregates or impurities are deliberately included in the statistical analysis, the shape of the lateral distribution histograms can deviate from the characteristic lognormal shape. Examples are shown in section 4.

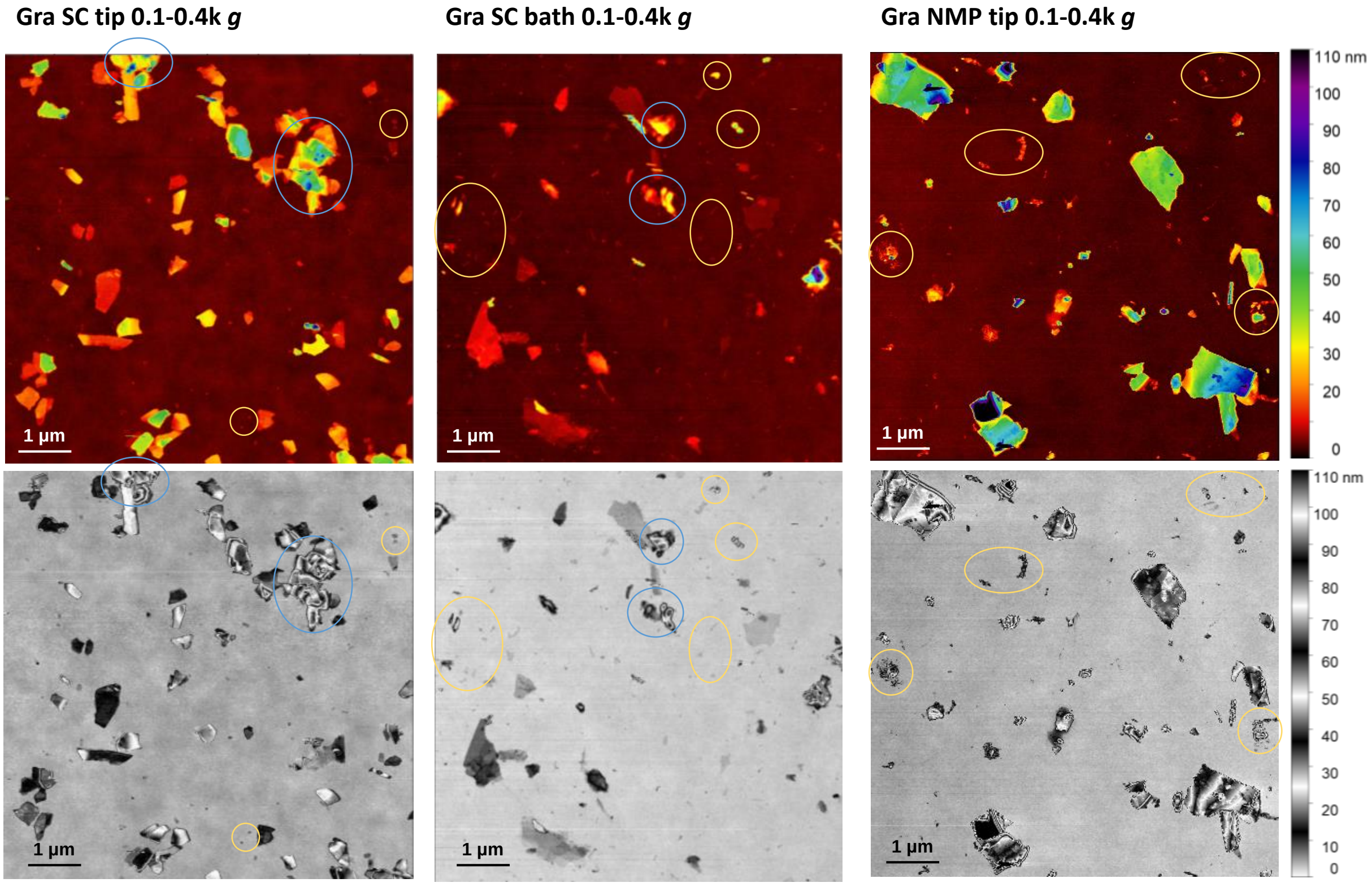

**Figure S3: Wide view AFM images of graphite isolated between 100 and 400 *g* in the centrifugation cascade.** The result of exfoliation in SC by tip sonication (left), bath sonication (middle) and tip sonication in NMP (right) is compared. Scale bars (including height) are identical. Images are shown in two different colour scales. Aggregates (blue circles) and impurities (orange circles) are indicated.

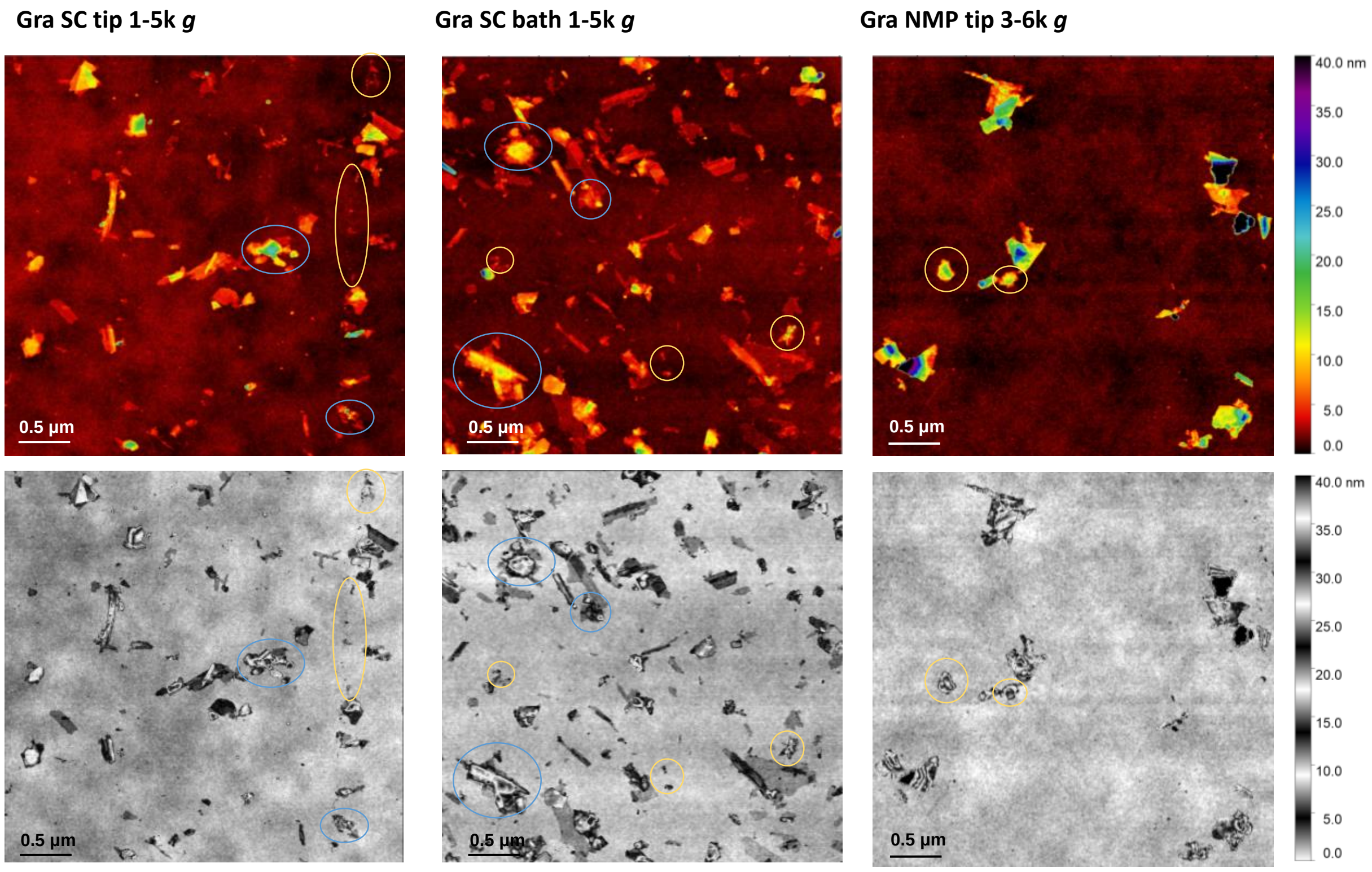

**Figure S4: Wide view AFM images of graphite isolated at intermediate centrifugal acceleration in cascade.** The result of exfoliation in SC by tip sonication (left), bath sonication (middle) and tip sonication in NMP (right) is compared. Scale bars (including height) are identical. Images are shown in two different colour scales. Aggregates (blue circles) and impurities (orange circles) are indicated.

## 1.3 Length-thickness relationship of $WS_2$ exfoliated in SC and NMP

The finding that the solvent chosen for LPE of graphite only has a minor impact on the quality of the dispersion, *i.e.* the relationship between nanosheet lateral dimensions and layer number is intriguing. Since it is important to confirm that this is also the case for other materials, we have re-analysed previously published data on exfoliation of $WS_2$ in SC and NMP.[1] In this case, lateral size and layer number of the NMP based dispersion was derived from spectroscopic metrics based on extinction spectroscopy.[1] To ensure that the metrics that were quantified for SC-based dispersions can be applied, the $WS_2$ was transferred from NMP to aqueous SC after the LCC-based size selection. The data is shown in figure S5. As presented by the plots of <L> and <N> versus the midpoint of the pair of centrifugal acceleration used in LCC (central RCF) in figure S5A-B, nanosheets are both laterally larger (figure S5A) and thicker (figure S5B) in NMP compared to SC. Furthermore, the exponents relating <L> and <N> to RCF are slightly different in NMP and SC. This is very similar to the graphite data presented in figure 3 of the main manuscript. Therefore, when plotting <L> as function of <N> (Fig. S5C), we again find the NMP data to be shifted towards larger and thicker nanosheets. As with graphite, the data is characterised by a powerlaw and fitting gives identical intercepts with N=1. Furthermore, the exponent relating <L> and <N> is smaller in NMP compared to SC in analogy to the graphite data. This suggests that this behaviour seems to be indeed universal or at least applicable to across a wide range of nanosheets/ materials.

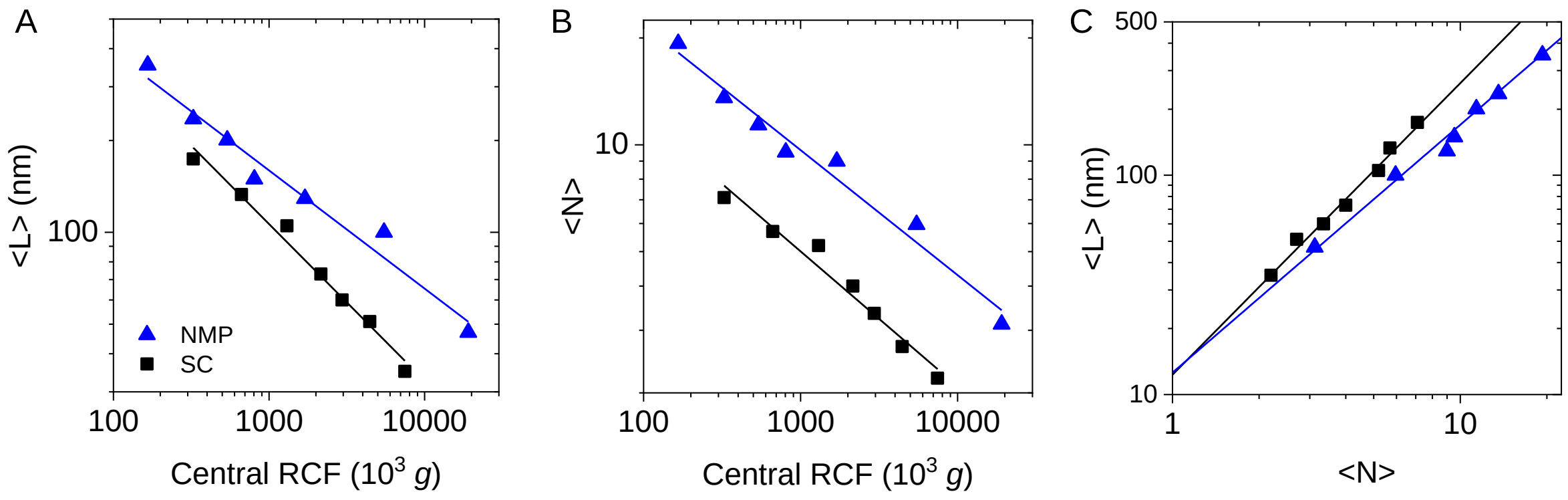

**Figure S5: Comparison of $WS_2$ exfoliated in aqueous SC and NMP.** A) Plot of mean nanosheet length (<L>) of the fractions as function of the central RCF. G) Plot of mean nanosheet layer number (<N>) as function of the central RCF. H) Plot of mean nanosheet length as function of layer number. The data was adapted from [1].

## 1.4 AFM images, lateral dimensions (L, W) and number of layer distribution histograms

In the following, representative AFM images of the size-selected fractions of the 12 materials under study are shown along with their length, width and layer number distribution histograms. We note that a huge amount of experimental data accumulated over years is compared in this work. In some cases, the sample characterisation was used in previous independent studies. In these cases, the respective reference is given in the figure caption.

### *1.4.1 Graphene in aqueous sodium cholate (tip sonication)*

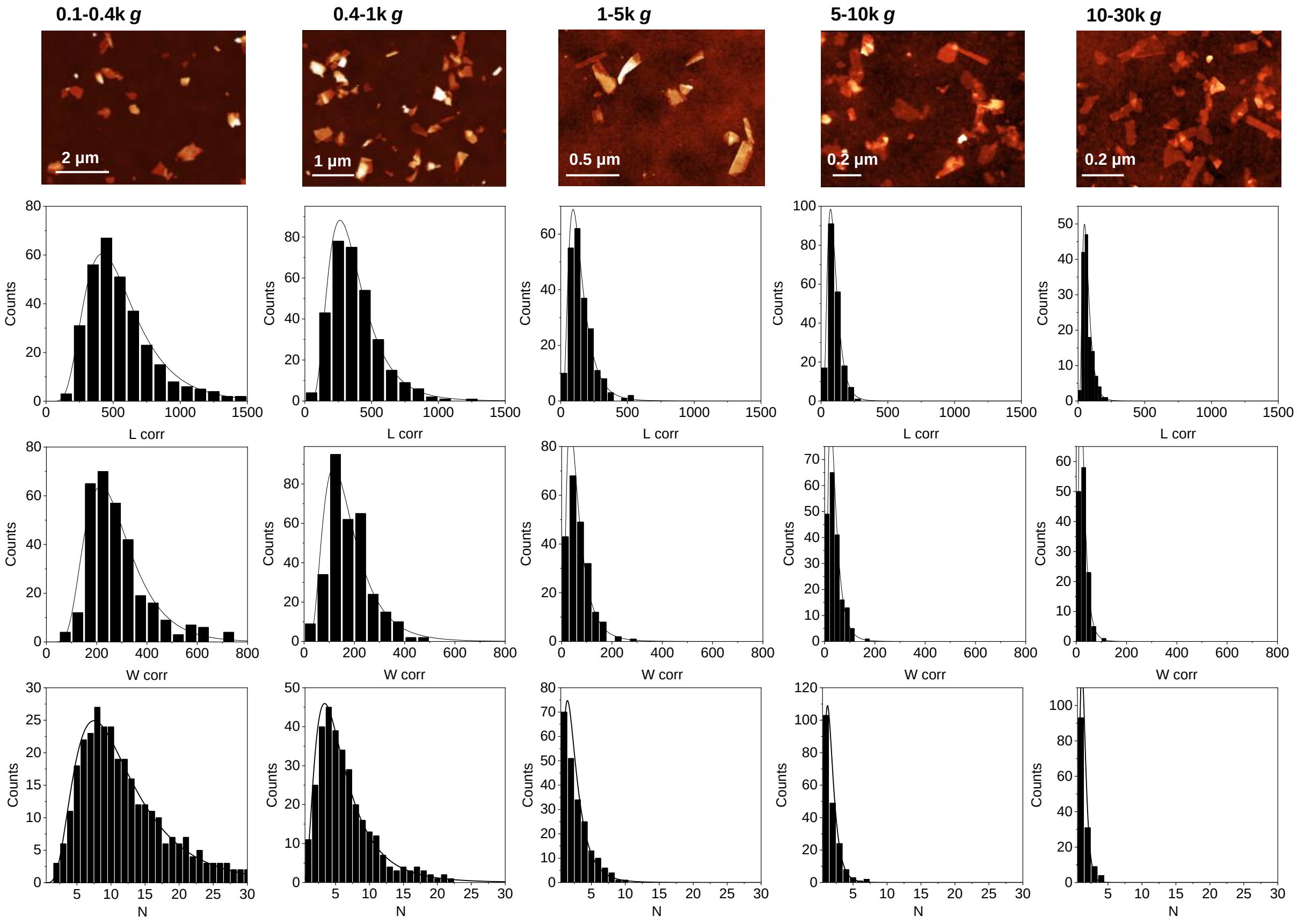

**Figure S6: AFM analysis of the graphene SC tip sonication fractions.** Top row: representative images of the fractions isolated at the centrifugal acceleration indicated. Distribution histograms of i) second row: longest lateral dimension, length, *L*, (in nm), ii) third row: Dimension perpendicular to *L*, termed the width *W* (in nm), iii), fourth row: layer number *N*.

### 1.4.2 *Graphene in aqueous sodium cholate (bath sonication)*

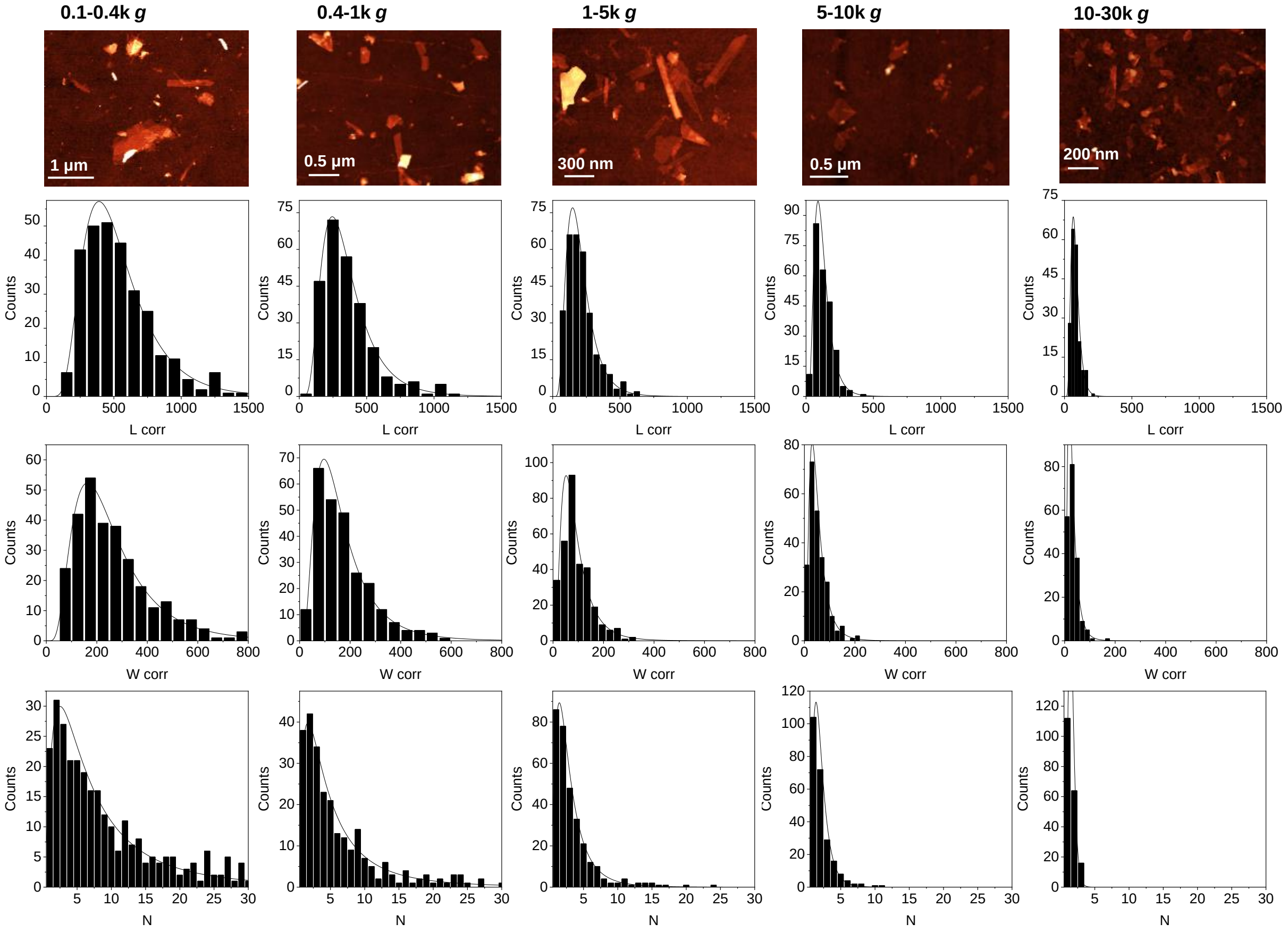

**Figure S7: AFM analysis of the graphene SC bath sonication fractions.** Top row: representative images of the fractions isolated at the centrifugal acceleration indicated. Distribution histograms of i) second row: longest lateral dimension, length, *L*, (in nm), ii) third row: Dimension perpendicular to *L*, termed the width *W* (in nm), iii), fourth row: layer number *N*.

### *1.4.3 Graphene in NMP (tip sonication)*

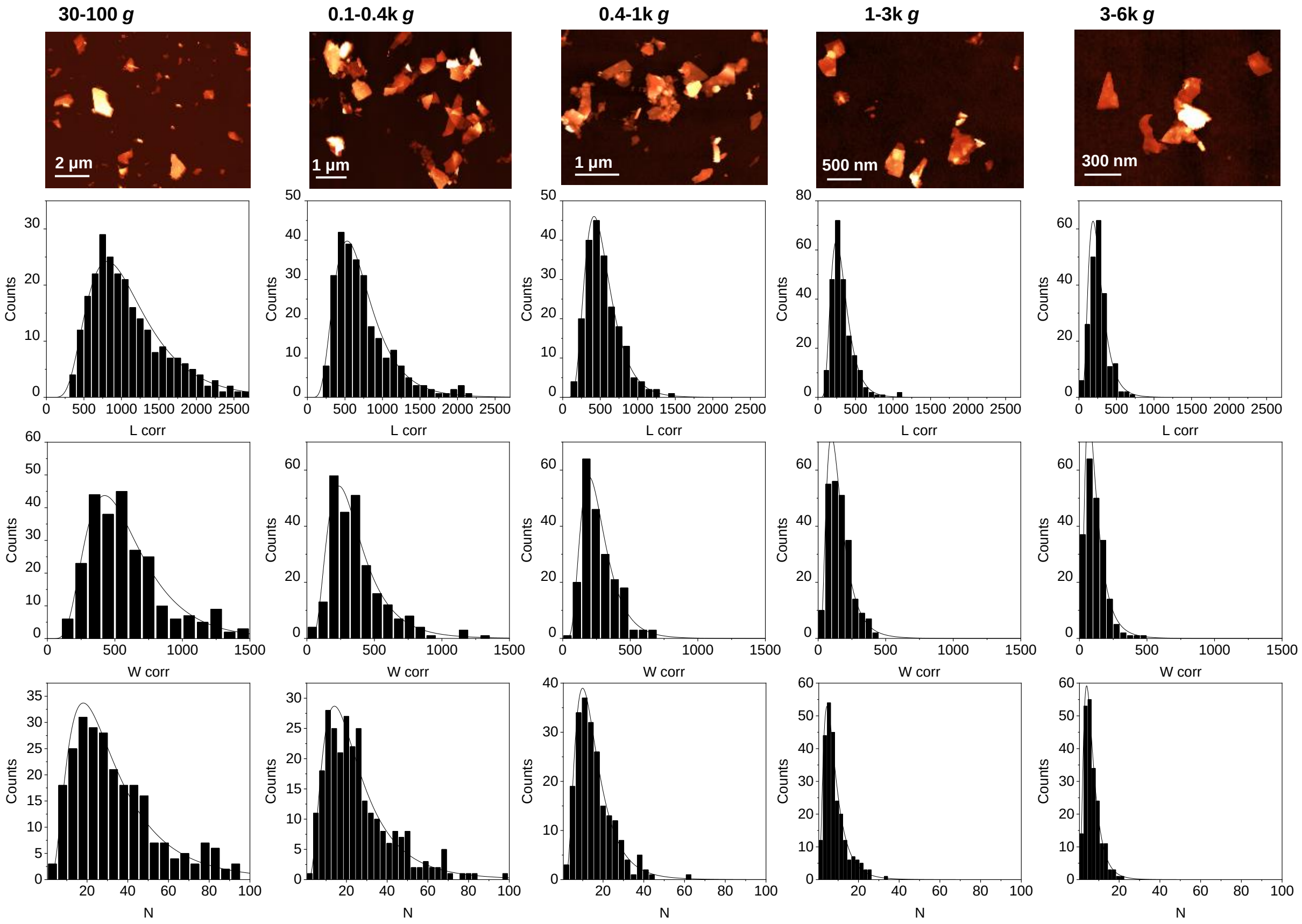

**Figure S8: AFM analysis of the graphene NMP tip sonication fractions.** Top row: representative images of the fractions isolated at the centrifugal acceleration indicated. Distribution histograms of i) second row: longest lateral dimension, length, *L*, (in nm), ii) third row: Dimension perpendicular to *L*, termed the width *W* (in nm), iii), fourth row: layer number *N*.

## *1.4.4 WS₂*

Two batches were produced and analysed.

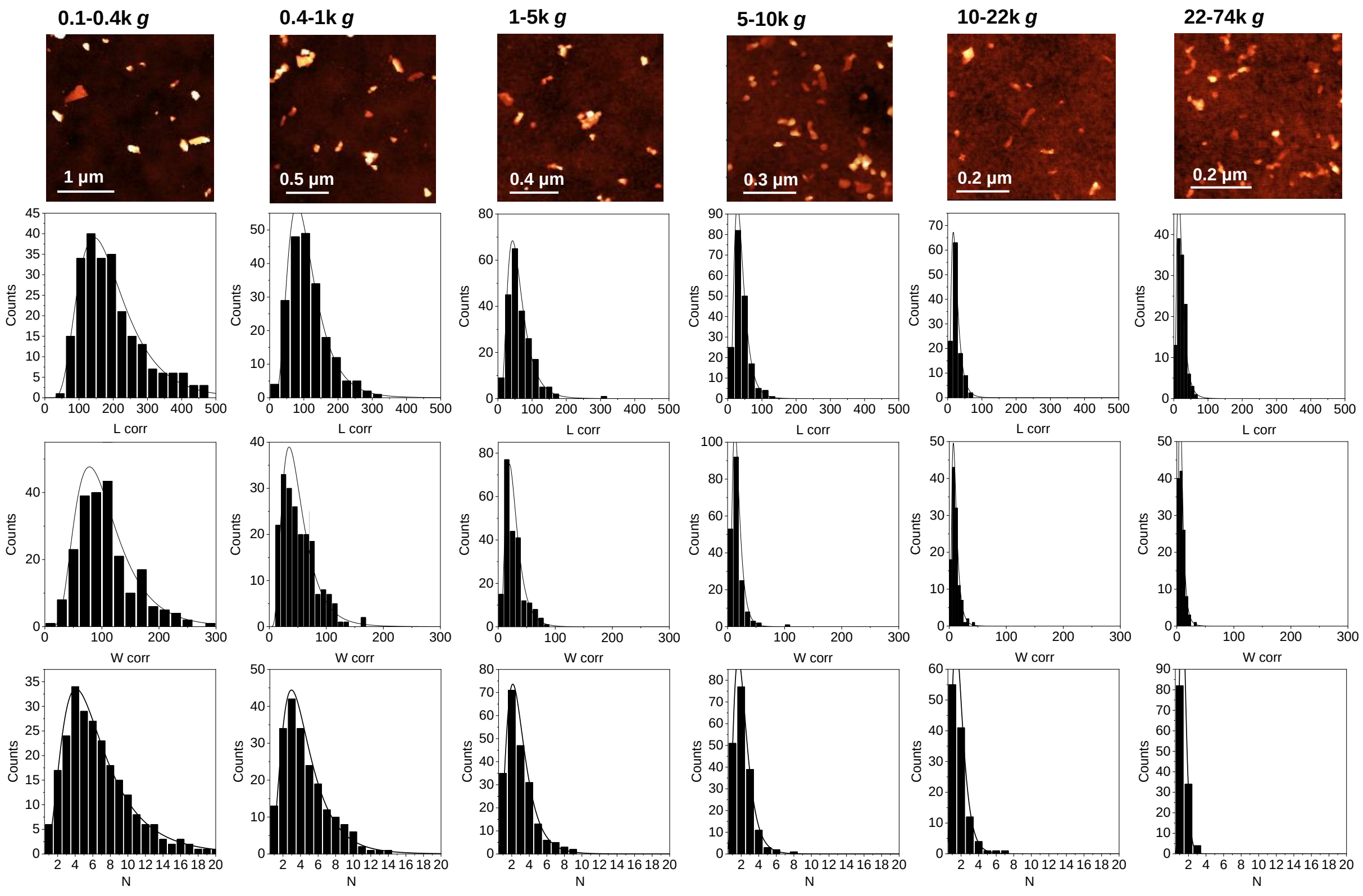

**Figure S9: AFM analysis of the $WS_2$ fractions- batch 1.** Top row: representative images of the fractions isolated at the centrifugal acceleration indicated. Distribution histograms of i) second row: longest lateral dimension, length, *L*, (in nm), ii) third row: Dimension perpendicular to *L*, termed the width *W* (in nm), iii), fourth row: layer number *N*.

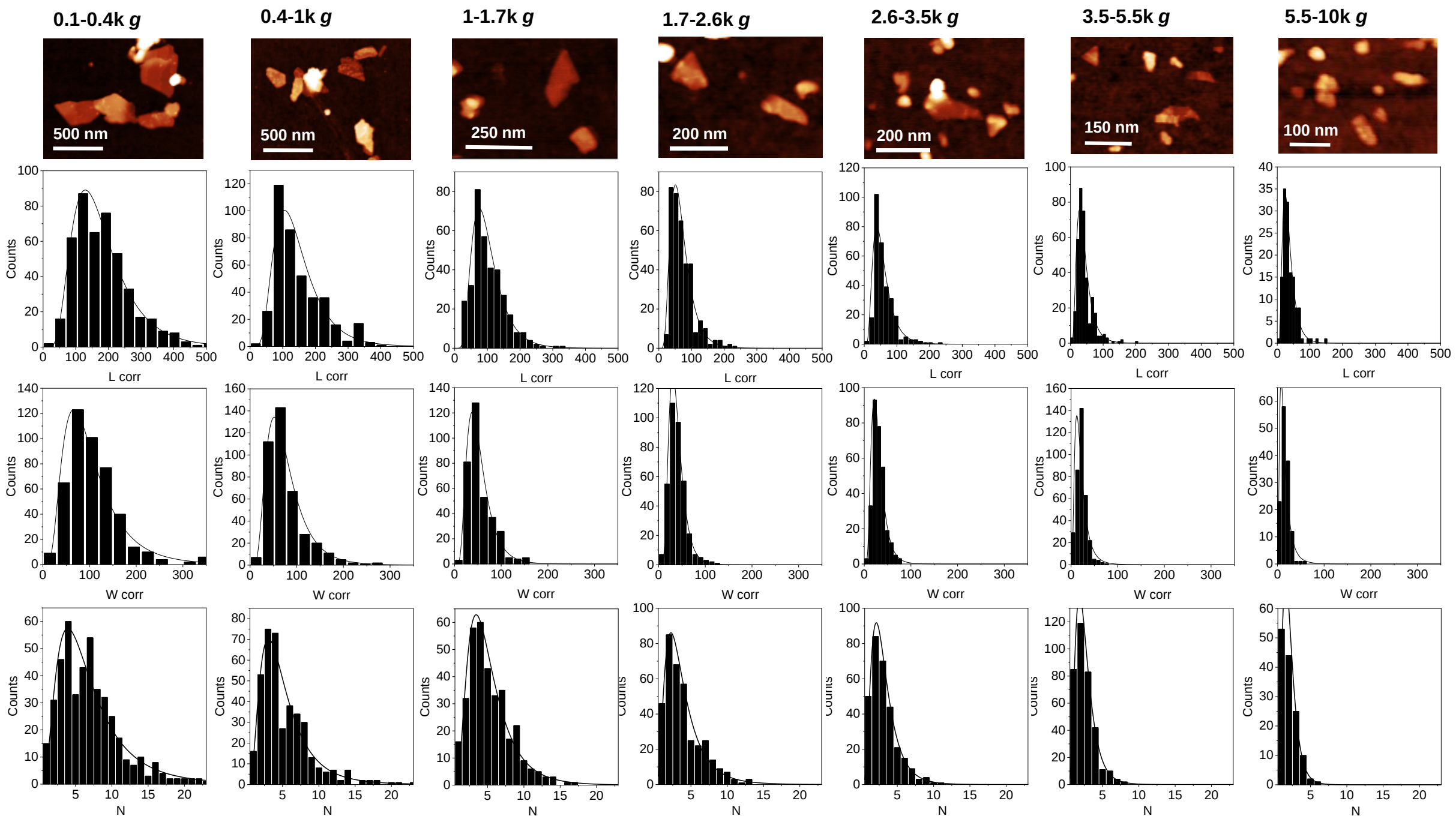

**Figure S10: AFM analysis of the $WS_2$ fractions- batch 2.**[1] Top row: representative images of the fractions isolated at the centrifugal acceleration indicated. Distribution histograms of i) second row: longest lateral dimension, length, *L*, (in nm), ii) third row: Dimension perpendicular to *L*, termed the width *W* (in nm), iii), fourth row: layer number *N*.

### *1.4.5 MoS₂*

Two batches were produced and analysed.

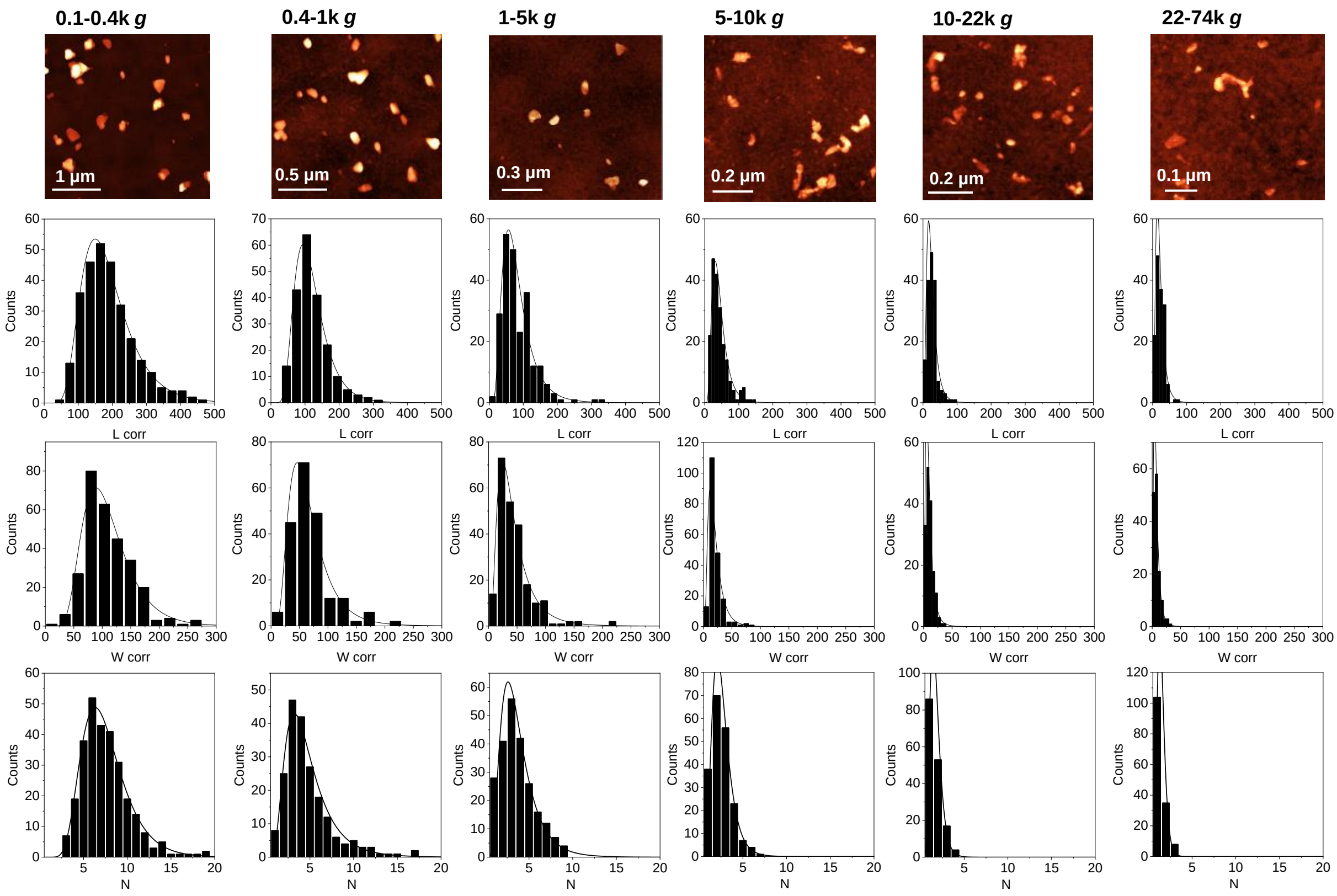

**Figure S11: AFM analysis of the $MoS_2$ fractions- batch 1.** Top row: representative images of the fractions isolated at the centrifugal acceleration indicated. Distribution histograms of i) second row: longest lateral dimension, length, *L*, (in nm), ii) third row: Dimension perpendicular to *L*, termed the width *W* (in nm), iii), fourth row: layer number *N*.

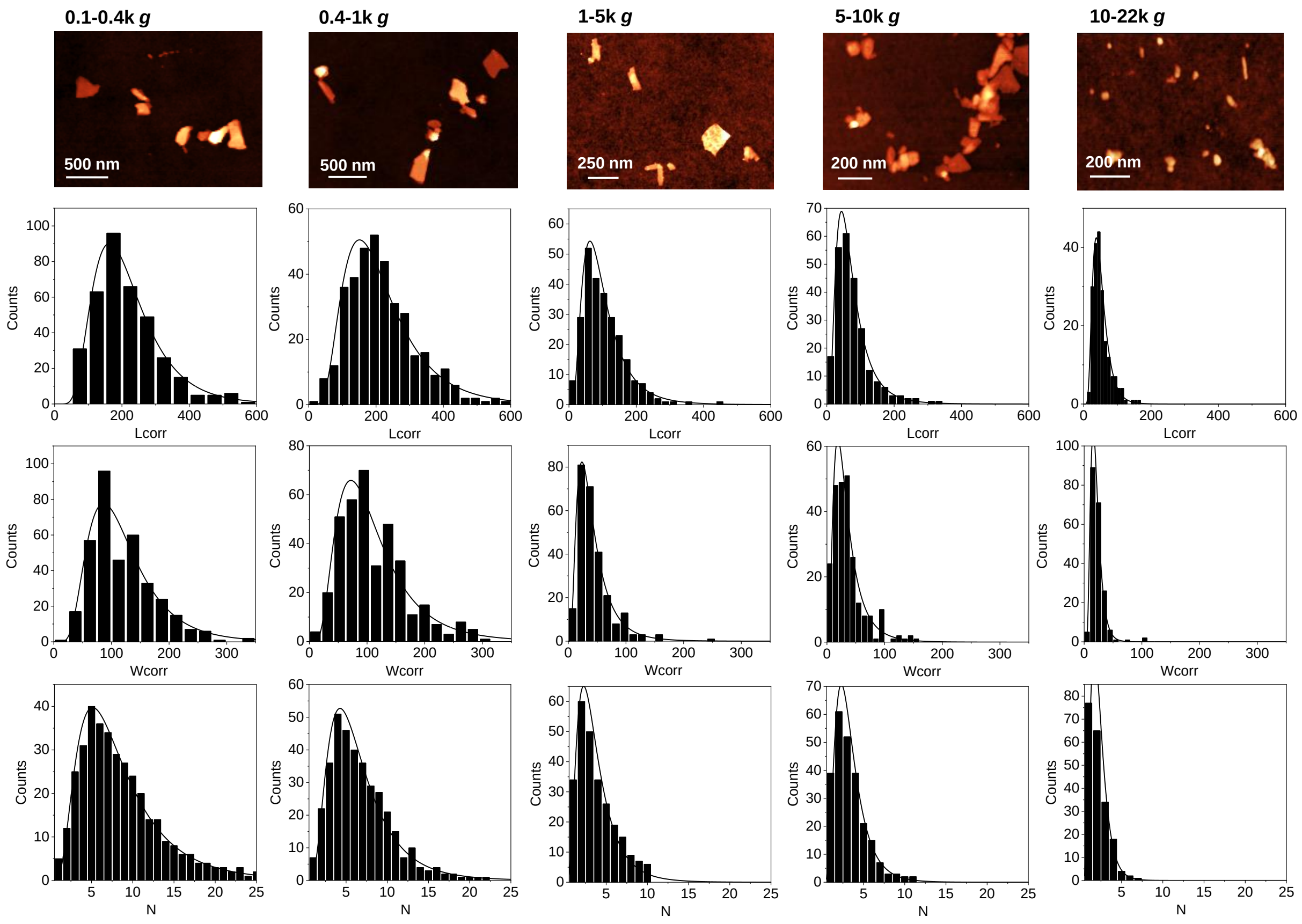

**Figure S12: AFM analysis of the $MoS_2$ fractions- batch 2.** Top row: representative images of the fractions isolated at the centrifugal acceleration indicated. Distribution histograms of i) second row: longest lateral dimension, length, *L*, (in nm), ii) third row: Dimension perpendicular to *L*, termed the width *W* (in nm), iii), fourth row: layer number *N*.

### 1.4.6 $MoSe_2$

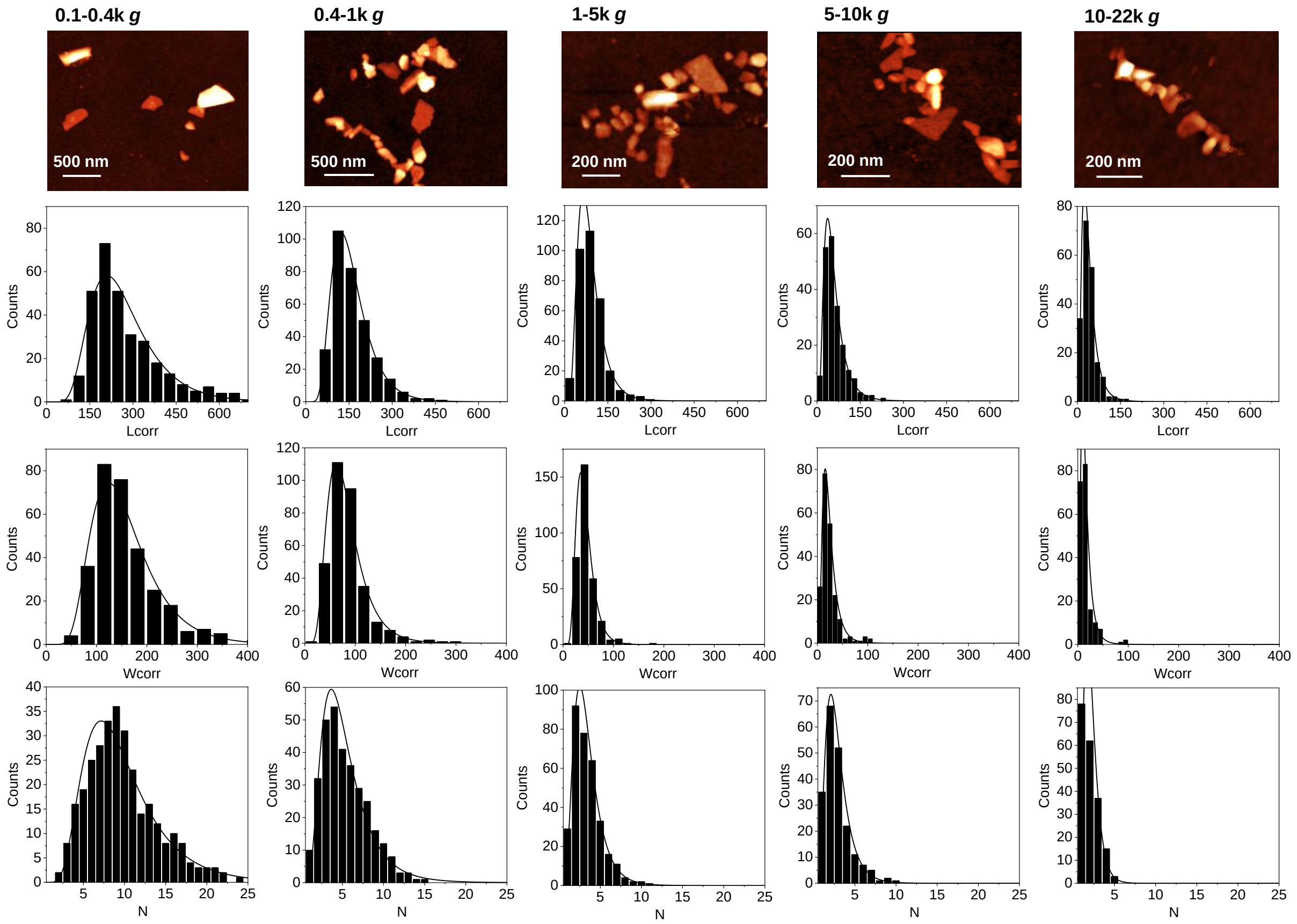

**Figure S13: AFM analysis of the $MoSe_2$ fractions.** Top row: representative images of the fractions isolated at the centrifugal acceleration indicated. Distribution histograms of i) second row: longest lateral dimension, length, *L*, (in nm), ii) third row: Dimension perpendicular to *L*, termed the width *W* (in nm), iii), fourth row: layer number *N*.

### *1.4.7 PtSe₂*

The apparent AFM heights of materials obtained via LPE are usually overestimated. This is likely due to residual solvent trapped between the layers. In order to convert the apparent measured AFM thickness to the actual number of layers, we have applied an approach termed step height analysis introduced previously for graphene, $MoS_2$, GaS, $WS_2$, and BP (see methods). This means that we measured the height of steps associated with terraces of incompletely exfoliated nanosheets on the nanosheet surface. In total 99 height profiles of incompletely exfoliated nanosheets were extracted. Values corresponding to the apparent height difference are plotted in ascending order in figure S14. The apparent height of each step is a multiple of a value representing the apparent height of one $PtSe_2$ monolayer seen in the AFM. The theoretical per-monolayer lattice constant is always much smaller. The step height of one LPE $PtSe_2$ monolayer was determined to be 2.05 nm. Therefore, to convert the measured apparent AFM height to number of layers, the measured height was divided by 2.05 nm throughout this manuscript.

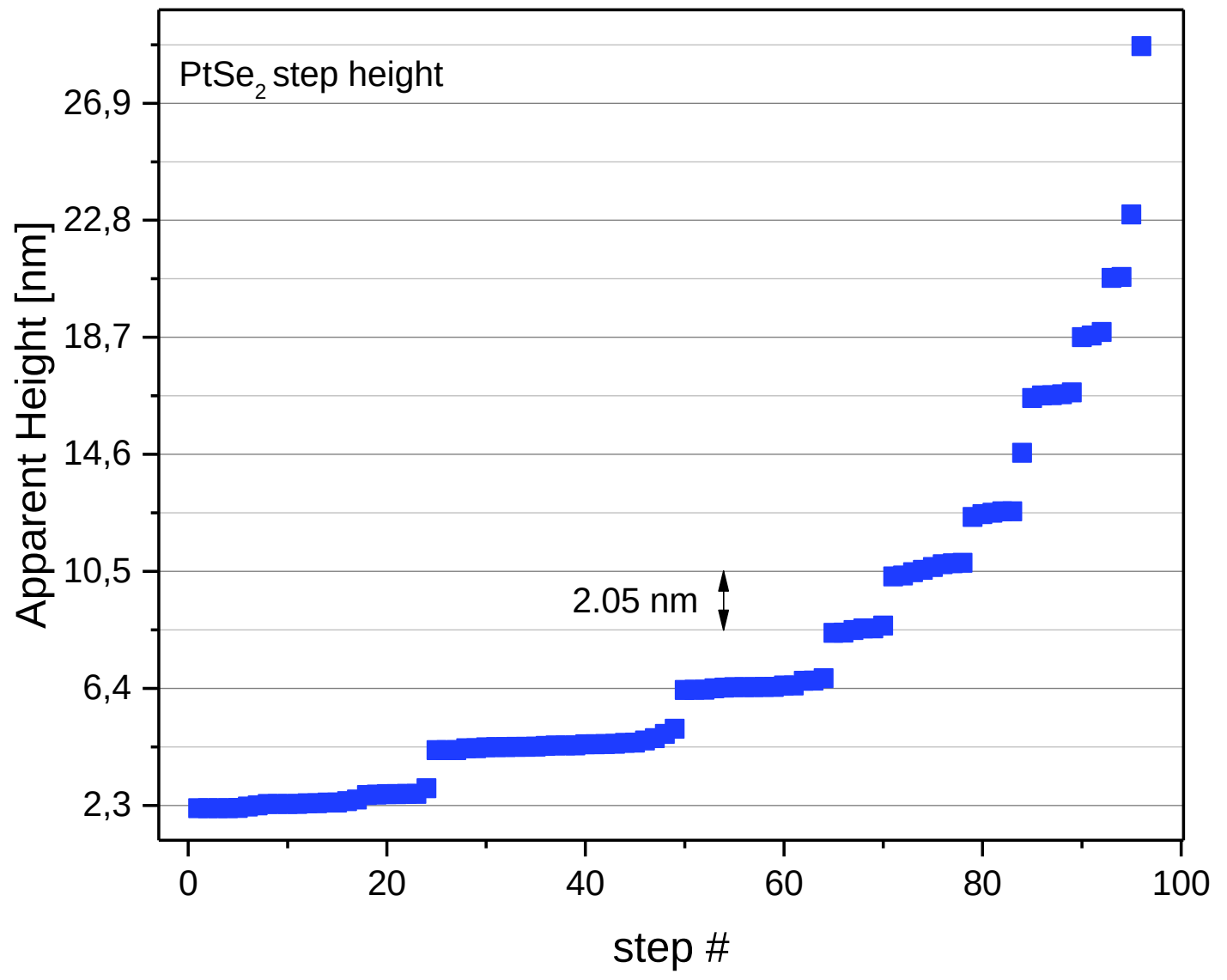

**Figure S14: Step height analysis of $PtSe_2$.** Heights of >95 steps of deposited $PtSe_2$ nanosheets in ascending order. The step height clustered in groups and was always found to be a multiple of 2.05 nm, which is the apparent height of one monolayer.

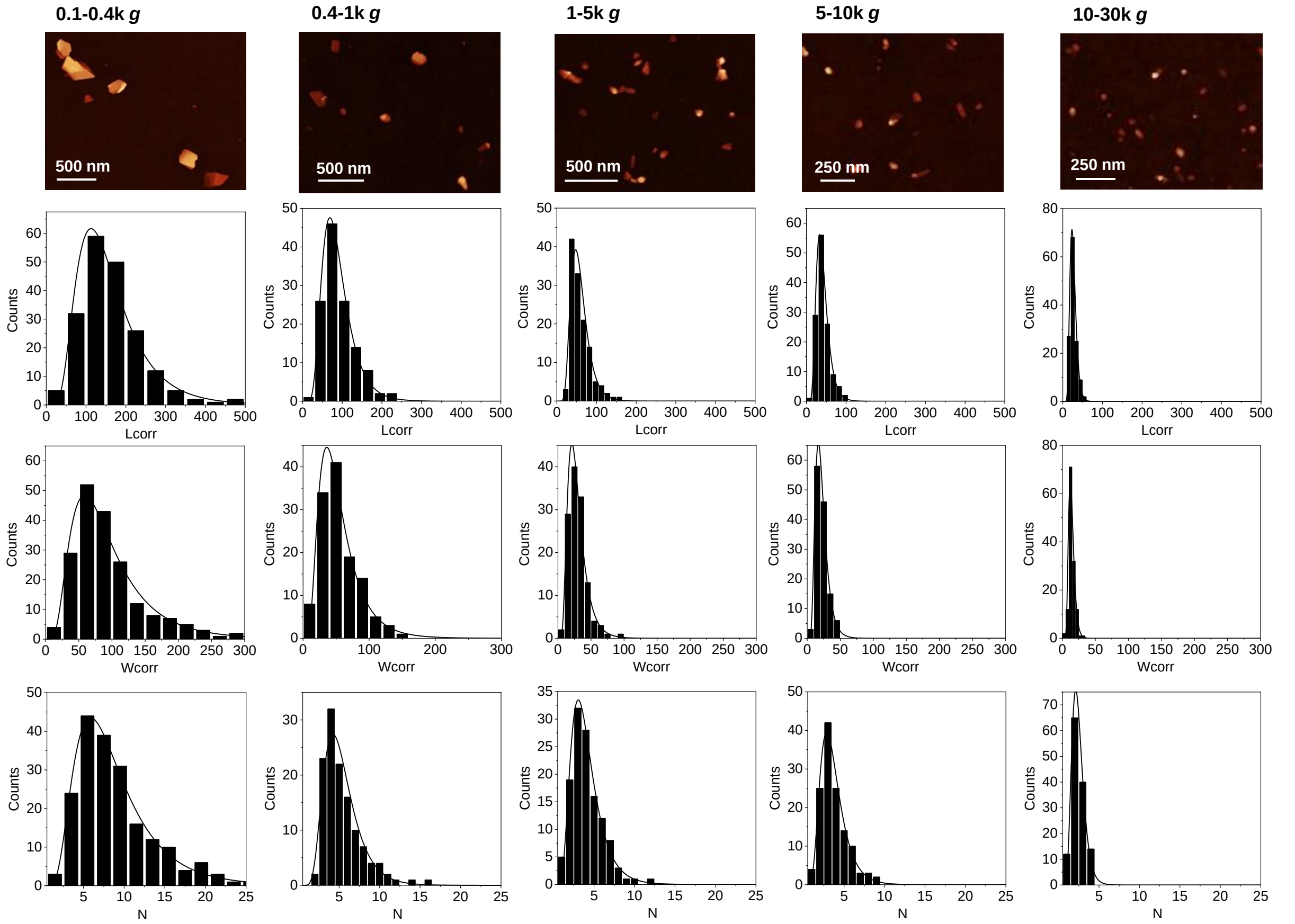

**Figure S15: AFM analysis of the $PtSe_2$ fractions.** Top row: representative images of the fractions isolated at the centrifugal acceleration indicated. Distribution histograms of i) second row: longest lateral dimension, length, *L*, (in nm), ii) third row: Dimension perpendicular to *L*, termed the width *W* (in nm), iii), fourth row: layer number *N*.

### *1.4.8 BN*

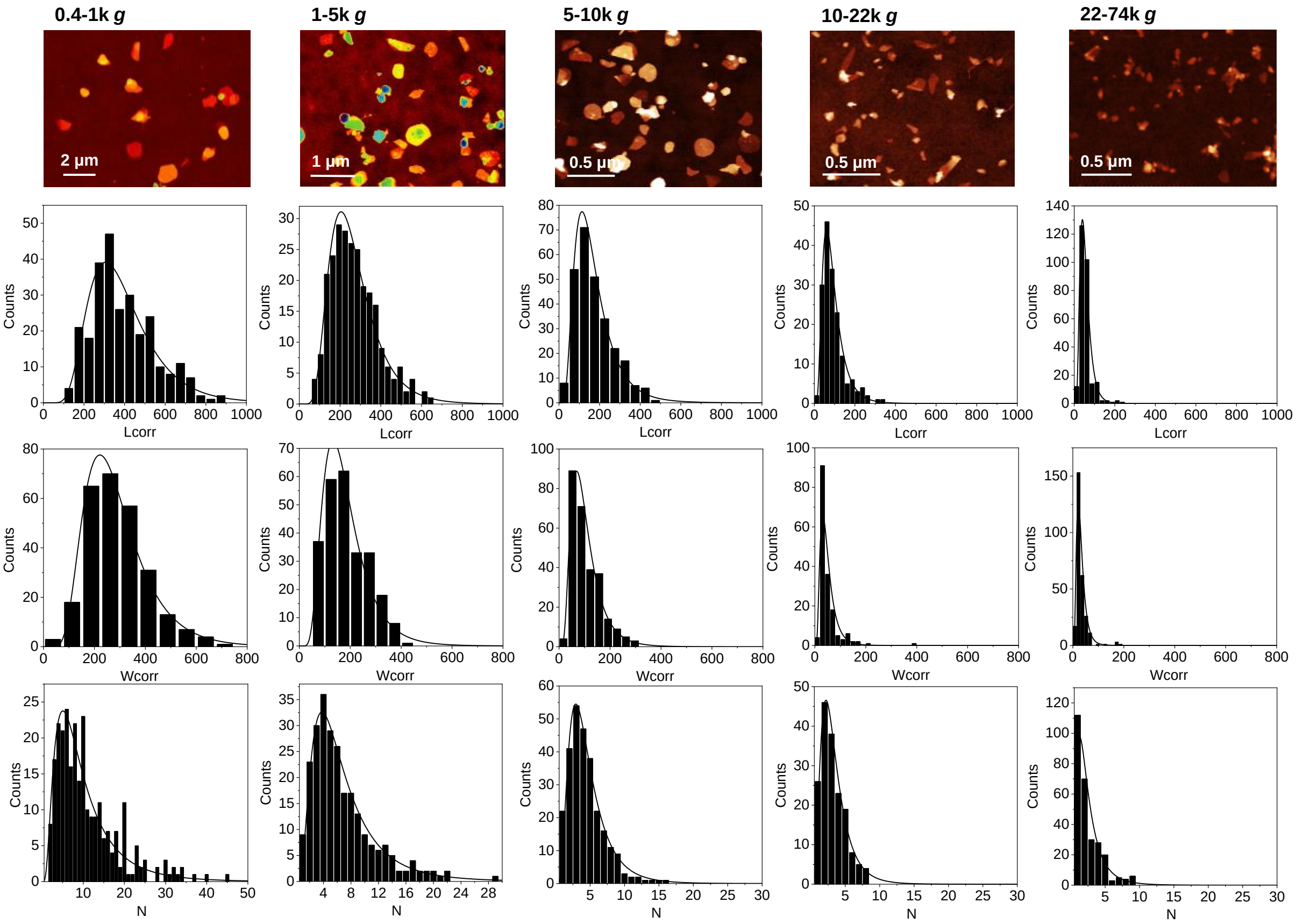

**Figure S16: AFM analysis of the BN fractions.**[2] Top row: representative images of the fractions isolated at the centrifugal acceleration indicated. Distribution histograms of i) second row: longest lateral dimension, length, *L*, (in nm), ii) third row: Dimension perpendicular to *L*, termed width the *W* (in nm), iii), fourth row: layer number *N*.

*1.4.9 GaS*

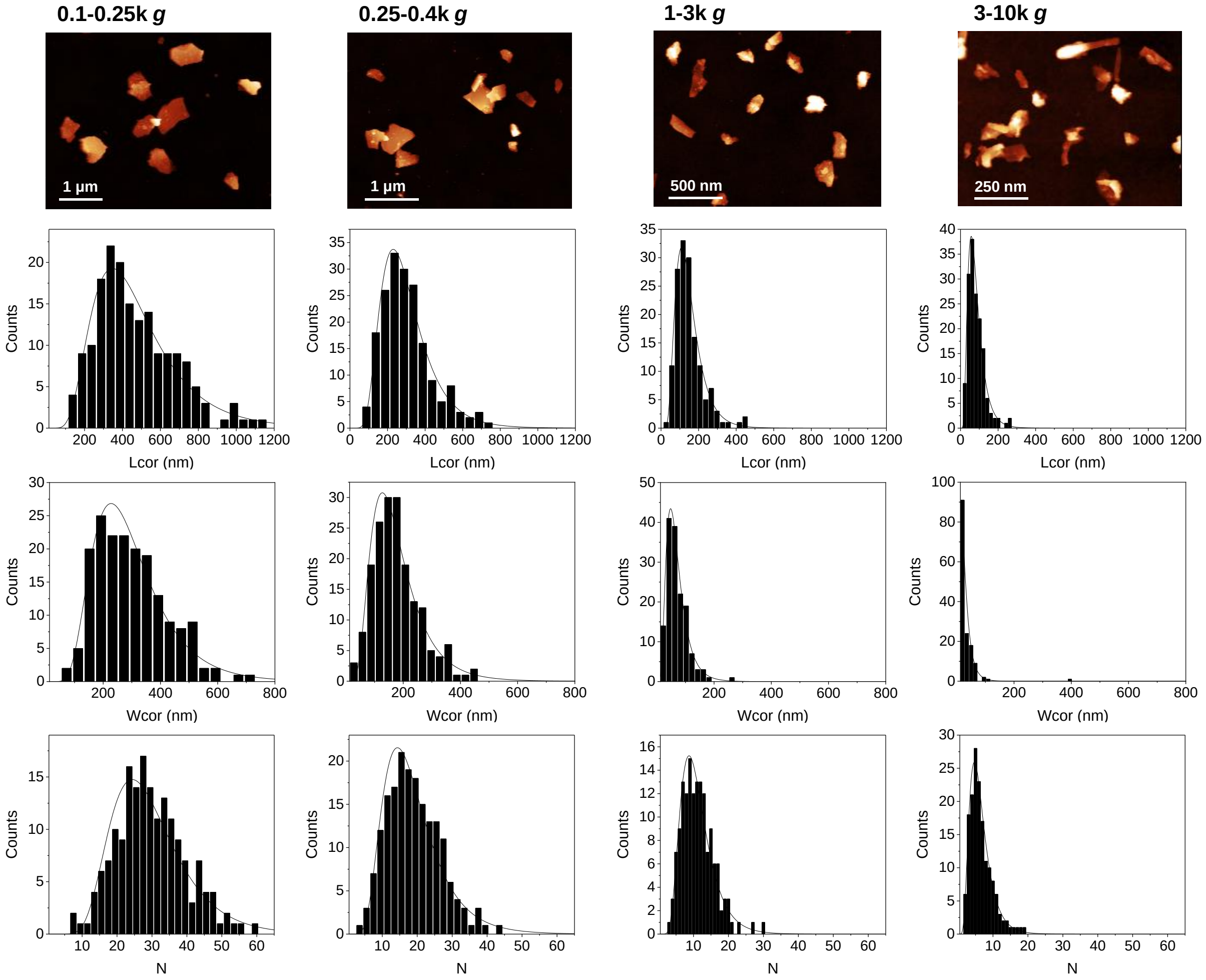

**Figure S17: AFM analysis of the GaS fractions.**[3] Top row: representative images of the fractions isolated at the centrifugal acceleration indicated. Distribution histograms of i) second row: longest lateral dimension, length, $L$, (in nm), ii) third row: Dimension perpendicular to $L$, termed the width $W$ (in nm), iii), fourth row: layer number $N$.

### 1.4.10 Talc

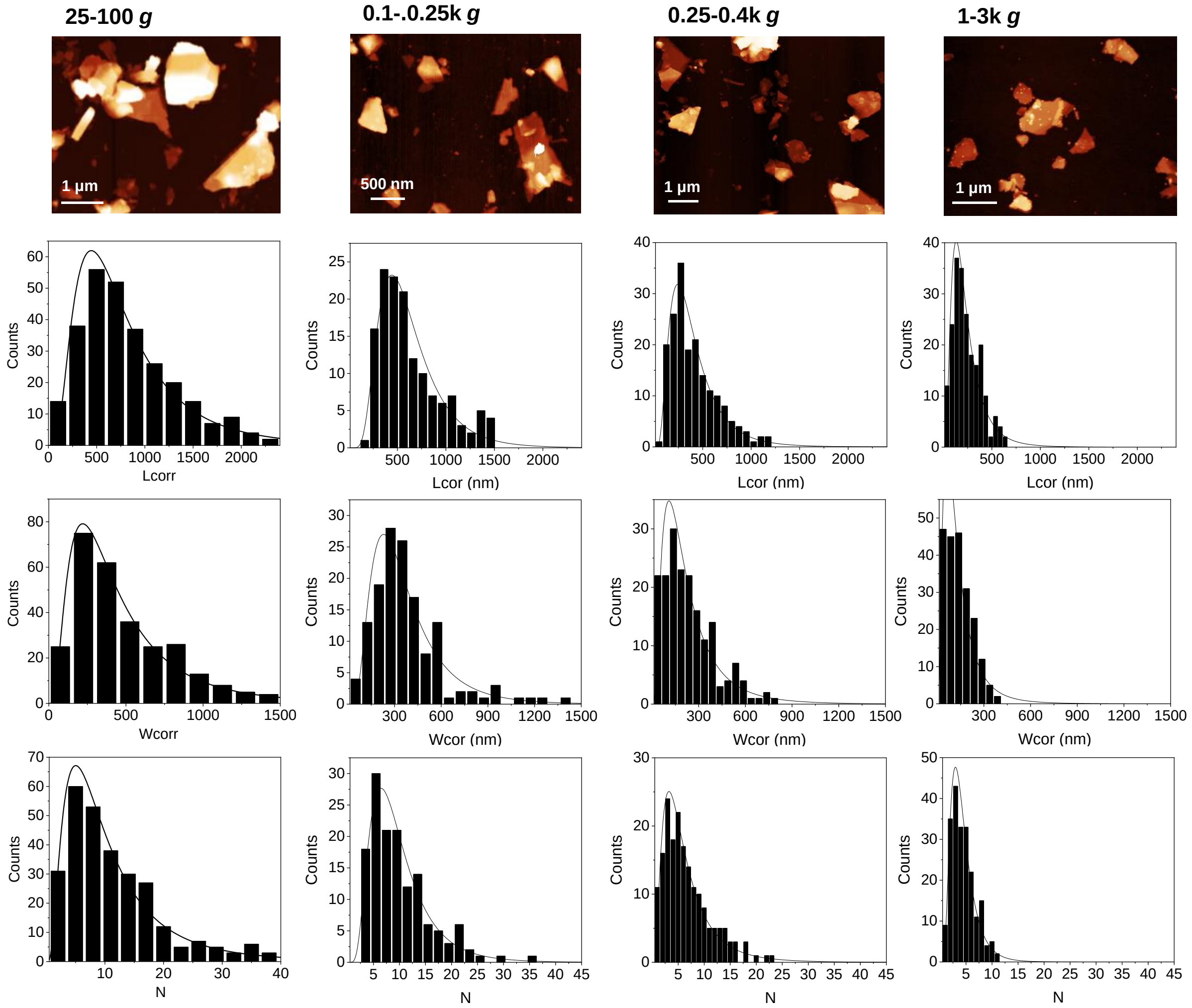

**Figure S18: AFM analysis of the Talc fractions.**[3] Top row: representative images of the fractions isolated at the centrifugal acceleration indicated. Distribution histograms of i) second row: longest lateral dimension, length, *L*, (in nm), ii) third row: Dimension perpendicular to *L*, termed the width *W* (in nm), iii), fourth row: layer number *N*.

## *1.4.11 Ni(OH)₂*

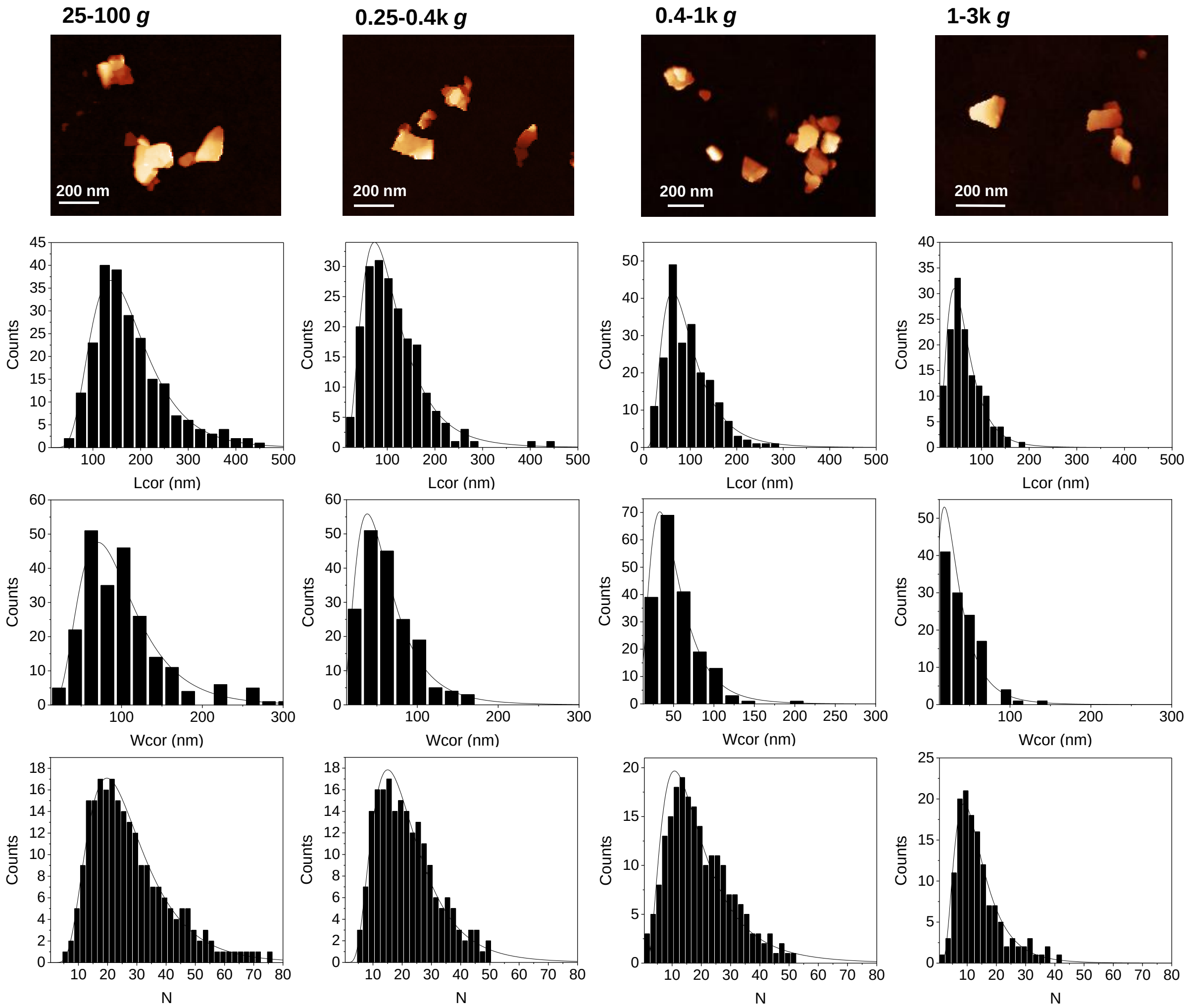

**Figure S19: AFM analysis of the $Ni(OH)_2$ fractions.**[3] Top row: representative images of the fractions isolated at the centrifugal acceleration indicated. Distribution histograms of i) second row: longest lateral dimension, length, *L*, (in nm), ii) third row: Dimension perpendicular to *L*, termed the width *W* (in nm), iii), fourth row: layer number *N*.

*1.4.12 $Mg(OH)_2$*

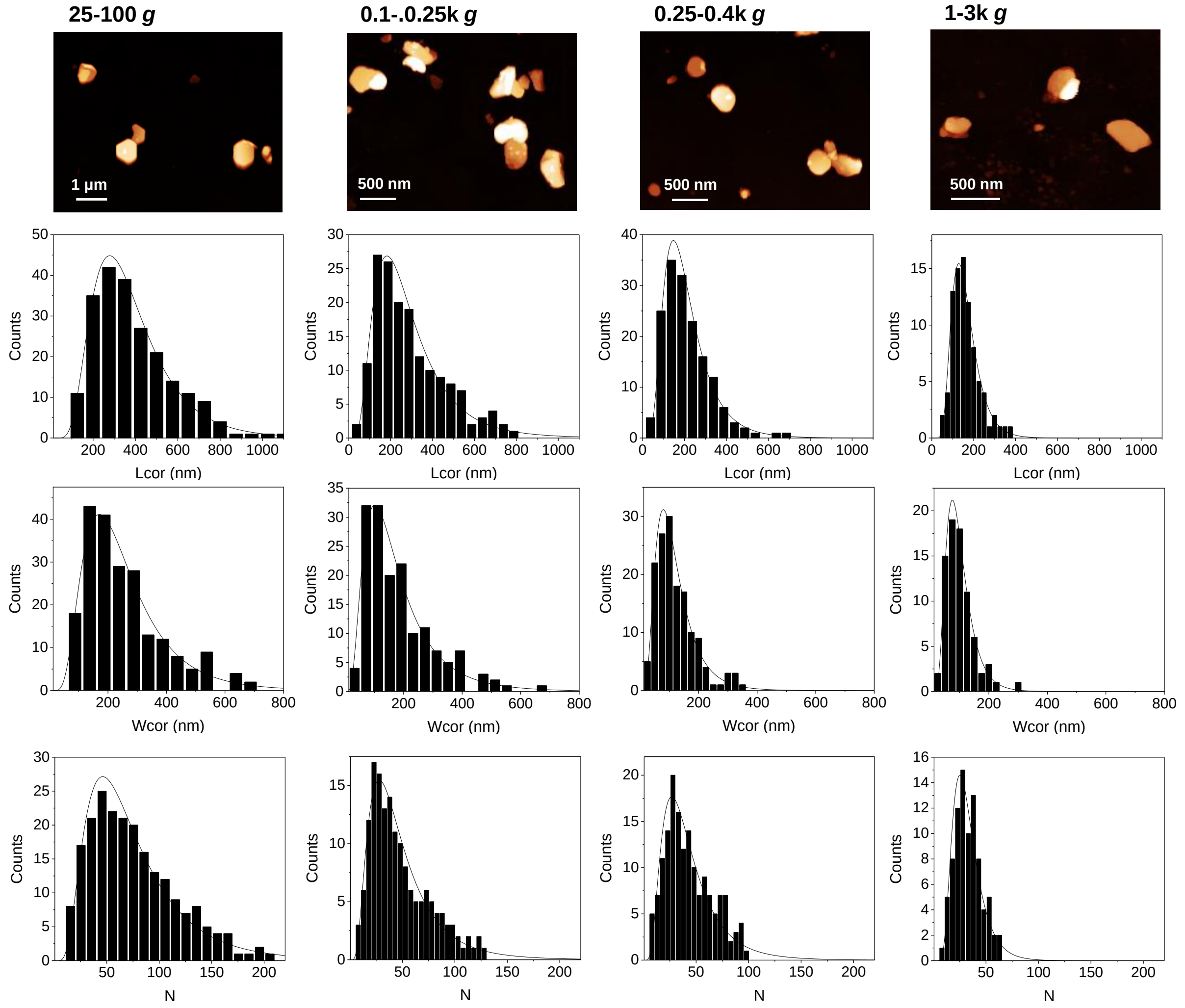

**Figure S20: AFM analysis of the $Mg(OH)_2$ fractions.**[3] Top row: representative images of the fractions isolated at the centrifugal acceleration indicated. Distribution histograms of i) second row: longest lateral dimension, length, *L*, (in nm), ii) third row: Dimension perpendicular to *L*, termed the width *W* (in nm), iii), fourth row: layer number *N*.

*1.4.13 Co(OH)₂*

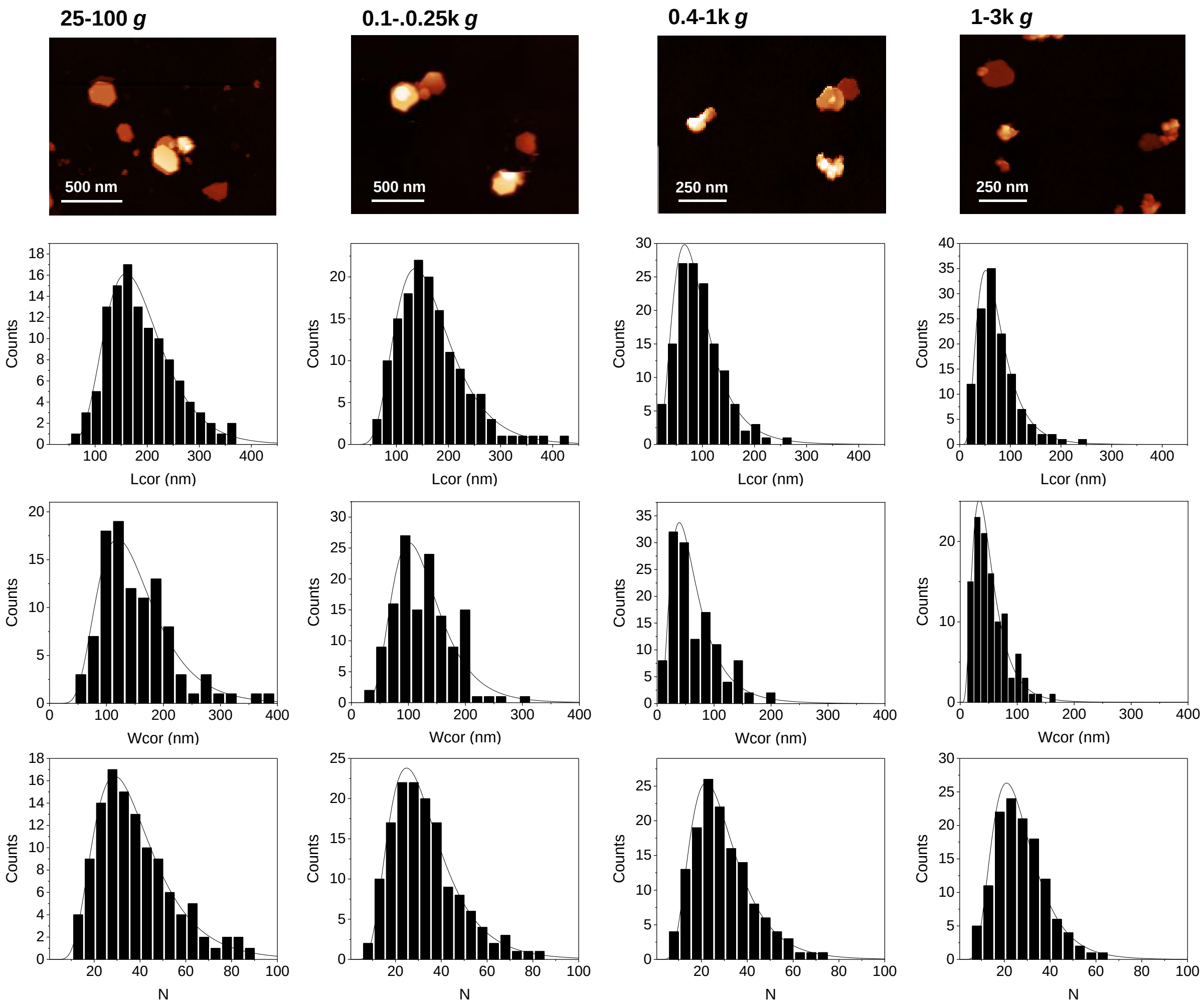

**Figure S21: AFM analysis of the $Co(OH)_2$ fractions.** Top row: representative images of the fractions isolated at the centrifugal acceleration indicated. Distribution histograms of i) second row: longest lateral dimension, length, *L*, (in nm), ii) third row: Dimension perpendicular to *L*, termed the width *W* (in nm), iii), fourth row: layer number *N*.

### 1.4.14 $Zn(OH)_2$

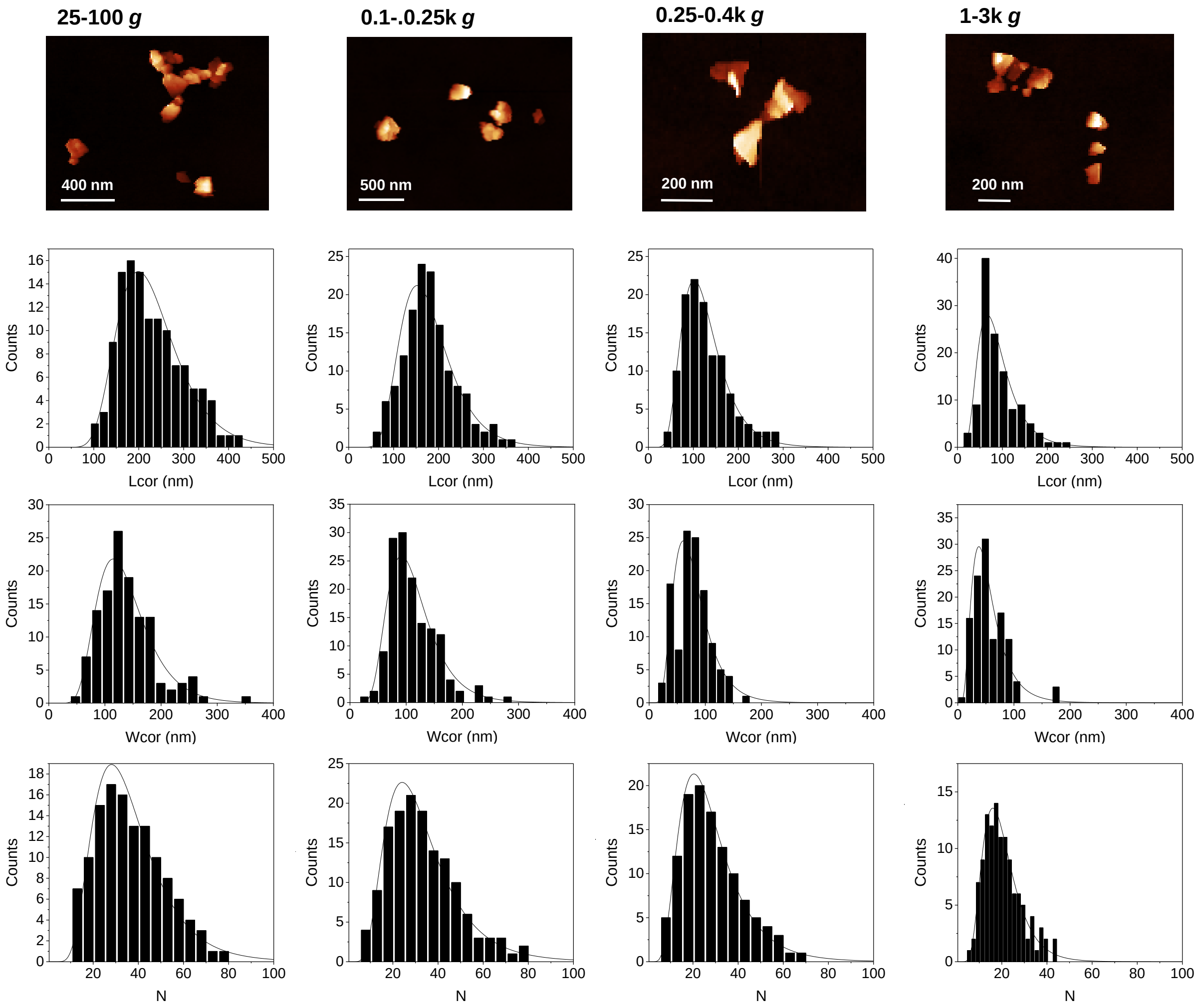

**Figure S22: AFM analysis of the $Zn(OH)_2$ fractions.** Top row: representative images of the fractions isolated at the centrifugal acceleration indicated. Distribution histograms of i) second row: longest lateral dimension, length, *L*, (in nm), ii) third row: Dimension perpendicular to *L*, termed the width *W* (in nm), iii), fourth row: layer number *N*.

## 1.5 Scaling of nanosheet dimensions *versus* centrifugal acceleration

In the following, we show the scaling of nanosheet dimensions in the fractions as functions of the midpoint of the pair of centrifugal acceleration values used in the liquid cascade centrifugation. For materials for which a significant portion of monolayers (ML) were isolated and identified, information on monolayer content and size is included. Monolayer contents are given as monolayer volume fractions.[1] While the ML data is of minor relevance for this work, we nonetheless include it as it may be of interesting for readers who may aim to produce monolayer-rich dispersions based on the findings of this study. In addition, this data is important to help evaluate whether the AFM statistics are robust, i.e., to verify whether a sufficiently large number of nanosheets was counted.

### *1.5.1 Graphene*

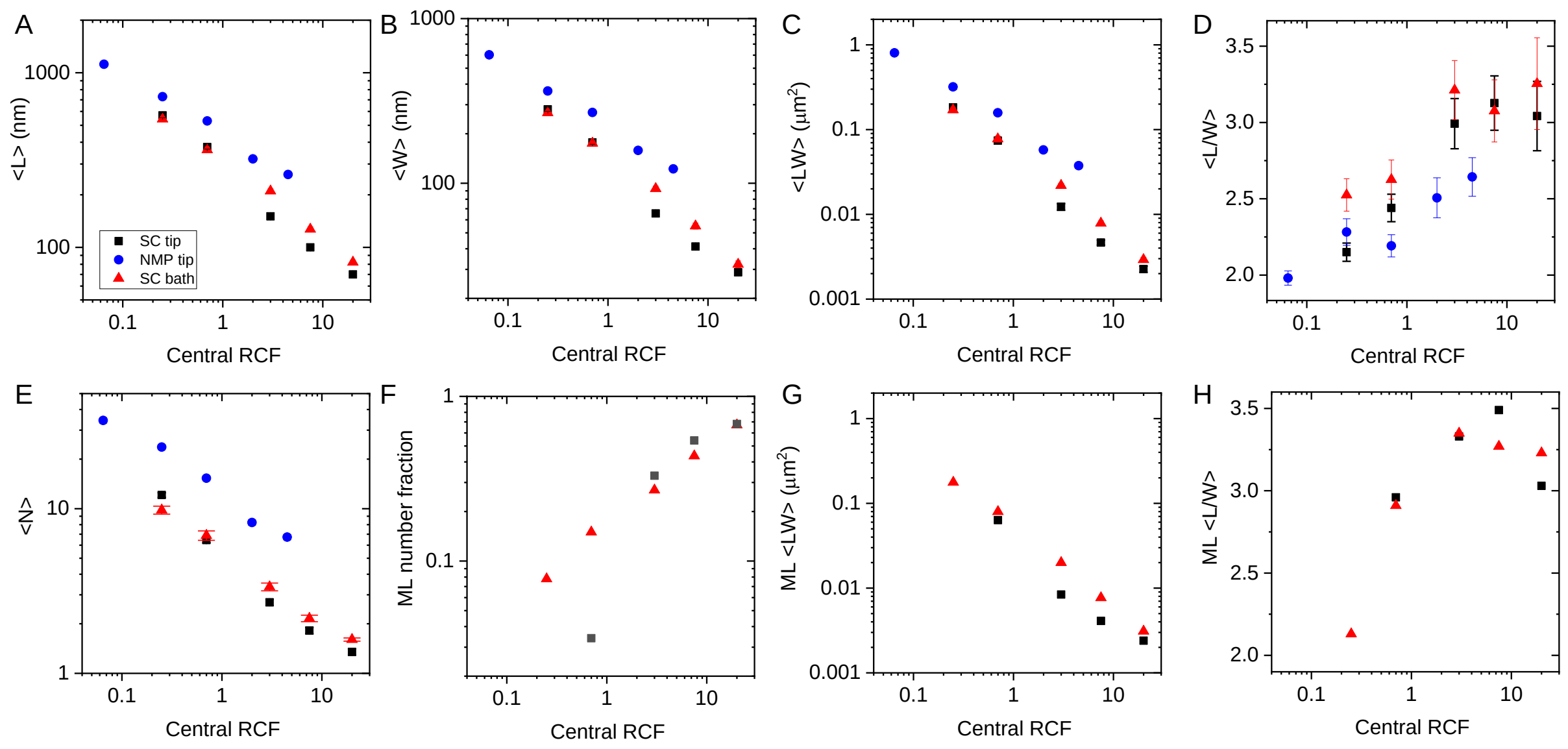

**Figure S23: Scaling of graphene nanosheet dimensions with centrifugation conditions.** All data is plotted as function of the midpoint of the pair of centrifugal accelerations (expressed as relative centrifugal force in $10^3$ *g*). A) Mean length, <*L*>, B) Mean width, <*W*>, C) Mean area expressed as <*LW*>, D) Mean length-width aspect ratio expressed as <*L*/*W*>, E) Mean layer number, <*N*>, F) Monolayer volume fraction, *ML Vf*, G) Mean monolayer area, *ML* <*LW*>, H) Monolayer length-width aspect ratio, *ML* <*L*/W>.

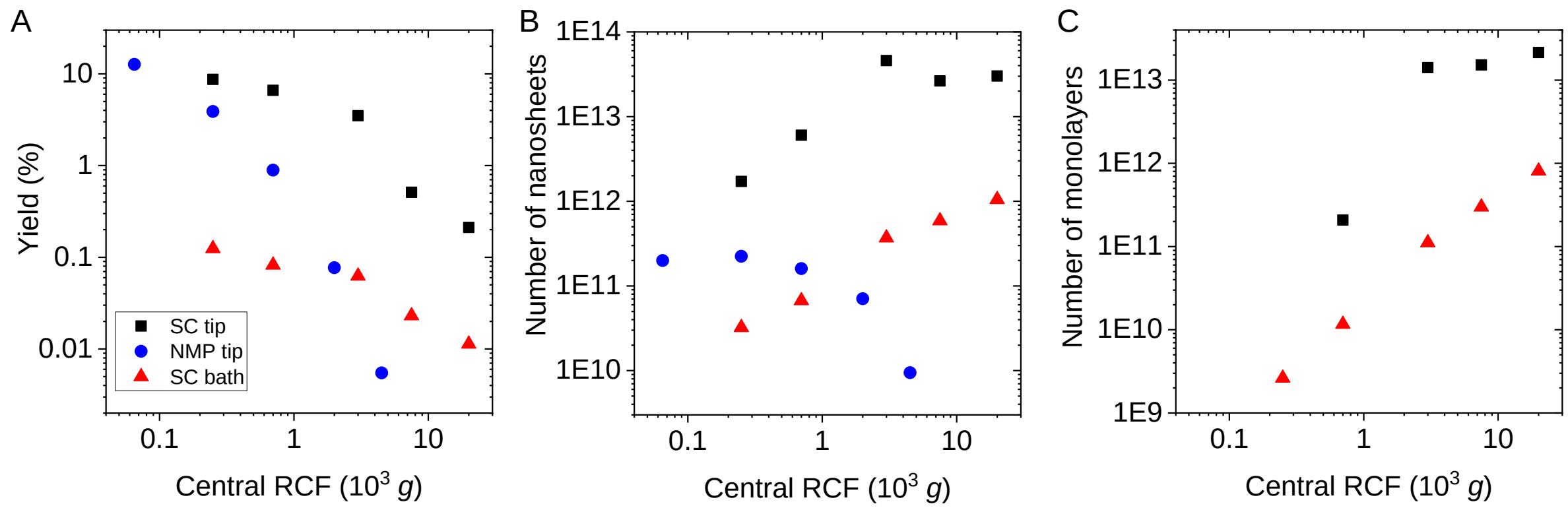

**Figure S24: Graphite yield and population of nanosheets in the LCC fractions.** Three different exfoliation conditions are compared: Tip sonication in aqueous SC, tip sonication in NMP and bath sonication in SC. A) Yield as function of RCF, B) Total number of nanosheets as function of RCF. In spite of the lower mass in fractions isolated at higher centrifugal accelerations, the number of nanosheets (calculated with knowledge of mass and dimensions) increases with increasing RCF in the case of the cholate samples. However, it decreases steeply for the NMP dispersions evidencing that only very few small and thin sheets are produced. C) Plot of number of monolayers as function of RCF. The monolayer population in NMP was too small to be statistically relevant.

### *1.5.2 TMDs*

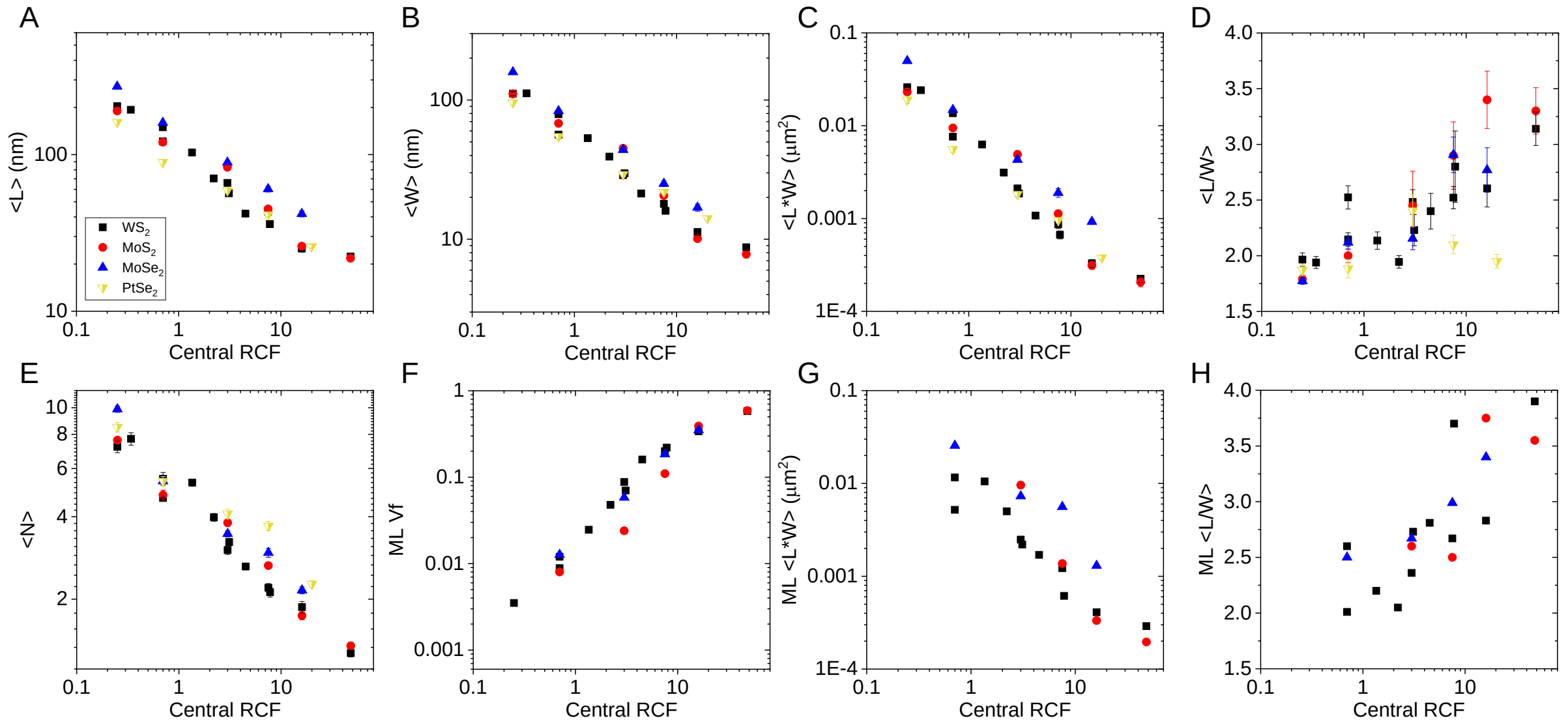

**Figure S25: Scaling of TMD nanosheet dimensions with centrifugation conditions.** All data is plotted as function of the midpoint of the pair of centrifugal accelerations (expressed as relative centrifugal force in $10^3$ *g*). A) Mean length, *<L>*, B) Mean width, *<W>*, C) Mean area expressed as *<LW>*, D) Mean length-width aspect ratio expressed as *<L/W>*, E) Mean layer number, *<N>*, F) Monolayer volume fraction, *ML Vf*, G) Mean monolayer area, *ML <LW>*, H) Monolayer length-width aspect ratio, *ML <L*/W>.

### *1.5.3 BN*

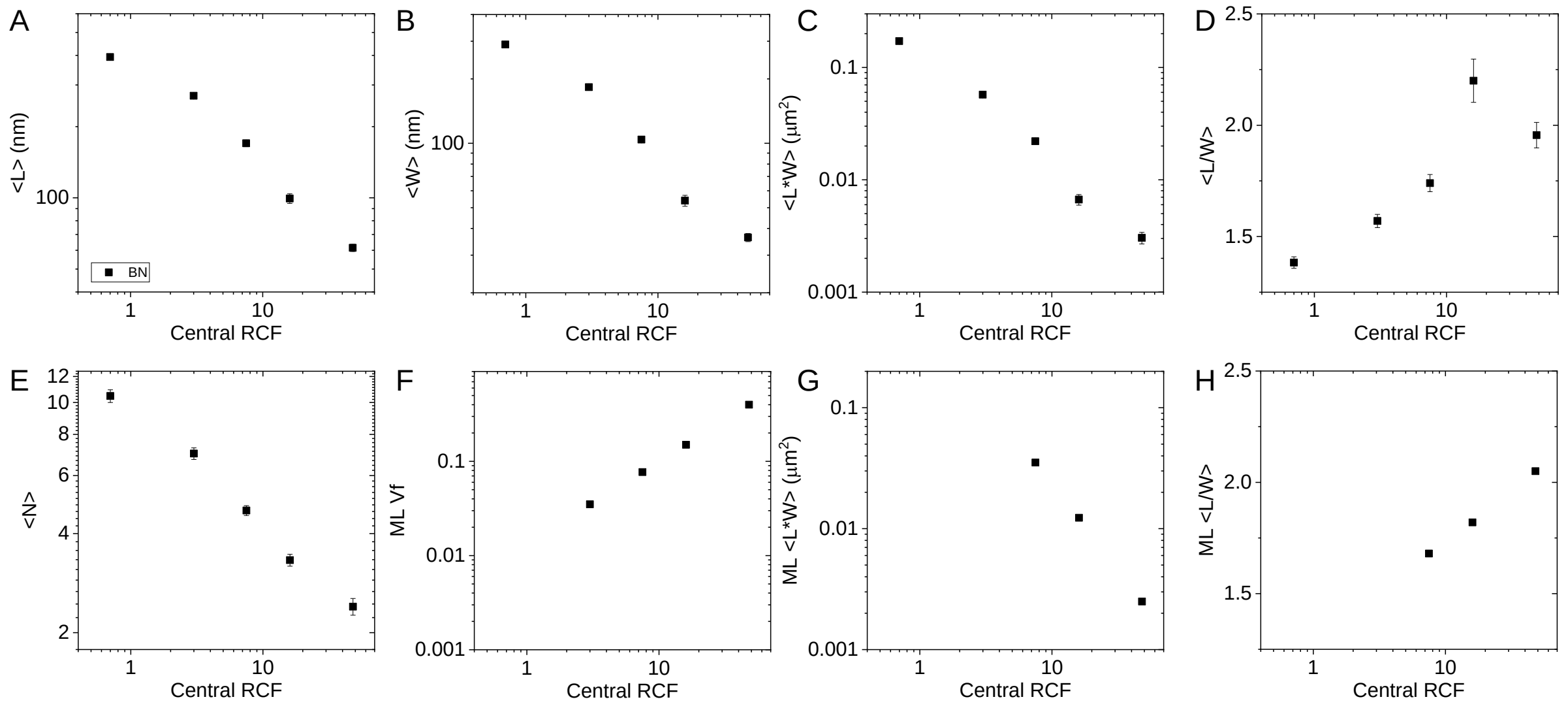

**Figure S26: Scaling of BN nanosheet dimensions with centrifugation conditions.** All data is plotted as function of the midpoint of the pair of centrifugal accelerations (expressed as relative centrifugal force in $10^3$ *g*). A) A) Mean length, <*L*>, B) Mean width, <*W*>, C) Mean area expressed as <*LW*>, D) Mean length-width aspect ratio expressed as <*L/W*>, E) Mean layer number, <*N*>, F) Monolayer volume fraction, *ML Vf*, G) Mean monolayer area, *ML* <*LW*>, H) Monolayer length-width aspect ratio, *ML* <*L/*W>.

### *1.5.4 GaS*

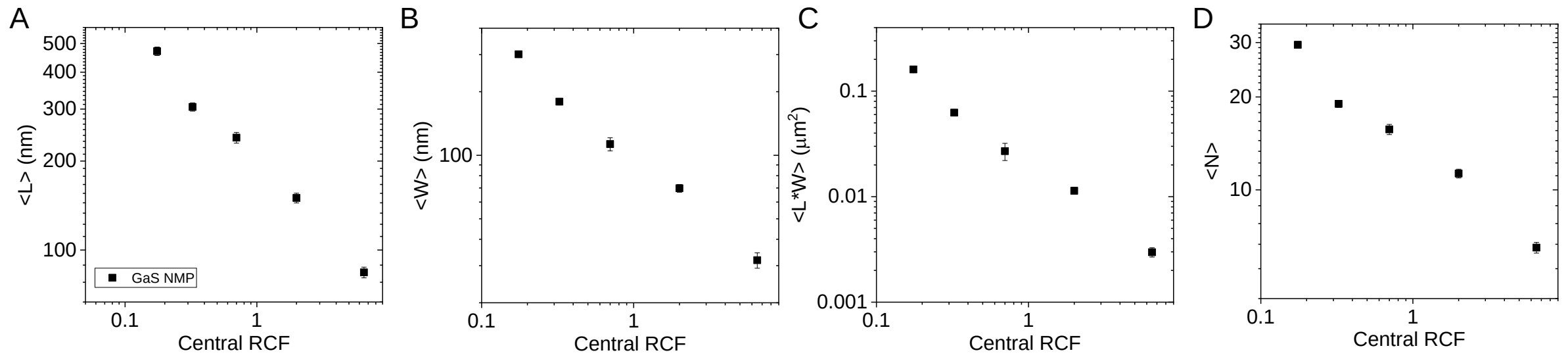

**Figure S27: Scaling of GaS nanosheet dimensions with centrifugation conditions.** All data is plotted as function of the midpoint of the pair of centrifugal accelerations (expressed as relative centrifugal force in $10^3$ *g*). A) Mean length, <*L*>, B) Mean width, <*W*>, C) Mean area expressed as <*LW*>, D) Mean layer number, <*N*>.

### 1.5.5 *Talc*

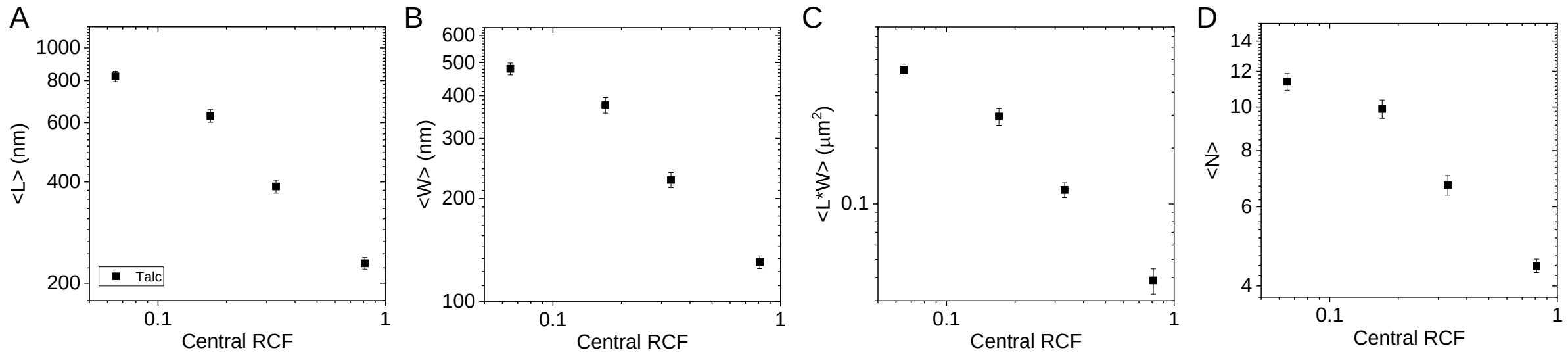

**Figure S28: Scaling of talc nanosheet dimensions with centrifugation conditions.** All data is plotted as function of the midpoint of the pair of centrifugal accelerations (expressed as relative centrifugal force in $10^3$ *g*). A) Mean length, <*L*>, B) Mean width, <*W*>, C) Mean area expressed as <*LW*>, D) Mean layer number, <*N*>.

### 1.5.6 *Hydroxides*

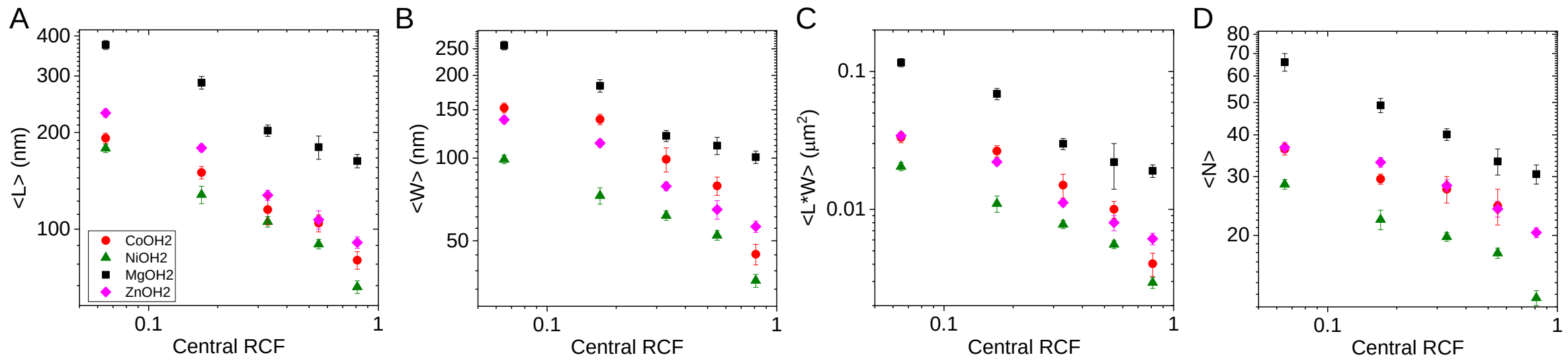

**Figure S29: Scaling of hydroxide nanosheet dimensions with centrifugation conditions.** All data is plotted as function of the midpoint of the pair of centrifugal accelerations (expressed as relative centrifugal force in $10^3$ *g*). A) Mean length, <*L*>, B) Mean width, <*W*>, C) Mean area expressed as <*LW*>, D) Mean layer number, <*N*>.

## 1.6 Scaling of nanosheet dimensions *versus* each other

As illustrated in figure 4 in the main manuscript and the plots above, <*L*>, <*W*>, and <*N*> change following power laws with increasing centrifugal acceleration. Since all quantities scale with the midpoint of the centrifugation cascade, the nanosheet dimensions are also expected to scale with each other. This is summarised in figure S30. <*L*>, <*W*>, and hence <*LW*> scale as a power law with <*N*> in a similar way (see Fig. S30A-C).

To develop the model in the main manuscript, we used <*LW*> as an approximation for the nanosheet area, rather than simply squaring the longest dimension <*L*>. Due to different length-width aspect ratios present, this is essential in order to compare different materials, where this quantity can vary significantly. While most of the materials under study have a <*L/W*> of 2-3 (Figure S30D), we deliberately included materials with lower <*L/W*> aspect ratios to test whether our model can sufficiently describe LPE of materials with varying crystallite shape. Examples are BN (disks, <*L/W*> < 2) and $Co(OH)_2$ or $Mg(OH)_2$ (hexagons, <*L/W*> < 2) respectively. In addition to these considerations, we note that <*L/W*> changes systematically with <*N*> within one material across the different fractions (Fig. S30D). This clearly is a manifestation of changes in shape that are a direct indication for sonication-induced scission of nanosheets. Interestingly, in all cases, <*L/W*> increases as the nanosheets become thinner (and smaller), i.e. as the nanosheets become more and more belt-like. As shown in figure S2B, this is also observed in stock dispersions and hence it is not a result of the centrifugation. The reason for this is currently unclear and we postulate that it reflects a tearing mechanism that does not occur at random positions of the nanosheets, but occurs instead in a zipper-type fashion.

Throughout the main manuscript, we used the arithmetic mean to describe the midpoint of the *L*, *W*, *N* distributions of the dispersions. While appropriate for this work, the arithmetic mean may not always be the best measure to describe the mean layer number, as – even without a scaling of area and layer number – thicker nanosheets make up a larger portion of the nanosheet mass than thinner nanosheets do. In such cases, it can be valuable to calculate the volume-fraction weighted mean layer number, <N>$_{Vf}$. Throughout this study, we realised that the calculation of <N>$_{Vf}$ can be extremely useful as a diagnostic tool, simply because it scales with the arithmetic <*N*> (Fig. S30E). Importantly, this is only the case if the statistics are robust, i.e., if a sufficiently large number of nanosheets were counted in each fraction. Within this study, sufficiently large proved to mean 200-300 nanosheets in polydisperse fractions isolated at low centrifugal accelerations, down to a minimum of 150 nanosheets for less polydisperse fractions isolated at the highest centrifugal accelerations. Empirically, we find that

$$\langle N \rangle = 0.38 + 0.65 \langle N \rangle_{Vf} \qquad \text{(S1)}$$

As shown in figure S30E, all materials under study collapse on a master-curve that can, in turn, be used to assess whether the AFM statistics are robust. For example, if outliers are not removed prior to calculating the arithmetic means, or conversely, if the long tail of the log-normal histograms is not sufficiently populated, a deviation from the mastercurve is observed and this indicates that more nanosheets need to be analysed.

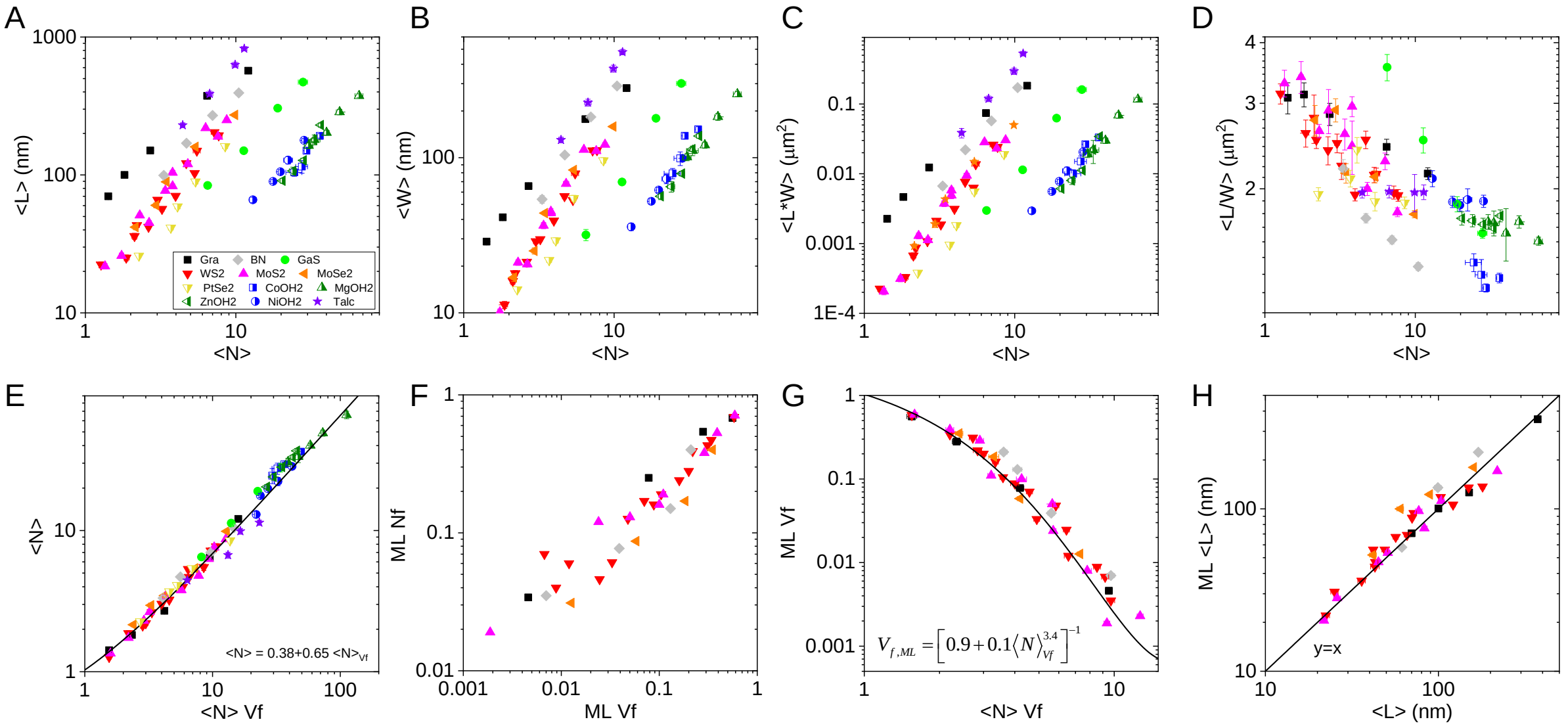

**Figure S30: Scaling of nanosheets dimensions *versus* each other.** The well-defined scaling can be used to test the robustness of the AFM statistics. For example, the master-curves shown in panels E-H are only obtained if a sufficient number of nanosheets are measured in each fraction. A-D) Plots of nanosheet lateral dimensions as functions of layer number, <*N*>. A) Mean length, <*L*>, as a function of <*N*>, B) Mean width, <*W*>, as a function of <*N*>, C) Mean nanosheet area <*LW*> as a function of <*N*> (same data as shown in figure 3A of the main manuscript), D) Mean nanosheet length-width aspect ratio, <*L/W*>, as a function of <*N*>. The changes in <*L/W*> are likely to reflect shape changes due to sonication-induced scission. E) Plot of the arithmetic mean layer number, <*N*>, as a function of the volume fraction weighted mean layer number, $\langle N \rangle_{Vf}$. A robust master-curve is obtained, with data points from all materials collapsing on the *same* curve. F) Similarly to the case of the mean layer number shown in (E), the monolayer number fractions (*ML Nf*) scale with the monolayer volume fraction (*ML Vf*). Note that only materials that contain sufficient monolayer contents are included. G) Combing the data in (E-F), we find that the monolayer volume fractions scale robustly with the volume-fraction weighted mean layer number (and hence with thearithmetic mean layer number). Again, a robust master-curve is obtained. H) Plot of monolayer length, *ML* <*L*>, as a function of the mean <*L*> of all nanosheets. The line is a guide for the eye of slope 1. It describes the experimental data very well.

Even though this study is not concerned with the production of dispersions of high monolayer content, this is an important aspect that can be extracted from our statistical AFM data and will therefore be summarised below for the interested reader. Similar to the correlation between arithmetic and volume-fraction weighted mean, we find a scaling of monolayer number ($ML_{Nf}$) and volume fraction ($ML_{Vf}$), and this is plotted in figure S28F. Only a subset of the materials (TMDs, graphene, BN) is included, as the monolayer content was not sufficient in the other cases (hydroxides, GaS, talc). Overall, we find a similar linear scaling as in the case of the mean layer number, albeit with more scatter (this could have been improved by counting more nanosheets).

This result implies that the monolayer content scales with the layer number. To illustrate this, we plot $ML_{Vf}$ as a function of $<N>_{Vf}$ in figure S30G. Empirically, we find that

$$ML_{Vf} = \left[0.9 + 0.1\ < N_{Vf} >^{3.4}\right]^{-1} \qquad \text{(S2)}$$

Similar to equation (1), this empirical scaling can be used to test the robustness of the AFM statistical analysis, in particular because we find a low scatter in this plot across all materials.

In figure S30H, we plot the monolayer length ($ML<L>$) as a function of the arithmetic mean length of all nanosheets. The line of slope 1 illustrates that the lateral size of the monolayers scales perfectly with the lateral size of all nanosheets. Practically speaking, to maximise the monolayer length, for example, by developing novel size selection or exfoliation schemes, it is important to maximise <*L*> in general, a quantity that is often accessible through the use of spectroscopic metrics.[1, 3-7]

## **1.7 Fits of <*LW*> versus <*N*> (also versus in Contents)**

The fits of <*LW*> *versus* <*N*> from the main manuscript are shown in more detail in figure S31 with each group of materials in a single panel to avoid clutter. Exponents are indicated for each material in the respective panel. Since materials within a certain material class (TMDs, hydroxides), sit very close beside each other on the <*LW*> *versus* <*N*> relationship, we examined them as a group and individually (Fig, S32) to test the variation in the exponents in each case. While exponents are very similar in the TMD family, there are non-negligible variations in the group of the hydroxides. We suspect that this is because the nanosheets differ significantly in their shape (opposed to TMDs), as indicated by the AFM images above: $Co(OH)_2$ and $Mg(OH)_2$ nanosheets are predominantly found as hexagons, and $Ni(OH)_2$ are similar in shape to the TMDs, often containing triangular nanosheets. We suspect that these variations in shape lead to a spread in exponents (see below) and thus treated all materials (also of the same class) individually for the remainder of the work.

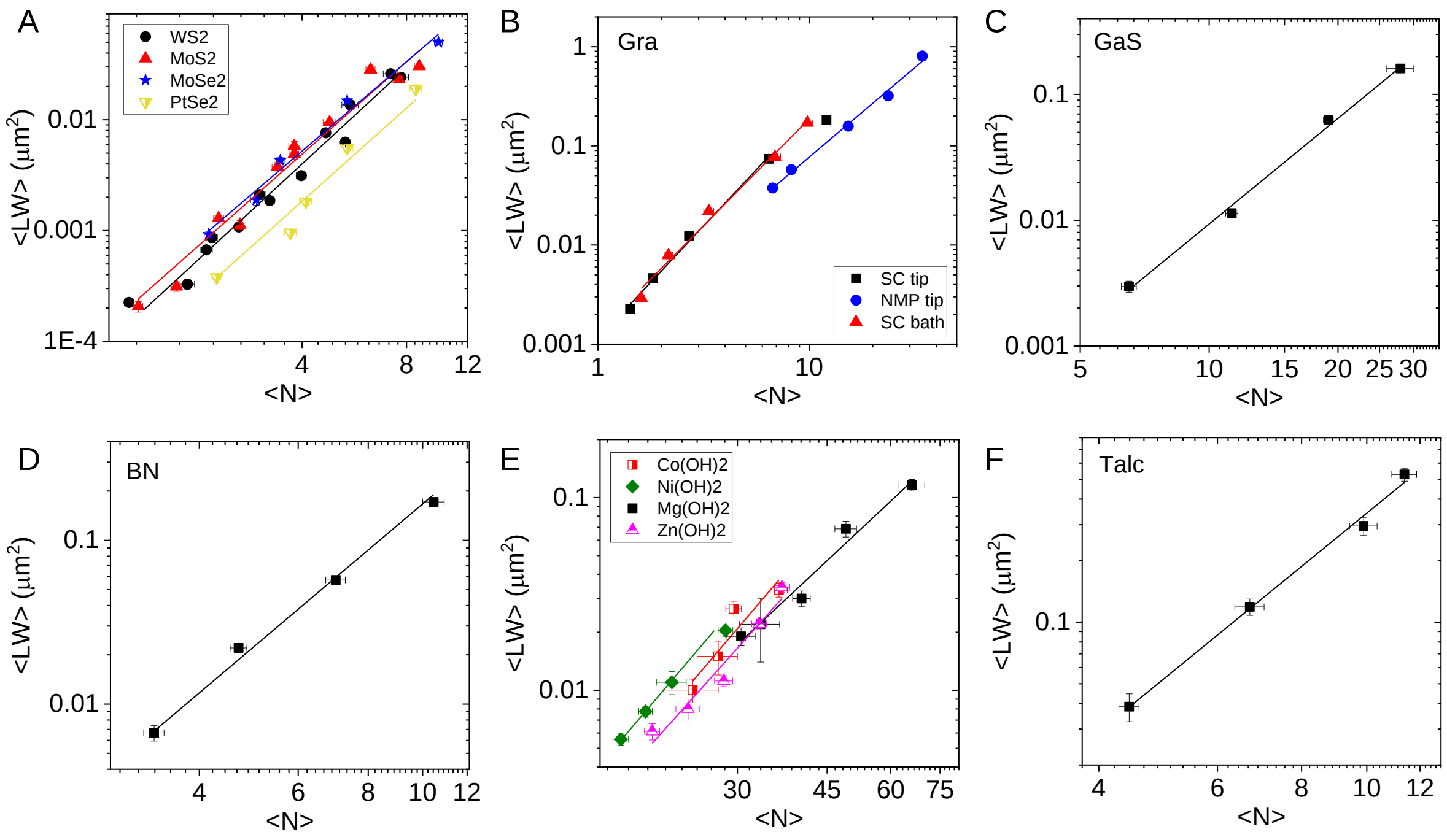

**Figure S31: Fits of <*LW*> *versus* <*N*> with materials in different panels for clarity.** The extracted fit parameters are summarized in table S1. A) TMDs; B) Graphene; C) GaS; D) BN; E) Hydroxides, and F) Talc.

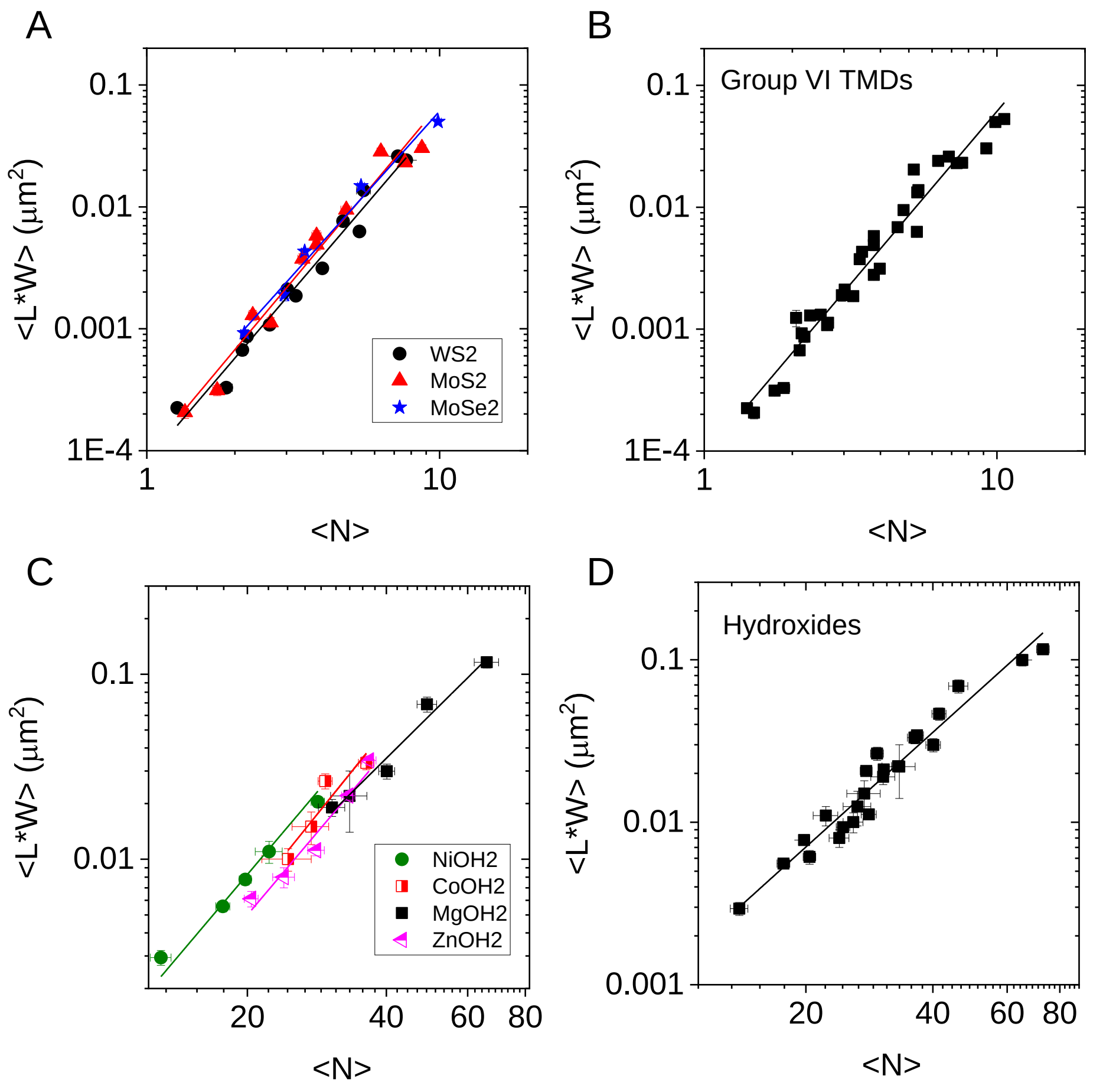

**Figure S32: Comparison of fits of <*LW*> *versus* <*N*> with materials of one class treated individually or as a group.** A-B) Group-VI TMDs, B-C) Hydroxides. In the case of the TMDs, the individual fits (A) are similar to the fit of the entire group (B). However, in case of the hydroxides, the exponent of the individual fits (C) vary, so that a grouped fit (D) is not appropriate. We attribute this to the significant variations on nanosheet shape in case of the hydroxides. Each material was therefore treated separately.

## 1.8 Validity of $\langle N(LW)^{0.5}\rangle = \langle LW\rangle^{0.5}\langle N\rangle^{b}$

An assumption used to simplify equation (6) in the main manuscript is that the *L*, *W*, *N* distribution data is consistent with the approximation

$$\left\langle N\sqrt{LW}\right\rangle = \sqrt{\langle LW\rangle}\,\langle N\rangle^{b} \quad . \qquad \text{(S3)}$$

This assumption holds, for example, if *L* and *W* can be modelled as being independently log-normally distributed, where the factor for geometric variances $\exp(-(\sigma_L^2+\sigma_W^2)/8)$, which results from pulling the square-root out of the expectation value, can be modelled as a weak thickness dependence $c\langle N\rangle^{b-1}$. If this is case, particularly if *L* and *W* depend on *N* with relatively narrow log-normal distributions, but noting that log-normal distributions are not the only possible valid distribution assumption *per se*, the idea is that we might ideally find that $\beta = 2b$ (with $\beta$ being the exponent of the <*LW*> versus <*N*> power law fits). We note, in passing, that (Eq. S3) is compatible with the medians of the log-normal distributions (i.e. geometric means exp($\mu_L$) and exp($\mu_W$)) being *arbitrarily* <*N*>-dependent, and that *N* itself does not need to be log-normally distributed.

To test this, we plot $\langle N(LW)^{0.5}\rangle/\langle LW\rangle^{0.5}$ as function of <*N*> and fit the data to a power law of the form $y = c\, x^b$ to extract the exponent *b* and the constant *c* which we expect to be closest to 1 for narrowest distributions. The fits are shown in figure S33.

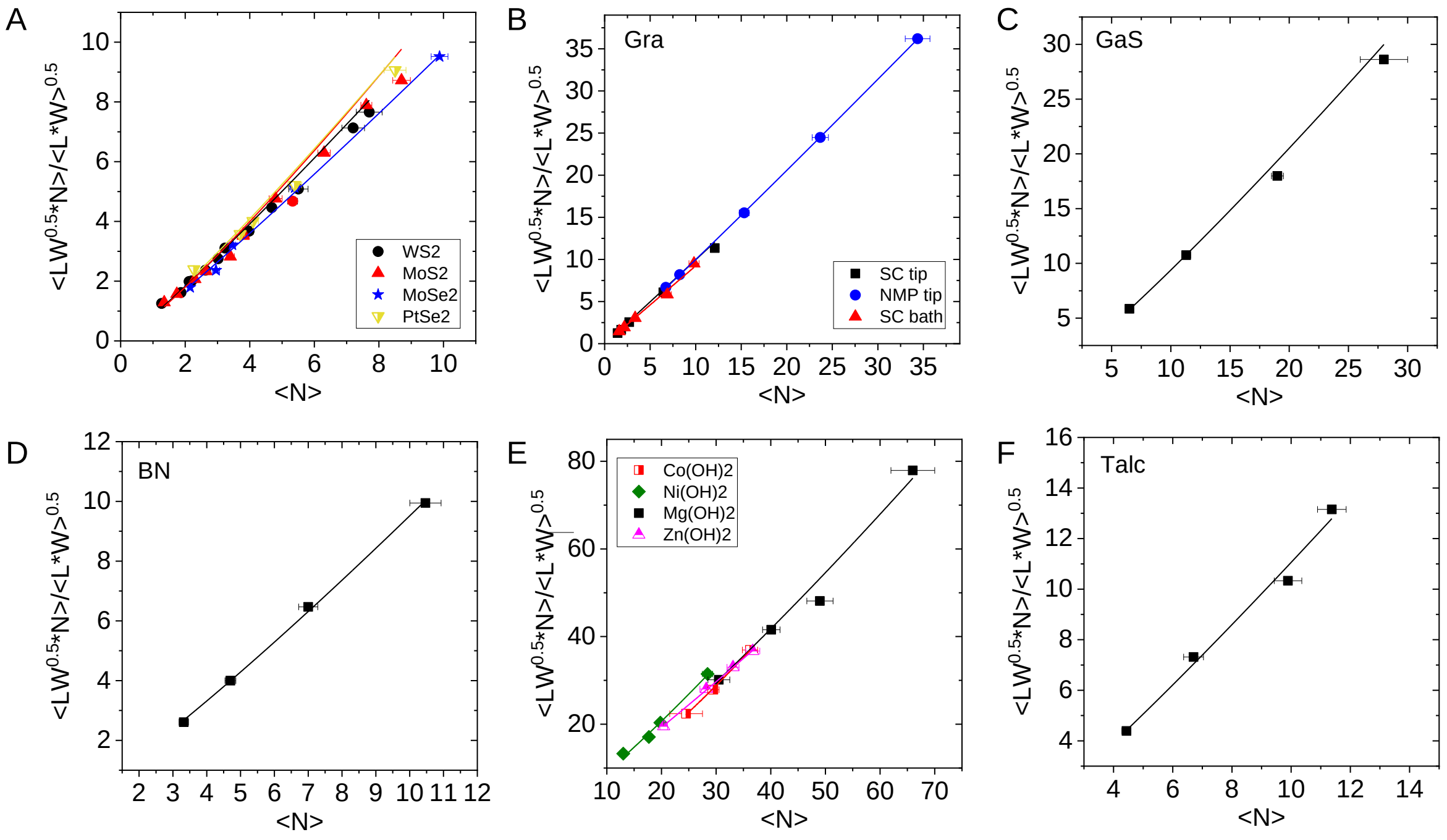

**Figure S33: Plots of $\langle N(LW)^{0.5}\rangle/\langle LW\rangle^{0.5}$ as function of <*N*> for the different materials to test the validity of equation (S3).** The data is fit to a power law to extract the exponent *b* and the constant *c* (see table S1). A) TMDs; B) Graphene; C) GaS; D) BN; E) Hydroxides, and F) Talc.

The extracted fit parameters *b* and *c* are plotted versus each other in figure S34A and are tabulated in table S1. Interestingly, we find that the parameters are correlated. For many materials (all TMDs, graphene, talc, $Zn(OH)_2$), we find that *c* is in the range 0.8-1. Only for two materials ($Mg(OH)_2$ and $Co(OH)_2$), *c* drops to values of 0.4-0.5. We believe that this is because the fit quality (figure S33E) is rather poor in these cases. In figure S34B, the exponent *b* from the shape fitting (figure S33) is plotted as function of the exponent $\beta$ from the nanoflake dimension power law fits (figure S31). In line with the model described in the main manuscript, we indeed find that the data is broadly consistent with $\beta$=2b, albeit rather scattered, with an offset of approximately 0.5 as indicated by the line. This is possibly due to a species-dependent ,systematic <*N*>-dependence in the nanoflake shape, which is assumed to be elliptical here. Given the nanosheet shape polydispersity, this is very encouraging.

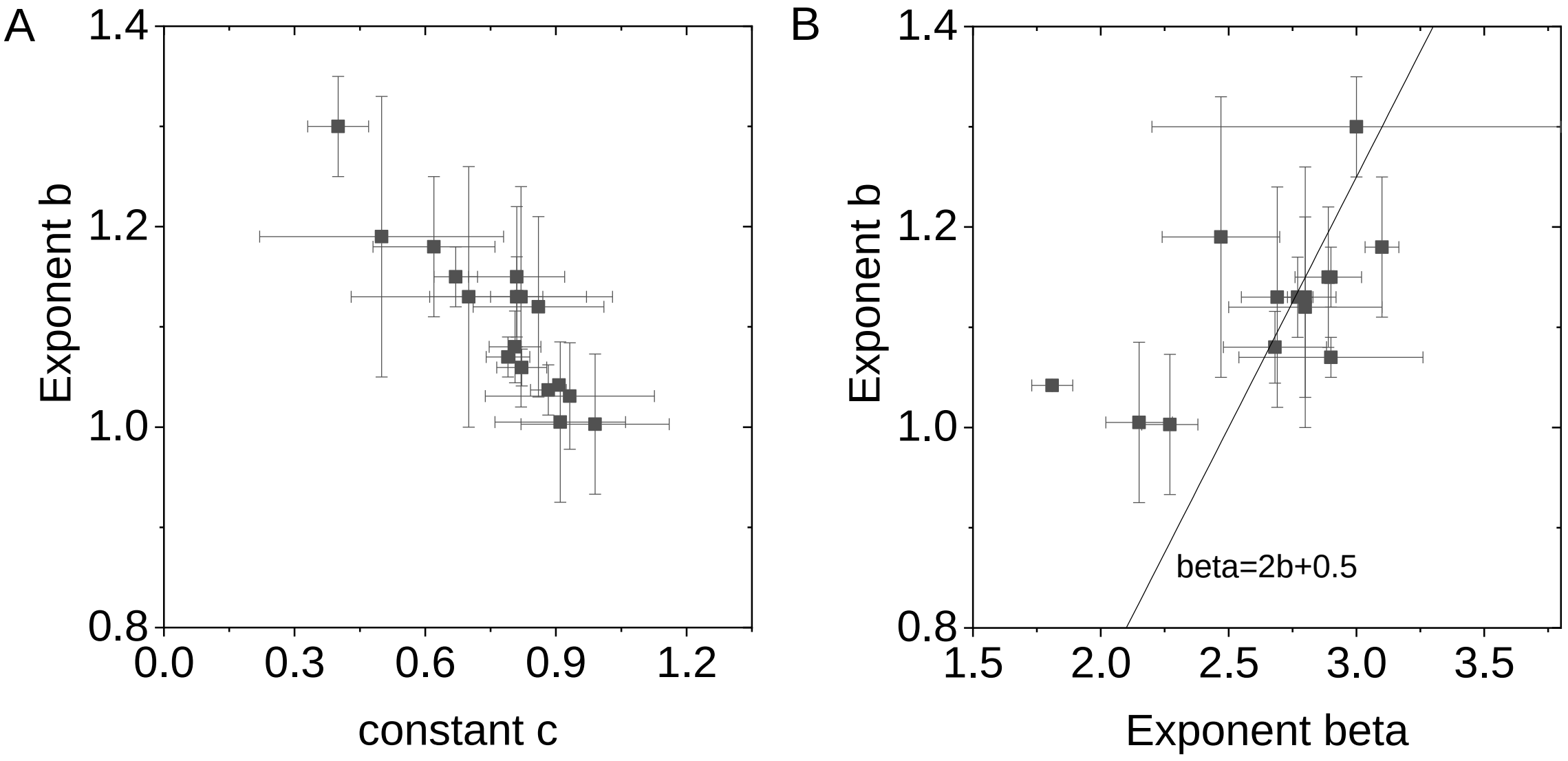

**Figure S34: Result of the shape fittings.** A) Plot of the parameters *b* and *c* extracted from the shape power law fittings *versus* each other. B) Plot of the exponents *b* as function of the <*LW*> versus <*N*> exponents (β). The model detailed in the main manuscript implies that the exponents are related to each other by a factor of 2 . Within error of the fits, this is broadly the case albeit with an offset.

Since the constant $c \neq 1$ is required to fit the shape power laws shown in figure S33, it should in principle be taken into account in the overall model. When implemented, we expect that

$$\frac{D_{ML}}{h_0} = 2ac\frac{E_E}{E_S} \qquad \text{(S4)}$$

should be more accurate. Therefore, we plot $D_{ML}/(h_0c)$ *versus* the in-plane to out-of-plane modulus ratio in figure S35. Since we found that c < 1, all data points move up slightly compared to figure 5E. However, the data is still mostly consistent with *a=1* within the combined errors of the fits (recalling that *a* is the sought-after deviation from equipartion factor, and that *a=1* implies that equipartition of energy holds during LPE). Given the uncertainty in both experimental data, shape assumptions, neglect of the offset, and the approximation of the in-to-out-of-plane modulus ratio from the first-principles calculated elastic constants (see below), the agreement with the model (including c) remains remarkable.

However, because we find that c≈1, to simplify the discussion, in the main text, we set c=1.

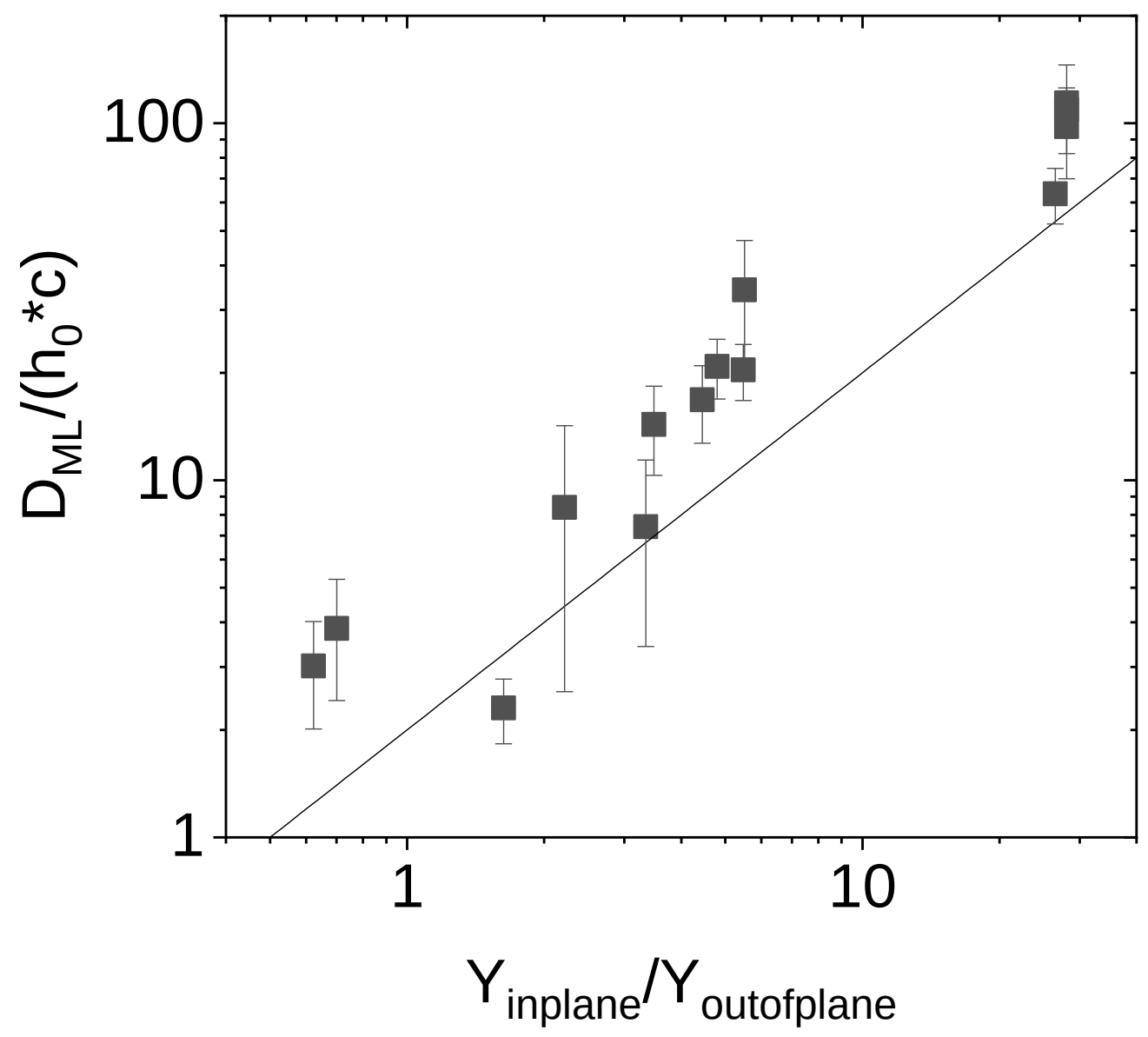

**Figure S35: Impact of the constant *c* extracted from the shape fittings.** Plot of $D_{ML}/(h_0c)$ as a function of the in-plane to out-of-plane Young's modulus ratio.

**Table S1: Parameters extracted from the fits shown in figure S26, 28.**

| Material | ***β*** | **err *β*** | **log Int. <L*W>-<N>** | **err log Int. <L*W>-<N>** | **$D_{ML}$ (nm)** | **err $D_{ML}$** | ***b*** | **err *b*** | ***c*** | **err *c*** |
|---|---|---|---|---|---|---|---|---|---|---|
| $MoS_2$ | 2.77 | 0.04 | -3.98 | 0.04 | 10.23 | 0.051 | 1.13 | 0.04 | 0.81 | 0.06 |
| $WS_2$ | 2.89 | 0.13 | -4.14 | 0.073 | 8.51 | 0.075 | 1.15 | 0.07 | 0.81 | 0.11 |
| $MoSe_2$ | 2.68 | 0.2 | -3.89 | 0.13 | 11.26 | 0.190 | 1.08 | 0.04 | 0.81 | 0.06 |
| Gra SC tip | 2.27 | 0.11 | -2.94 | 0.079 | 33.88 | 0.457 | 1.00 | 0.07 | 0.99 | 0.17 |
| Gra SC bath | 2.15 | 0.13 | -2.88 | 0.08 | 2.15 | 0.13 | 1.005 | 0.08 | 0.91 | 0.15 |
| Gra NMP tip | 1.81 | 0.08 | -2.92 | 0.097 | 1.81 | 0.08 | 1.042 | 0.003 | 0.907 | 0.011 |
| GaS | 2.80 | 0.12 | -4.82 | 0.5 | 3.89 | 0.201 | 1.13 | 0.13 | 0.7 | 0.27 |
| BN | 2.90 | 0.0035 | -3.68 | 0.02 | 14.45 | 0.039 | 1.15 | 0.03 | 0.67 | 0.05 |
| $Ni(OH)_2$ | 3.10 | 0.066 | -6.13 | 0.09 | 0.86 | 0.006 | 1.18 | 0.07 | 0.62 | 0.14 |
| $Mg(OH)_2$ | 2.47 | 0.23 | -5.41 | 0.38 | 1.97 | 0.069 | 1.19 | 0.14 | 0.5 | 0.28 |
| $Zn(OH)_2$ | 2.90 | 0.36 | -6.1 | 0.52 | 0.89 | 0.037 | 1.07 | 0.02 | 0.79 | 0.05 |
| $Co(OH)_2$ | 3.00 | 0.8 | -6.3 | 1.22 | 0.71 | 0.068 | 1.30 | 0.05 | 0.4 | 0.07 |
| Talc | 2.69 | 0.14 | -3.16 | 0.13 | 25.3 | 0.541 | 1.13 | 0.11 | 0.82 | 0.21 |
| $PtSe_2$ | 2.80 | 0.3 | -4.42 | 0.05 | 6.16 | 0.034 | 1.12 | 0.09 | 0.86 | 0.15 |

## 1.9 Other shapes beyond ellipses

Starting from equation 3 in the main text:

$$\langle A_S \rangle = \frac{ah_0}{2}\frac{E_E}{E_S}\langle PN \rangle$$

It is possible to work out expressions relating mean nanosheet area and thickness for shapes other than ellipses.

In the case of square nanosheets we have:

$\langle A_S \rangle = \langle LW \rangle$ and $\langle PN \rangle = 2\langle N(L+W) \rangle$

For diamonds:

$\langle A_S \rangle = \langle LW \rangle / 2$ and $\langle PN \rangle = 2\left\langle N\sqrt{L^2+W^2} \right\rangle$

Combining with the equation above:

Squares: $\langle LW \rangle = ah_0 \frac{E_E}{E_S}\langle N(L+W) \rangle$

Diamonds: $\langle LW \rangle = 2ah_0 \frac{E_E}{E_S}\left\langle N\sqrt{L^2+W^2} \right\rangle$

Neither of these equations gives a simple relationship between flake area and thickness.

To achieve this, we must make a further assumption, that the nanosheet aspect ratio is fixed: W/L=$A_R$. Then:

Squares: $\langle L^2 \rangle = ah_0 \frac{E_E}{E_S}\frac{(1+A_R)}{A_R}\langle NL \rangle$

Diamonds: $\langle L^2 \rangle = 2ah_0 \frac{E_E}{E_S}\frac{\sqrt{1+A_R^2}}{A_R}\langle NL \rangle$

In both cases, we can make the assumption: $\langle NL \rangle = \sqrt{\langle L^2 \rangle}\langle N \rangle^{\gamma}$

Squares: $\langle L^2 \rangle = \left(2ah_0 \frac{E_E}{E_S}\frac{(1+A_R)}{2A_R}\right)^2 \langle N \rangle^{2\gamma}$

Diamonds: $\left\langle L^2 \right\rangle = \left( 2ah_0 \frac{E_E}{E_S} \frac{\sqrt{1+A_R^2}}{A_R} \right)^2 \left\langle N \right\rangle^{2\gamma}$

Clearly, these expressions are similar each other with the main difference that in both cases, they contain the aspect ratio.

These expressions can be compared with the equivalent expression for an ellipse.

Cf: Ellipse: $\left\langle LW \right\rangle = \left( 2ah_0 \frac{E_E}{E_S} \right)^2 \left\langle N \right\rangle^{2b}$

Taking W/L=$A_R$, gives an equivalent expression for the Ellipse: $\left\langle L^2 \right\rangle = \left( \frac{2ah_0}{\sqrt{A_R}} \frac{E_E}{E_S} \right)^2 \left\langle N \right\rangle^{2b}$

All three expressions have the same form with the only differences being small numerical factors and the detailed dependence on $A_R$. The advantage of using the ellipse, is that equation 8 (main text) could be generated without assuming a fixed aspect ratio.

# 2 Monolayer thickness

**Table S2: Tabulated crystallographic interlayer distance.** The data is extracted from literature as indicated and was used as monolayer thickness $h_0$.

| **Gra** | **BN** | **GaS** | **Talc** | **$WS_2$** | **$MoS_2$** | **$MoSe_2$** | **$PtSe_2$** |
|---|---|---|---|---|---|---|---|
| 0.35 nm[8] | 0.34 nm[9] | 0.75 nm[10] | 0.94 nm[11] | 0.63 nm[12] | 0.62 nm[13] | 0.67 nm[14] | 0.5 nm[15] |

| **$Ni(OH)_2$** | **$Co(OH)_2$** | **$Mg(OH)_2$** | **$Zn(OH)_2$** |
|---|---|---|---|
| 0.46 nm[16] | 0.46 nm[17] | 0.47 nm[18] | 0.49 nm[19] |

**Table S3: Tabulated calculated interlayer distance ($h_{0\text{-theory}}$).** Figure S36 demonstrates excellent agreement with literature.

| **BN** | **GaS** | **Talc** | **$WS_2$** | **$MoS_2$** | **$MoSe_2$** | **$PtSe_2$** |
|---|---|---|---|---|---|---|
| 0.35 nm | 0.77 nm | 0.93 nm | 0.62 nm | 0.62 nm | 0.66 nm | 0.49 m, |

| **$Ni(OH)_2$** | **$Co(OH)_2$** | **$Mg(OH)_2$** | **$Zn(OH)_2$** |
|---|---|---|---|
| 0.43 nm | 0.43 nm | 0.46 nm | 0.46 nm |

Since the thickness of one layer, *i.e.* the interlayer distance of the materials under study is an important parameter to validate our model, it deserved further consideration. The comparison of experimental (crystallographic) data to the computed values ($h_{0\text{-theory}}$) is of particular interest, as the calculated elastic constants rely on an accurate theoretical description of the materials' structures. In figure S36A, the interlayer distance from the computations is plotted as function of the crystallographic monolayer thickness (tabulated above). We find excellent agreement between experiment and theory as indicated by the line (y=x). As a result, the plot of $D_{ML}/h_{0\text{-theory}}$ *versus* in plane and out of plane modulus ratio (figure S36B) is equally compelling as that using the experimental interlayer distances shown in the main manuscript (figure 5E).

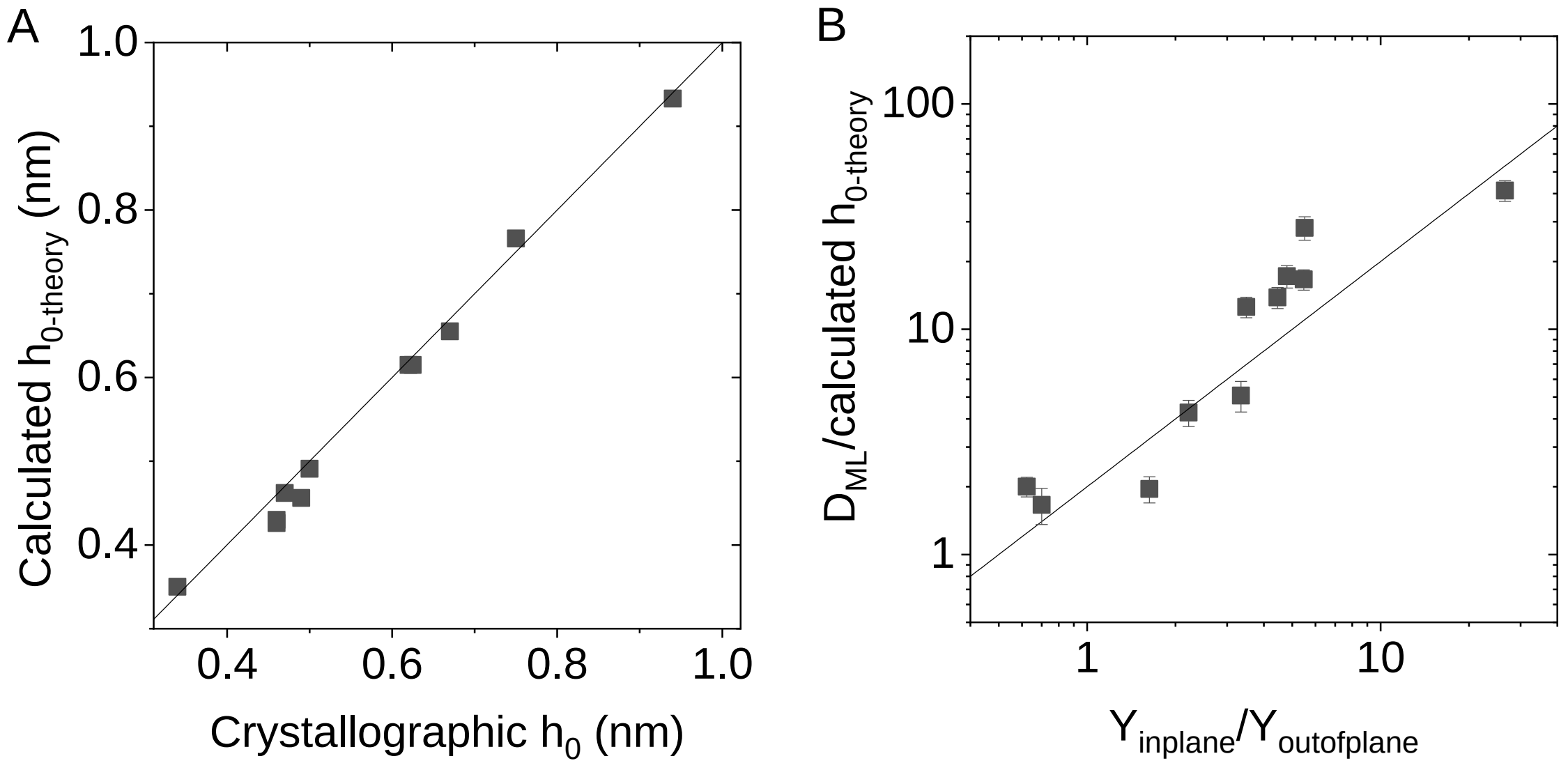

**Figure S36: Calculated interlayer distance.** A) Plot of the interlayer distance calculated within this work as function of the crystallographic data (see table S2-3). The agreement is excellent as indicated by the line (y=x). B) Plot of $D_{ML}/(h_{0\text{-theory}})$ as function of the in plane to out of plane modulus ratio. The line is y=2x.

## 3 Computed binding energies and elastic constants

### 3.1 Graphite

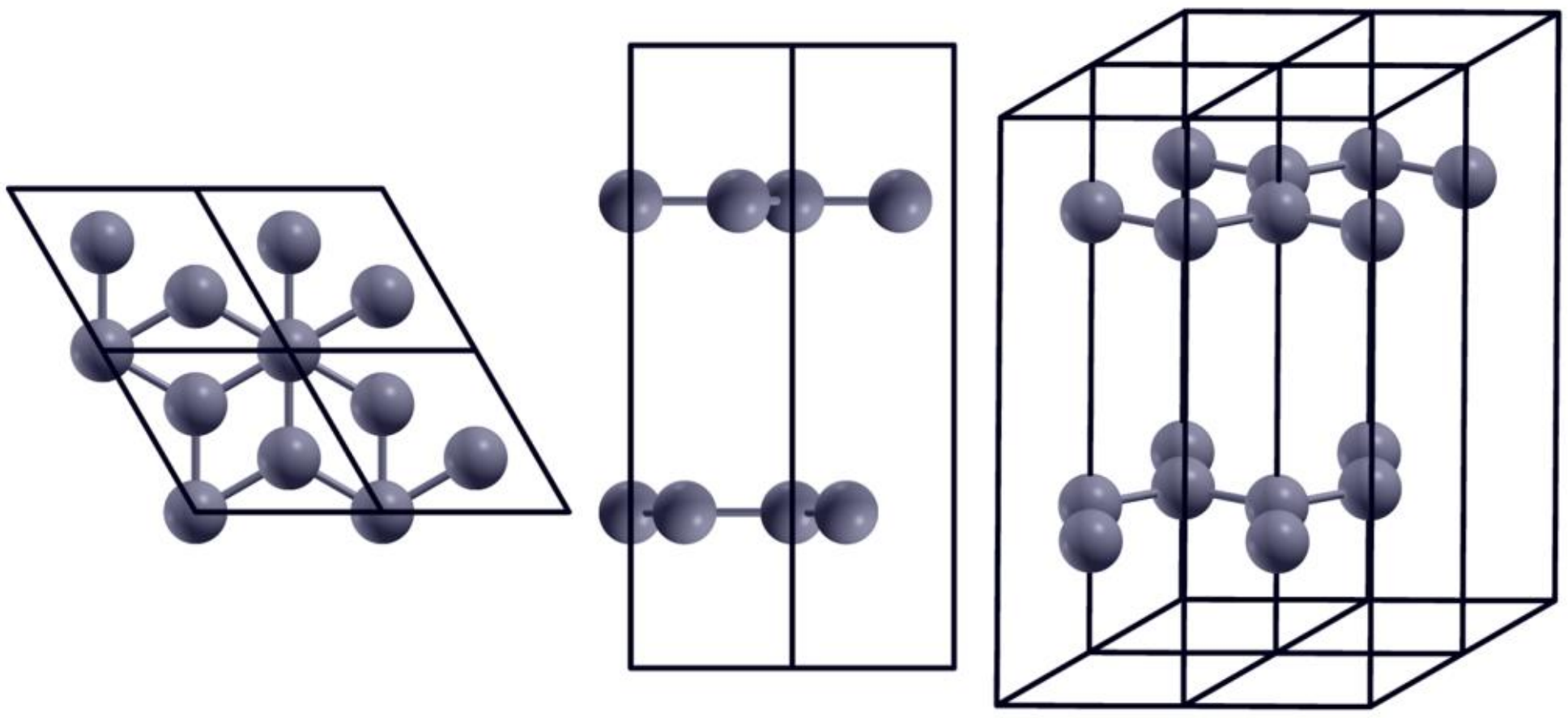

**Table S4: Tabulated calculated binding energy, ICOHP in plane covalent bond strength and elastic constant for AB graphite**

| | **vdW-df2-c09** | **rvv10** |
|---|---|---|
| **Structural parameters (Å)** | | |
| **a** | 2.464 | 2.468 |
| **c** | 6.711 | 6.687 |
| **Binding energy (eV/Å$^2$)** | | |
| | -0.0203 | -0.0255 |
| **Elastic constants (GPa)** | | |
| C11 | 1077 | 1052 |
| C12 | 194.1 | 186.2 |
| C13 | -4.1 | -3.7 |
| C33 | 37.1 | 44.2 |
| C44 | 5.9 | 3.9 |

| | |
|---|---|
| **ICOHP in plane bond strength (vdW-df2-c09) (eV)** | |
| C-C | -11.73 |
| **ICOHP in plane bond strength per edge (eV/Å)** | |
| | -4.756 |
| **ICOHP in plane bond strength per edge /$h_0$ (eV/Å$^2$)** | |
| | -1.422 |

## 3.2 $WS_2$

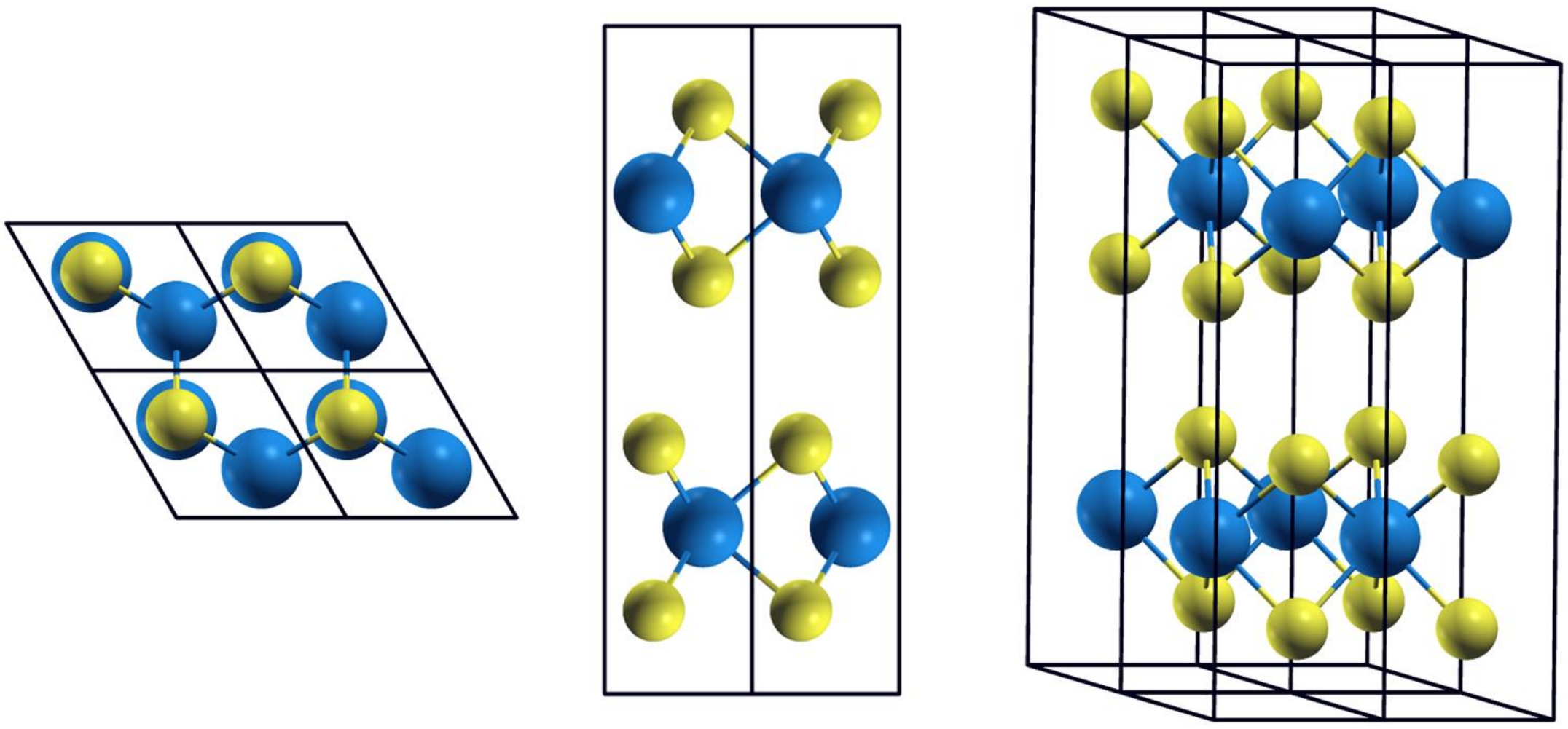

**Table S5: Tabulated calculated binding energy, ICOHP in plane covalent bond strength and elastic constant for 2H-$WS_2$**

| | vdW-df2-c09 | rvv10 |
|---|---|---|
| **Structural parameters (Å)** | | |
| **a** | 3.156 | 3.213 |
| **c** | 12.297 | 12.462 |
| **Binding energy (eV/Å$^2$)** | | |
| | -0.0226 | -0.0299 |
| **Elastic constants (GPa)** | | |
| C11 | 250 | 226.7 |
| C12 | 53.5 | 52.4 |
| C13 | 10.2 | 13.5 |
| C33 | 53.6 | 59.3 |
| C44 | 19.7 | 15.4 |

| | |
|---|---|
| **ICOHP in plane bond strength (vdW-df2-c09) (eV)** | |
| W-S | -5.34 |
| **ICOHP in plane bond strength per edge (eV/Å)** | |
| | -3.384 |
| **ICOHP in plane bond strength per edge /$h_0$ (eV/Å$^2$)** | |
| | -0.550 |

### 3.3 $MoS_2$

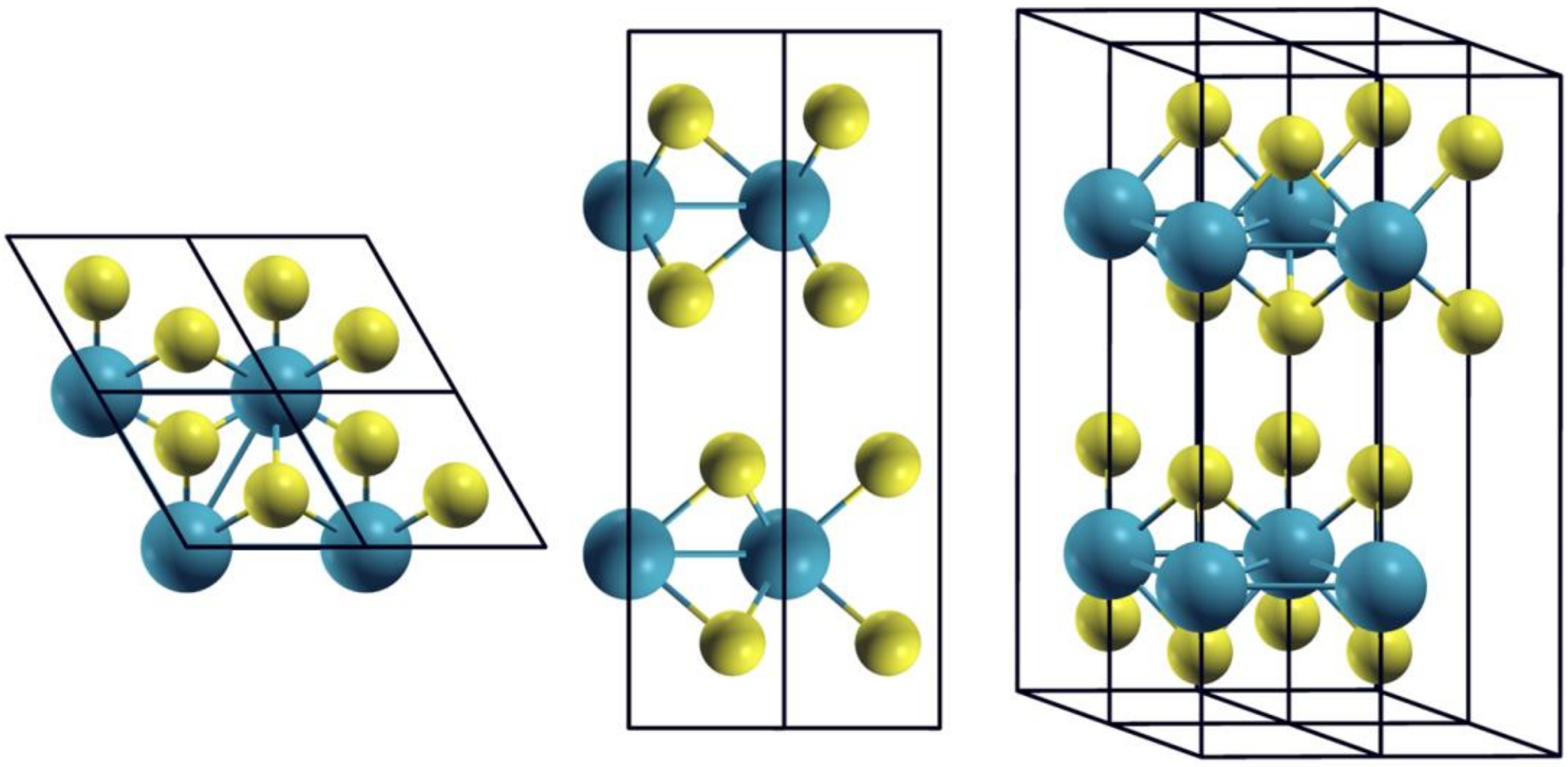

**Table S6: Tabulated calculated binding energy, ICOHP in plane covalent bond strength and elastic constant for 2H-$MoS_2$**

| | **vdW-df2-c09** | **rvv10** |
|---|---|---|
| **Structural parameters (Å)** | | |
| **a** | 3.158 | 3.221 |
| **c** | 12.304 | 12.406 |
| **Binding energy (eV/Å$^2$)** | | |
| | -0.0216 | -0.0287 |
| **Elastic constants (GPa)** | | |
| C11 | 227.1 | 204.7 |
| C12 | 58.9 | 56.8 |
| C13 | 16.8 | 19 |
| C33 | 38.7 | 47.1 |
| C44 | 12.4 | 10.4 |

| | |
|---|---|
| **ICOHP in plane bond strength (vdW-df2-c09) (eV)** | |
| Mo-S | -3.30 |
| **ICOHP in plane bond strength per edge (eV/Å)** | |
| | -2.087 |
| **ICOHP in plane bond strength per edge /$h_0$ (eV/Å$^2$)** | |
| | -0.339 |

### 3.4 $MoSe_2$

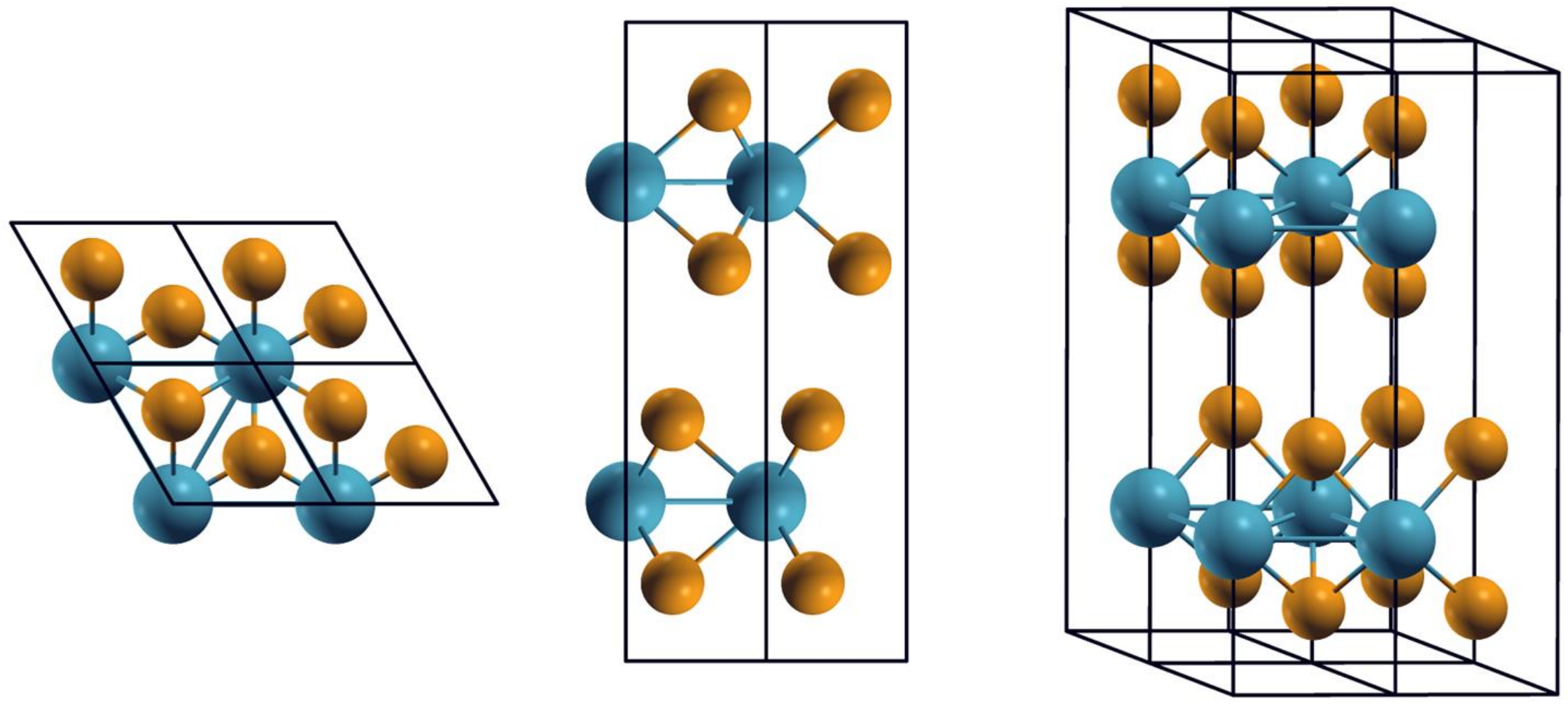

**Table S7: Tabulated calculated binding energy, ICOHP in plane covalent bond strength and elastic constant for 2H-$MoSe_2$**

| | **vdW-df2-c09** | **rvv10** |
|---|---|---|
| **Structural parameters (Å)** | | |
| **a** | 3.285 | 3.358 |
| **c** | 13.100 | 13.117 |
| **Binding energy (eV/Å$^2$)** | | |
| | -0.0204 | -0.0284 |
| **Elastic constants (GPa)** | | |
| C11 | 180.5 | 160.6 |
| C12 | 44.1 | 44.8 |
| C13 | 15.7 | 19.9 |
| C33 | 35.4 | 47.7 |
| C44 | 8.7 | 9.3 |

| | |
|---|---|
| **ICOHP in plane bond strength (vdW-df2-c09) (eV)** | |
| Mo-Se | -3.03 |
| **ICOHP in plane bond strength per edge (eV/Å)** | |
| | -1.844 |
| **ICOHP in plane bond strength per edge /$h_0$ (eV/Å$^2$)** | |
| | -0.281 |

## 3.5 $PtSe_2$

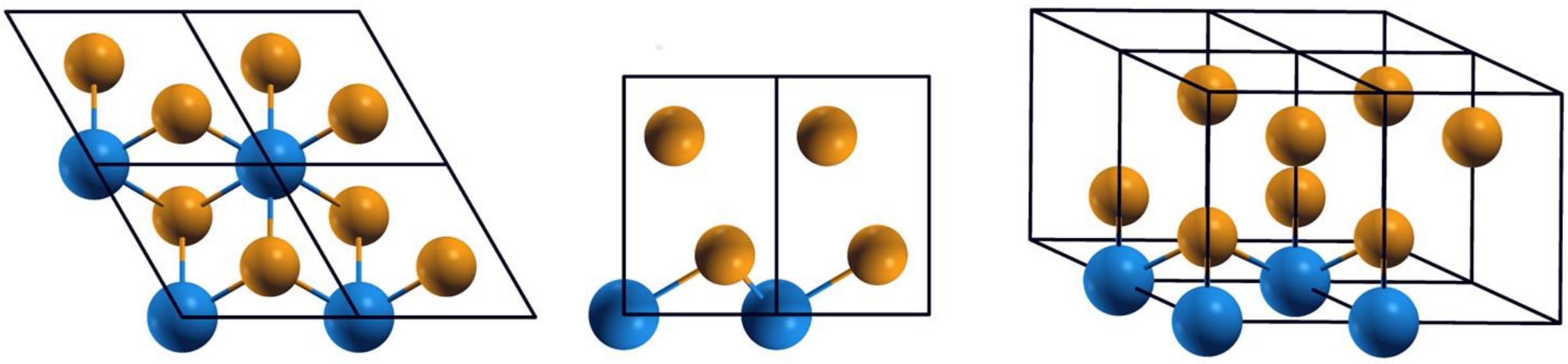

**Table S8: Tabulated calculated binding energy, ICOHP in plane covalent bond strength and elastic constant for 2H-$PtSe_2$**

| | **vdW-df2-c09** | **rvv10** |
|---|---|---|
| **Structural parameters (Å)** | | |
| **a** | 3.761 | 3.819 |
| **c** | 4.909 | 5.128 |
| **Binding energy (eV/Å$^2$)** | | |
| | -0.0296 | -0.0351 |
| **Elastic constants (GPa)** | | |
| C11 | 186.6 | 162.2 |
| C12 | 61.1 | 53.3 |
| C13 | 38.7 | 35.9 |
| C14 | -23.4 | -17.7 |
| C33 | 47.8 | 43.3 |
| C44 | 40.0 | 28.3 |

| | |
|---|---|
| **ICOHP in plane bond strength (vdW-df2-c09) (eV)** | |
| Pt-Se | -2.27 |
| **ICOHP in plane bond strength per edge (eV/Å)** | |
| | -1.212 |
| **ICOHP in plane bond strength per edge /$h_0$ (eV/Å$^2$)** | |
| | -0.246 |

## 3.6 BN

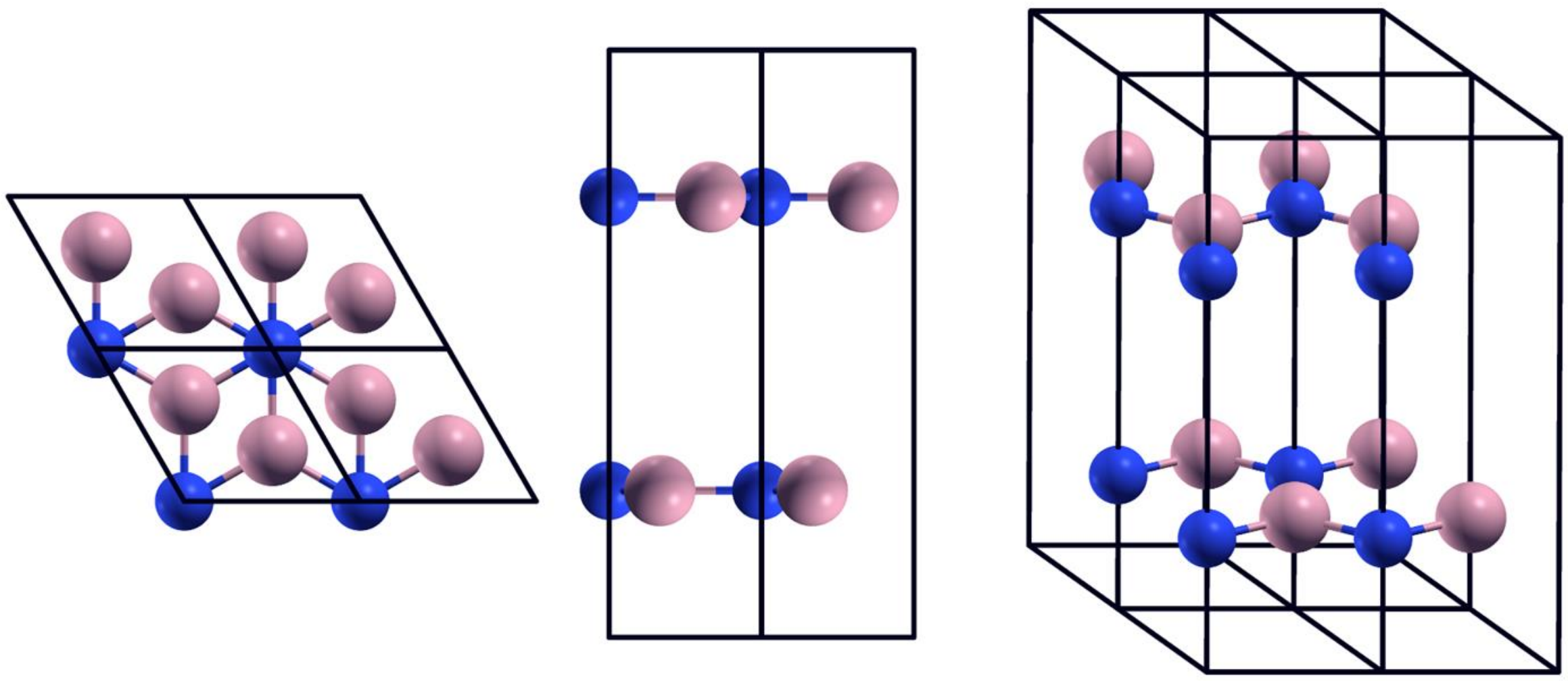

**Table S9: Tabulated calculated binding energy, ICOHP in plane covalent bond strength and elastic constant for hexagonal BN**

| | **vdW-df2-c09** | **rvv10** |
|---|---|---|
| **Structural parameters (Å)** | | |
| **a** | 2.510 | 2.520 |
| **c** | 6.691 | 6.801 |
| **Binding energy (eV/Å$^2$)** | | |
| | -0.0194 | -0.0244 |
| **Elastic constants (GPa)** | | |
| C11 | 834.5 | 813.7 |
| C12 | 186.5 | 171.9 |
| C13 | -4.3 | -4.7 |
| C33 | 29.9 | 35.8 |
| C44 | -5.1 | -4.9 |

| | |
|---|---|
| **ICOHP in plane bond strength (vdW-df2-c09) (eV)** | |
| B-N | -10.64 |
| **ICOHP in plane bond strength per edge (eV/Å)** | |
| | -4.239 |
| **ICOHP in plane bond strength per edge /$h_0$ (eV/Å$^2$)** | |
| | -1.267 |

## 3.7 GaS

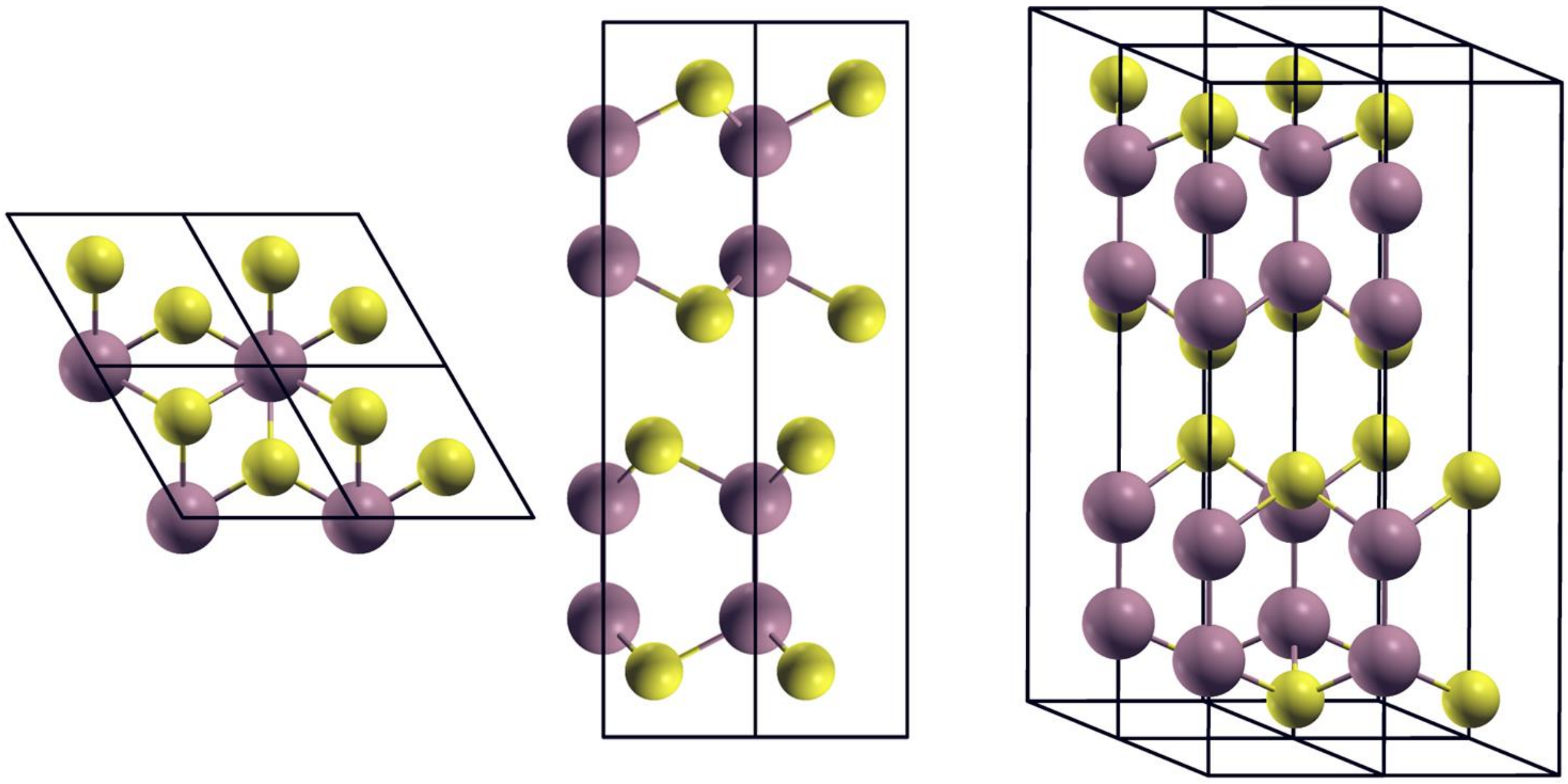

**Table S10: Tabulated calculated binding energy, ICOHP in plane covalent bond strength and elastic constant for $Ga_2S_2$**

| | **vdW-df2-c09** | **rvv10** |
|---|---|---|
| **Structural parameters (Å)** | | |
| **a** | 3.590 | 3.655 |
| **c** | 15.327 | 15.359 |
| **Binding energy (eV/Å$^2$)** | | |
| | -0.01605 | -0.0208 |
| **Elastic constants (GPa)** | | |
| C11 | 112.8 | 107.0 |
| C12 | 30.3 | 30.2 |
| C13 | 11.7 | 14.0 |
| C33 | 31.3 | 38.4 |
| C44 | 11.7 | 11.4 |

| | |
|---|---|
| **ICOHP in plane bond strength (vdW-df2-c09) (eV)** | |
| Ga-S | -4.52 |
| **ICOHP in plane bond strength per edge (eV/Å)** | |
| | -2.519 |
| **ICOHP in plane bond strength per edge /$h_0$ (eV/Å$^2$)** | |
| | -0.328 |

### 3.8 Talc

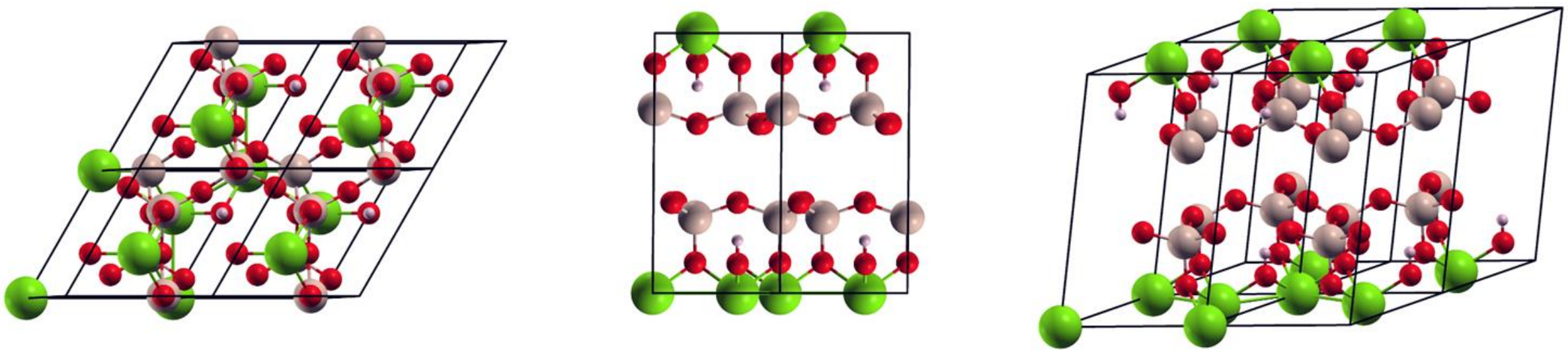

**Table S11: Tabulated calculated binding energy, ICOHP in plane covalent bond strength and elastic constant for Talc**

| | vdW-df2-c09 | rvv10 |
|---|---|---|
| **Structural parameters (Å)** | | |
| **a1** | (1.553,0.112,9.338) | (1.559,0.113,9.301) |
| **a2** | (2.644,4.600,0.051) | (2.656,4.625,0.051) |
| **a3** | (5.306,-0.002,0.068) | (5.333,0.003,0.068) |
| **Binding energy (eV/Å$^2$)** | | |
| | -0.0103 | -0.0131 |
| **Elastic constants (GPa)** | | |
| C11 | 210.1 | 225.9 |
| C12 | 48.2 | 63.5 |
| C13 | -5.4 | 8.5 |
| C14 | -1.5 | -2.2 |
| C15 | 6.5 | 0.7 |
| C16 | 0.7 | 3.0 |
| C22 | 207.1 | 220.9 |
| C23 | -1.2 | 13.3 |
| C24 | -3.1 | -2.9 |
| C25 | 4.4 | -2.3 |
| C26 | 0.5 | -3.0 |
| C33 | 35.9 | 64.8 |
| C34 | 0.2 | -0.8 |
| C35 | 3.2 | -2.4 |
| C36 | -2.9 | -7.2 |
| C44 | 19.4 | 24.2 |
| C45 | -8.0 | -11.9 |
| C46 | 0.6 | 0.4 |
| C55 | 8.7 | 14.8 |
| C56 | -0.1 | 0.6 |
| C66 | 80.8 | 81.2 |

| ICOHP in plane bond strength (vdW-df2-c09) (eV) | |
|---|---|
| Si-O | -8.03 |
| **ICOHP in plane bond strength per edge (eV/Å)** | |
| | -6.952 |
| **ICOHP in plane bond strength per edge /$h_0$ (eV/Å$^2$)** | |
| | -0.744 |

## 3.9 $Ni(OH)_2$

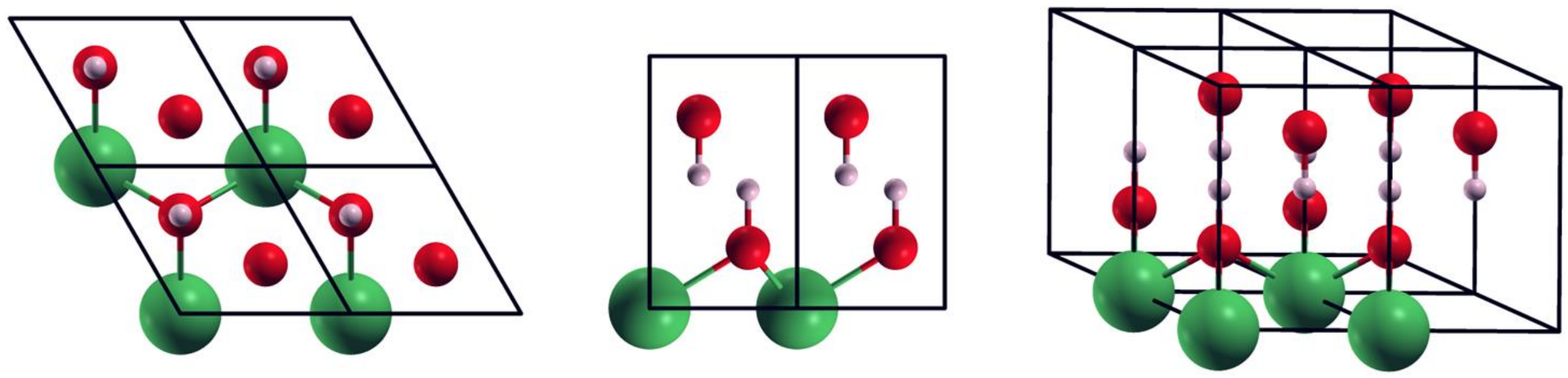

**Table S12: Tabulated calculated binding energy, ICOHP in plane covalent bond strength and elastic constant for $Ni(OH)_2$**

| | vdW-df2-c09 | rvv10 |
|---|---|---|
| **Structural parameters (Å)** | | |
| **a** | 3.175 | 3.223 |
| **c** | 4.284 | 4.383 |
| **Binding energy (eV/Å$^2$)** | | |
| | -0.0322 | -0.0396 |
| **Elastic constants (GPa)** | | |
| C11 | 140.4 | 134.1 |
| C12 | 120.7 | 114.2 |
| C13 | 31.2 | 30.1 |
| C14 | 4.6 | 5.2 |
| C33 | 58.8 | 62.4 |
| C44 | 31.8 | 27.3 |

| | |
|---|---|
| **ICOHP in plane bond strength (vdW-df2-c09) (eV)** | |
| Ni-O | -3.08 |
| **ICOHP in plane bond strength per edge (eV/Å)** | |
| | -1.935 |
| **ICOHP in plane bond strength per edge /$h_0$ (eV/Å$^2$)** | |
| | -0.452 |

## 3.10 $Mg(OH)_2$

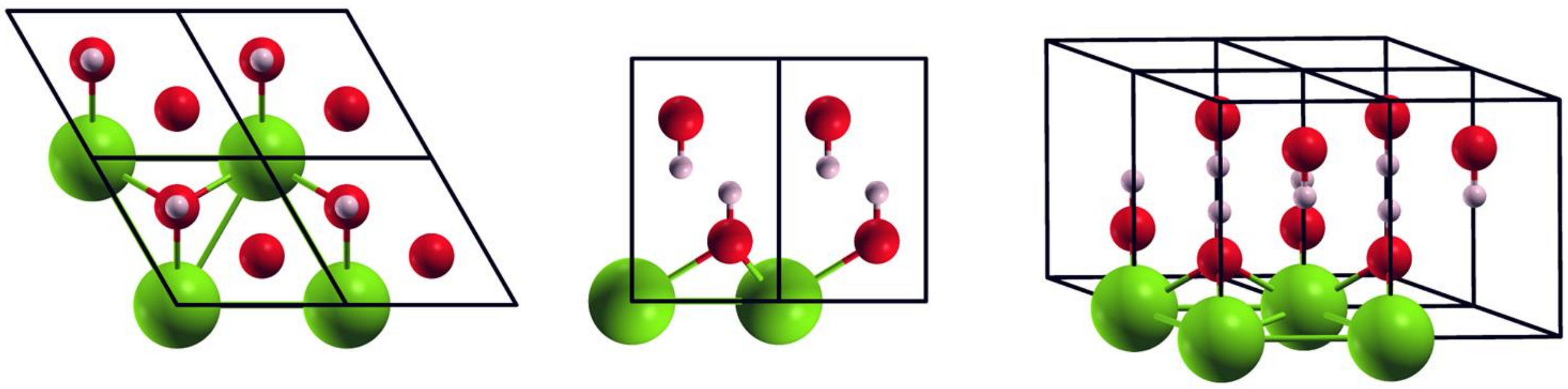

**Table S13: Tabulated calculated binding energy, ICOHP in plane covalent bond strength and elastic constant for $Mg(OH)_2$**

| | **vdW-df2-c09** | **rvv10** |
|---|---|---|
| **Structural parameters (Å)** | | |
| **a** | 3.163 | 3.176 |
| **c** | 4.622 | 4.664 |
| **Binding energy (eV/Å$^2$)** | | |
| | -0.0299 | -0.0365 |
| **Elastic constants (GPa)** | | |
| C11 | 160.4 | 161.4 |
| C12 | 44.3 | 47.1 |
| C13 | 13.6 | 17.5 |
| C14 | -1.3 | -1.1 |
| C33 | 66.8 | 73.4 |
| C44 | 27.9 | 26.6 |

| | |
|---|---|
| **ICOHP in plane bond strength (vdW-df2-c09) (eV)** | |
| Mg-O | -2.86 |
| **ICOHP in plane bond strength per edge (eV/Å)** | |
| | -1.808 |
| **ICOHP in plane bond strength per edge /$h_0$ (eV/Å$^2$)** | |
| | -0.391 |

## 3.11 $Co(OH)_2$

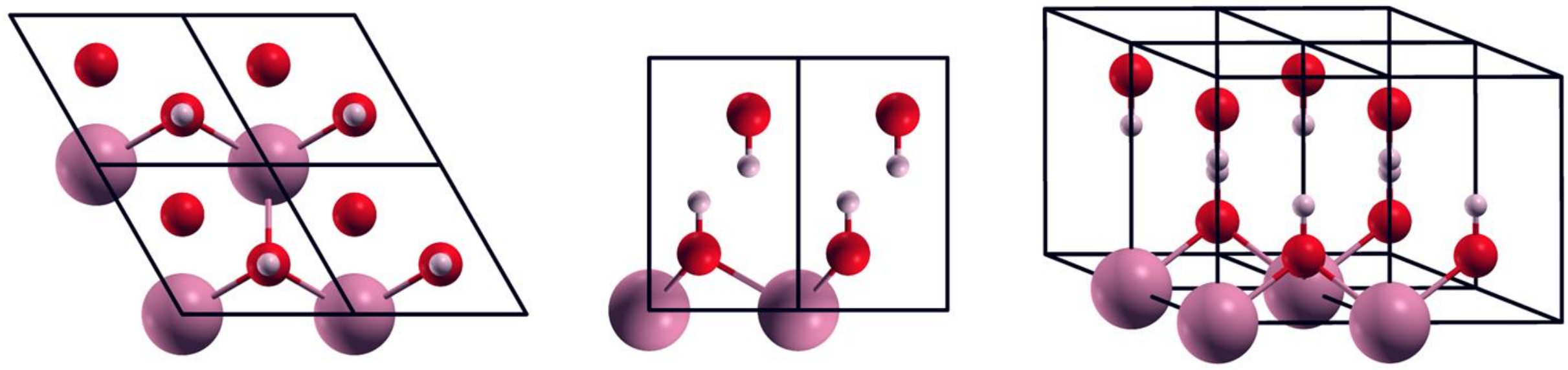

**Table S14: Tabulated calculated binding energy, ICOHP in plane covalent bond strength and elastic constant for $Co(OH)_2$.**

| | **vdW-df2-c09** | **rvv10** |
|---|---|---|
| **Structural parameters (Å)** | | |
| **a** | 3.101 (3.106) | 3.162 (3.163) |
| **c** | 4.753 (4.258) | 4.582 (4.351) |
| **Binding energy (eV/Å$^2$)** | | |
| | -0.0341 | -0.0409 |
| **Elastic constants (GPa)** | | |
| C11 | 149.3 | 139.7 |
| C12 | 125.9 | 115.6 |
| C13 | 35.8 | 35.6 |
| C14 | 2.2 | 2.2 |
| C33 | 61.6 | 59.3 |
| C44 | 39.5 | 31.8 |

| | |
|---|---|
| **ICOHP in plane bond strength (vdW-df2-c09) (eV)** | |
| Co-O | -3.35 |
| **ICOHP in plane bond strength per edge (eV/Å)** | |
| | -2.159 |
| **ICOHP in plane bond strength per edge /$h_0$ (eV/Å$^2$)** | |
| | -0.454 |

### 3.12 $Zn(OH)_2$

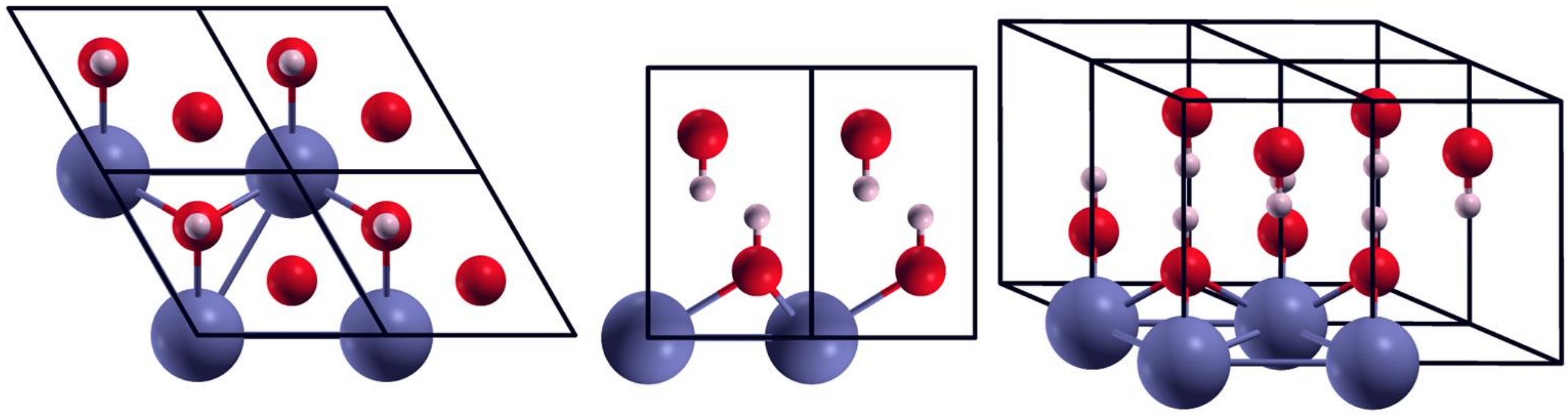

**Table S15: Tabulated calculated binding energy, ICOHP in plane covalent bond strength and elastic constant for $Zn(OH)_2$.**

| | vdW-df2-c09 | rvv10 |
|---|---|---|
| **Structural parameters (Å)** | | |
| **a** | 3.198 | 3.198 |
| **c** | 4.557 | 4.556 |
| **Binding energy (eV/Å$^2$)** | | |
| | -0.0305 | -0.0391 |
| **Elastic constants (GPa)** | | |
| C11 | 147.7 | 146.1 |
| C12 | 73.0 | 72.6 |
| C13 | 21.1 | 24.2 |
| C14 | 3.4 | 3.6 |
| C33 | 68.5 | 73.2 |
| C44 | 27.4 | 26.3 |

| | |
|---|---|
| **ICOHP in plane bond strength (vdW-df2-c09) (eV)** | |
| Zn-O | -2.97 |
| **ICOHP in plane bond strength per edge (eV/Å)** | |
| | -1.857 |
| **ICOHP in plane bond strength per edge /$h_0$ (eV/Å$^2$)** | |
| | -0.407 |

# 4 Relationship between in-plane/out-of-plane binding energy and modulus ratios

Here we develop an approximate model for the relationship between the ratio of in-plane to out-of-plane binding energies and the ratio of in-plane to out-of-plane moduli. Let us assume that the out-of-plane (OOP) interlayer van der Waals interaction potential is given by a generalised Lennard-Jones (LJ) function:

$$E_{OOP} = \frac{k_1}{r^a} - \frac{k_2}{r^b}$$

Here, the first term represents Pauli repulsion and the second term represents the interlayer van der Waals attraction. In the standard inter-atomic Lennard-Jones potential, *a* is generally taken as 12, while for interatomic interactions, *b*=6.[20] However, for a very thin nanosheet interacting with a thick stack, the total van der Waals attraction is the sum over all inter-atomic attractions.[20] This may result in a *b*-parameter which deviates from 6.[20] In addition, real systems can display a-parameters which deviate somewhat from 12.

It is simple to show that the LJ-like potential given above can be written in a general form which, when representing the inter-sheet (out-of-plane) interaction energy versus interlayer separation ($r_{IL}$), is given by:

$$E_{OOP}(r) = \varepsilon_{vdW}\left[\frac{b}{a-b}\left(\frac{r_{vdW}}{r_{IL}}\right)^a - \frac{a}{a-b}\left(\frac{r_{vdW}}{r_{IL}}\right)^b\right]$$

where $r_{vdW}$ is the equilibrium separation and $\varepsilon_{vdW}$ is the binding curve well depth.

By definition, the spring constant associated with the intersheet interaction is

$$k_{vdW} = \left.\frac{d^2V_{OOP}}{dr_{IL}^2}\right|_{r_{IL}=r_{vdW}}$$

which works out as $k = \frac{\varepsilon_{vdW}ab}{r_{vdW}^2}$. Using the standard definition, we can convert *k* to a modulus, *Y*: $Y_{Out-of-plane} = \frac{kr_{vdW}}{A_{vdW}}$ where $A_{vdW}$ is a characteristic area associated with the interaction. Combining the latter two expressions, we find that

$$Y_{out-of-plane} = \frac{ab}{r_{vdW}}\left(\frac{\varepsilon_{vdW}}{A_{vdW}}\right)$$

Now, considering the in-plane (IP), intra-sheet chemical bonding, we can perform a similar calculation. Typically, one would use something like the Morse potential for such a calculation. However, for simplicity, and as it is functionally similar to the Morse Potential, we use an generalized LJ potential for the in-plane binding also. Then, for the inter-sheet (out-of-plane) interaction energy versus interlayer separation ($r_{IA}$) we have:

$$E_{IP} = \varepsilon_{bond}\left[\frac{\beta}{\alpha-\beta}\left(\frac{r_{bond}}{r_{IA}}\right)^{\alpha} - \frac{\alpha}{\alpha-\beta}\left(\frac{r_{bond}}{r_{IA}}\right)^{\beta}\right]$$

where $r_{bond}$ is the equilibrium separation and $\varepsilon_{bond}$ is the in-plane binding curve well depth. Here, the exponents are not generally known and we will deal with them below.

Completing a similar calculation to that above, we find that:

$$Y_{In-plane} \approx \frac{\alpha\beta}{l_{bond}}\left(\frac{\varepsilon_{bond}}{A_{bond}}\right)$$

where $l_{bond}$ is the in-plane bond length and $A_{bond}$ is a characteristic area associated with the interaction.

Combining our two equations for elastic moduli, we arrive at:

$$\frac{Y_{In-plane}}{Y_{Out-of-plane}} \approx \frac{\alpha\beta r_{vdW}}{abl_{bond}}\frac{(\varepsilon_{bond}/A_{bond})}{(\varepsilon_{vdW}/A_{vdW})}$$

Finally, we note that we can combine the terminology used here with the terminology in the main paper, namely:

$$\frac{(\varepsilon_{bond}/A_{bond})}{(\varepsilon_{vdW}/A_{vdW})} = \frac{E_E}{E_S}$$

whereupon:

$$\frac{Y_{In-plane}}{Y_{Out-of-plane}} \approx \frac{\alpha\beta r_{vdW}}{abl_{bond}}\frac{E_E}{E_S}$$

To estimate the relationship between the modulus ratio and the energy ratio, it is necessary to estimate both equilibrium spacings and $a$ and $b$ as well as $\alpha$ and $\beta$. To estimate these parameters, we used DFT to calculate $E_{OOP}$ versus $r_{IL}$ for graphite. In addition, we extracted data for the bond-stretching potential energy, $E_{IP}$, as a function of the interatomic distance between bonded carbon atoms in graphene layers, $r_{IA}$, calculated using DFT from ref [21]. We then fit this data to LJ-like equations as shown above. We found that, in order to achieve good fits, one of the exponents had to be fixed. Thus, we performed a number of fittings, each time fixing *b* (for OOP) and $\beta$ ( for IP) at a different fixed value (see table S16) finding reasonably good fits in all cases (Fig. S37). For both OOP and IP situations we found roughly constant values of $ab/r_{vdW}$ and $\alpha\beta/r_{bond}$ irrespective of the fixed values of *b* and $\beta$ used in fitting. Considering the mean and spread, we find $ab/r_{vdW}$=14.3±1 and $\alpha\beta/r_{bond}$=11.7±1. This yields a value of $\alpha\beta r_{vdW} / abl_{bond}$ =0.82±0.16 which is very close to 1, allowing us to approximate

$$\frac{Y_{In-plane}}{Y_{Out-of-plane}} \approx \frac{E_E}{E_S}$$

which supports our idea that the modulus ratio is a good proxy for the energy ratio.

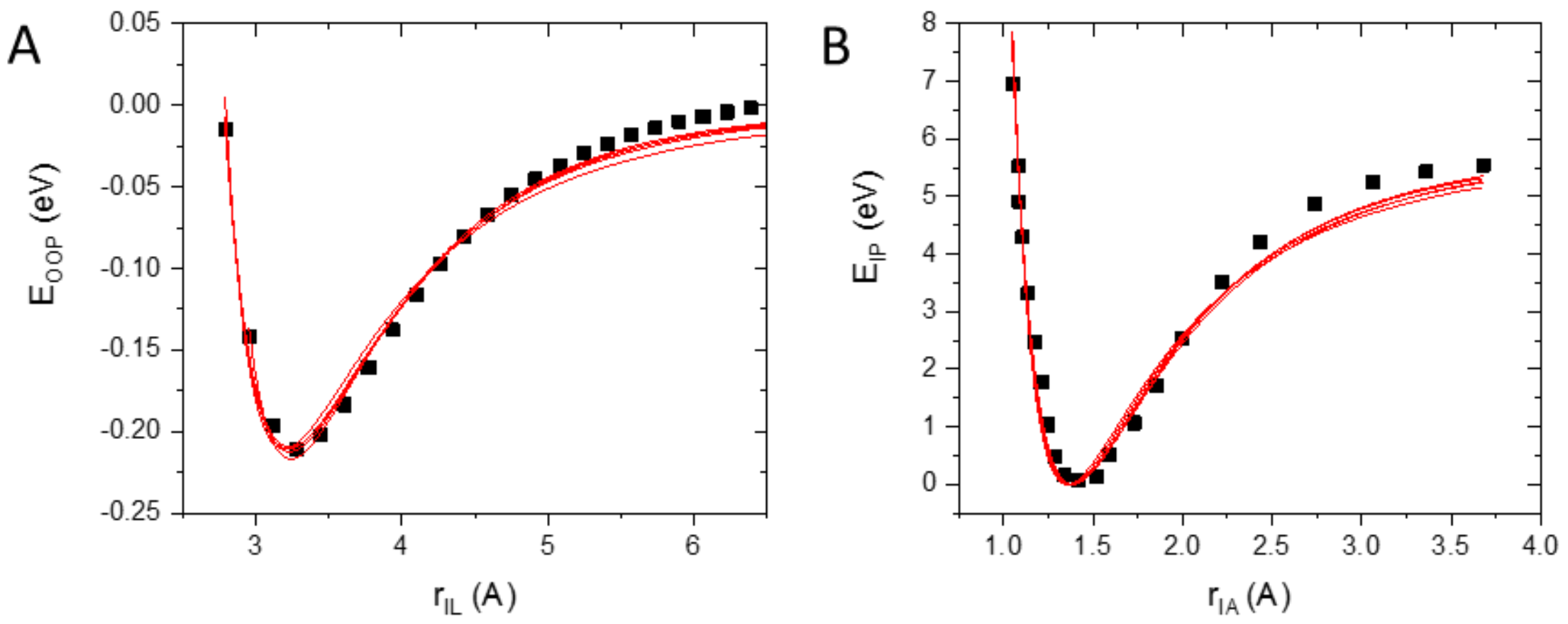

**Fig S37**: A) Bond energy versus spacing data for the inter-sheet (out-of-plane) interaction (calculated here). B) Bond energy versus spacing data for the intra-sheet (in-plane) interaction (as extracted from ref [21]). The lines are fits to LJ-like potentials as described in the text. The multiple fits were performed by fitting the b (in panel A) and β (in panel B) parameters at different values between 4 and 8. In B, because of the way the data was presented, a fixed offset of $E_{bond}$=6eV was added.

**Table S16: Fit parameters found by fitting the data in figure S37 to LJ-like functions.**

| | $E_{vdW}$ (eV) | $r_{vdW}$ (A) | a | b | Adj $R^2$ | $ab/r_{vdW}$ ($A^{-1}$) |
|---|---|---|---|---|---|---|
| | 0.211 | 3.19 | 12.46 | 4 | 0.996 | 15.6 |
| | 0.211 | 3.22 | 9.09 | 5 | 0.983 | 14.1 |
| | 0.21 | 3.23 | 7.25 | 6 | 0.999 | 13.5 |
| | 0.212 | 3.23 | 6.44 | 7 | 0.989 | 14.0 |
| | 0.213 | 3.22 | 5.75 | 8 | 0.988 | 14.3 |
| | | | | | | |
| | $E_{bond}$ (eV) | $R_{bond}$ (A) | $\alpha$ | $\beta$ | Adj $R^2$ | $\alpha\beta/r_{bond}$ ($A^{-1}$) |
| | 6 | 1.39 | 3.79 | 4 | 0.979 | 10.9 |
| | 6 | 1.38 | 3.11 | 5 | 0.975 | 11.3 |
| | 6 | 1.369 | 2.68 | 6 | 0.965 | 11.7 |
| | 6 | 1.36 | 2.39 | 7 | 0.951 | 12.3 |
| | 6 | 1.34 | 2.18 | 8 | 0.933 | 13.0 |

# 5 Detailed protocol for AFM statistics

In case of queries: backes@uni-heidelberg.de

1. **Deposition**
   - Clean wafers (recommendation $Si/SiO_2$ with ~ 200 nm oxide) by rinsing with ~ 5 mL of clean isopropanol. Blow off solvent - do not let it dry.
   - Check the cleanliness of the water and IPA, as contaminations can often be found especially when solvents are stored in plastic bottles. To do this, dropcast $H_2O$ and IPA on a wafer and let the solvent dry. Then check the wafer for stains which are often visible with the naked eye. If these are present, this will result in artefacts apparent both on the wafer surface and at the sheets plane. Therefore they must be avoided. If clean solvent is not available then distillation Is recommended
   - Dilute the sample to an optical extinction of ~ 0.4 (per cm) at the extinction maximum with IPA for solvent, or pure (checked as described before) DI water for surfactant dispersions respectively. Sonicate ~ 2 min in the sonic-bath to refresh the sample both before and after dilution. For deposition from high boiling point solvents such as NMP and CHP, first transfer the sample to IPA prior to dilution and deposition. This can be done by centrifuging the dispersion at relatively high RCF, so the material is spun down, then collected and re-dispersed in IPA.
   - Pre-heat the wafer on a hotplate to ~ 50°C above the boiling point of the dominant solvent in the dispersion. NB: the ideal temp. depends on the hotplate.
   - Drop 20 µL of the dispersion per 1x1 $cm^2$ pre-heated wafer. The solvent should flash-evaporate to allow the even deposition of the material. For a visualisation, see here [22]
   - If evaporation is too slow, increase the temperature of the hotplate. If the drop roles off from the surface, decrease the temperature of the hotplate.
   - Immediately after deposition, wash the wafer with 5 mL water (for surfactant-based samples only) and 5 mL IPA (in both cases). Blow off the solvents, do not let them dry.

2. **Imaging**
   - Identify promising regions via the optical contrast. This is strongly material and size-dependent. For TMDs, gra, BN examples are shown here. [23]
   - The recommended resolution during the measurement depends on the lateral sizes of the nanosheets. Adjust the field of view accordingly. Ideally, you should have minimal pixilation when zooming into an area that is subjected to the counting (see example below), but measuring an area as large as possible to accumulate sufficient counts. Usually, 1024 lines in a 10x10 $\mu m^2$ is a good starting point for 2D objects where lateral size is distributed as follows: 100 nm $<$ L $>$ 1 µm.
   - Depending on the scan head, it is often better to increase the number of lines to improve the resolution rather than zooming in further. For example, with a 70x70 $\mu m^2$ scanner, images are naturally blurry for zoom-in below 4 µm, while you can still improve the resolution below 2 µm with a 10x10 $\mu m^2$ scanner.

- If you observe objects that have the same shape repeatedly in an area, change the cantilever. These are artefacts even if they look like nanosheets some time. Also change the cantilever when objects appear doubled or when a shadow is visible on one (same) side.
- Target imaging 200-300 of individually deposited nanosheets per sample to facilitate sufficient number of counts and therefore a liable statistical results.

## 3. Size measurement

- For a visualization: see here [22]
- First crop the image to a size your eyes feel comfortable with. Measure the length, L (longest dimension in the centre of the flake) as indicated here:

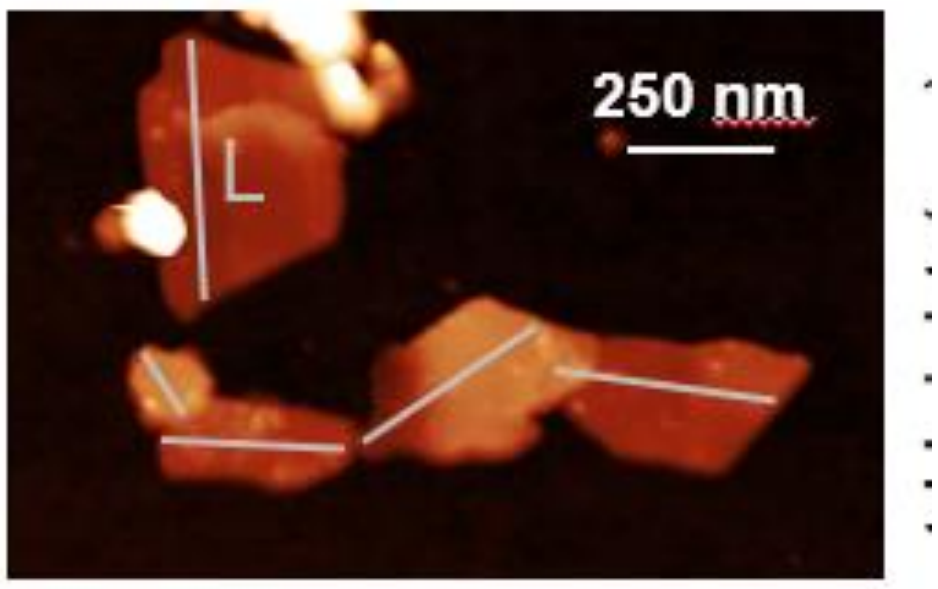

  If needed for your analysis measure width, W in the same manner (perpendicular to L).
- Play with the contrast: often rainbow or spectral scale is better when the samples are polydisperse. For less polydisperse samples, grey scale (or black/white) works well to identify the shapes of the flakes better.
- The thickness of a flake should be averaged across the flake, if its thickness is not homogeneous. Don't take the thinner or thicker part, just an approximate average and be consistent across all the flakes and samples.
- Collect L, W, and t in separate columns in origin (or excel).

## 4. Corrections

**Step height analysis is used to convert apparent thickness to layer number for unknown samples.**

- Locate incompletely exfoliates flakes and identify steps associated with terraces. This is often best seen in the black/white contrast.
- Be careful to distinguish between real steps, flakes sitting on top of each other and folded flakes.
- Measure the height of the steps from the line profiles, as for the apparent height measurement before.
- Plot them in ascending order. They will be a multiple of a certain number which is the apparent AFM thickness of one layer.

- To calculate the layer number N corresponding to individual nanosheet from the apparent AFM thickness, divide the measured thickness by the step height. Subtract additionally 0.5 nm to address surfactant/solvent trapped underneath the flake.

**L correction to account for tip broadening and pixilation**

The correction will be dependent on the instrument and especially tip radius of the cantilever. For the Bruker ICON with OLTESPA-R3 cantilevers, it is reported in [24].

- Correction functions can be established by comparing mean <L> from AFM and TEM for a number of samples. Alternatively, use nanowires and measure their width and thickness. In this case, width=thickness therefore calibration is accessible fairly quickly.

## 5. <u>Evaluation</u>

- After collecting ~ 60-100 counts (depending on sample polydispersity), plot N and L histograms for a first quality check. Choose the bins appropriately. Note: To estimate an arithmetic mean, the data quality below would be sufficient. However, this is not the case, when volumes need to be considered in the analysis (volume fraction weighted mean <N>, monolayer volume fraction etc) or when universal size dependences are to be identified (such as in this work).

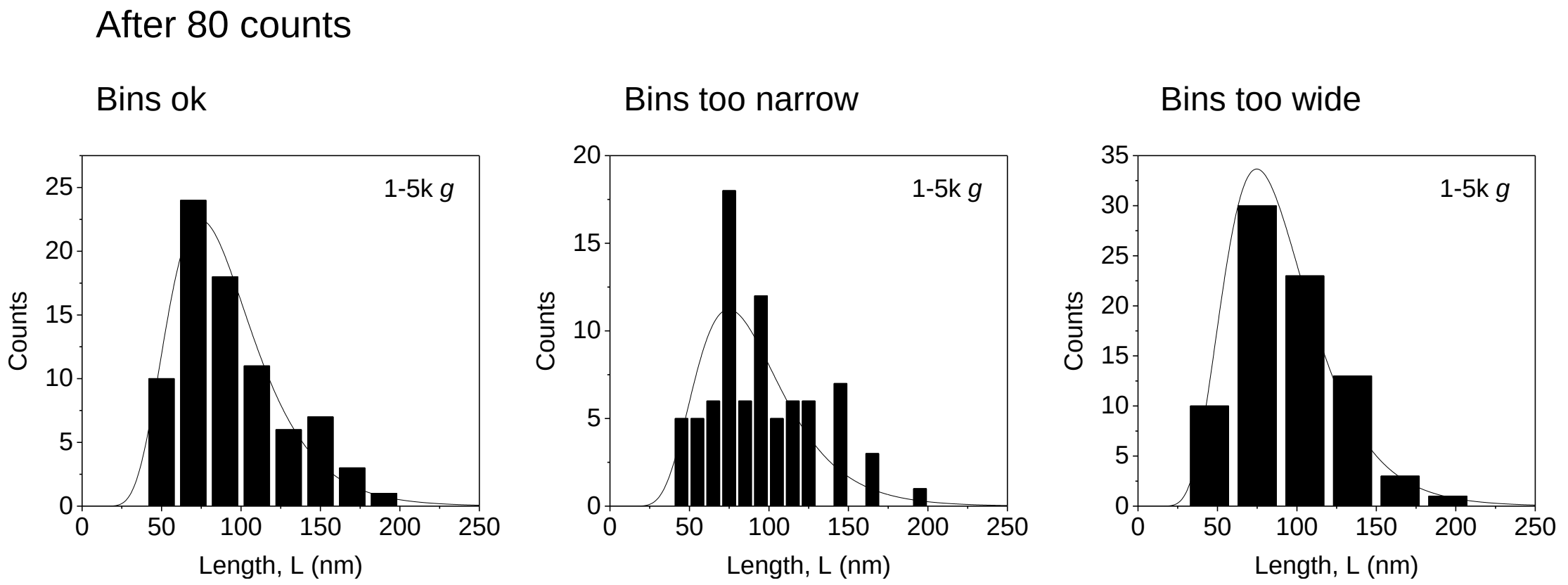

- For high quality statistics, keep counting until both N and L histograms have nice lognormal shape with a long tail, but no outliers (these should be excluded). Examples see below:

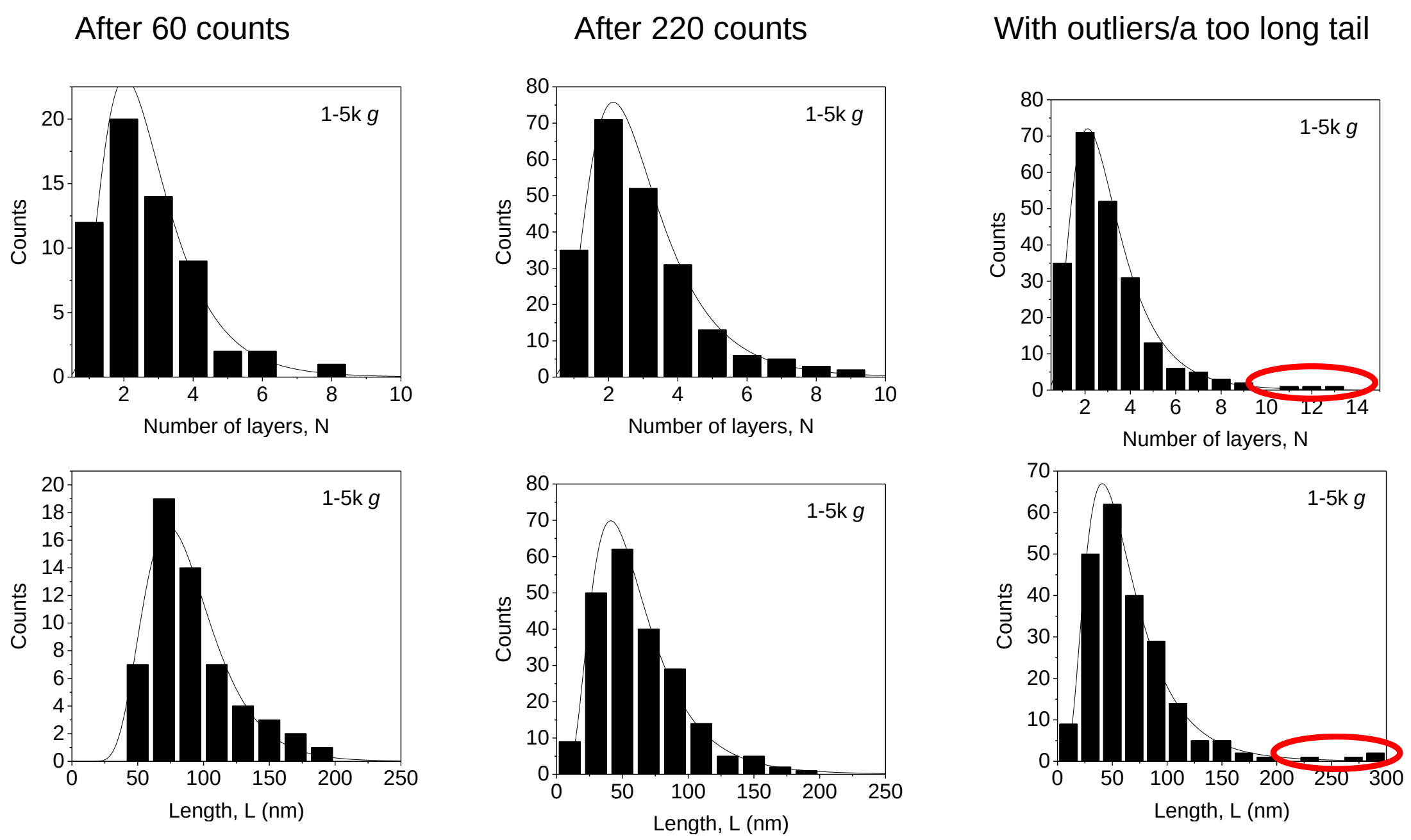

- The histogram shape is a good guideline whether measurement and counting is performed properly: if reaggregated nanosheets or impurities are counted, a deviation from the lognormal shape is observed. Examples see below. In this case, ideally make new wafers and try again.

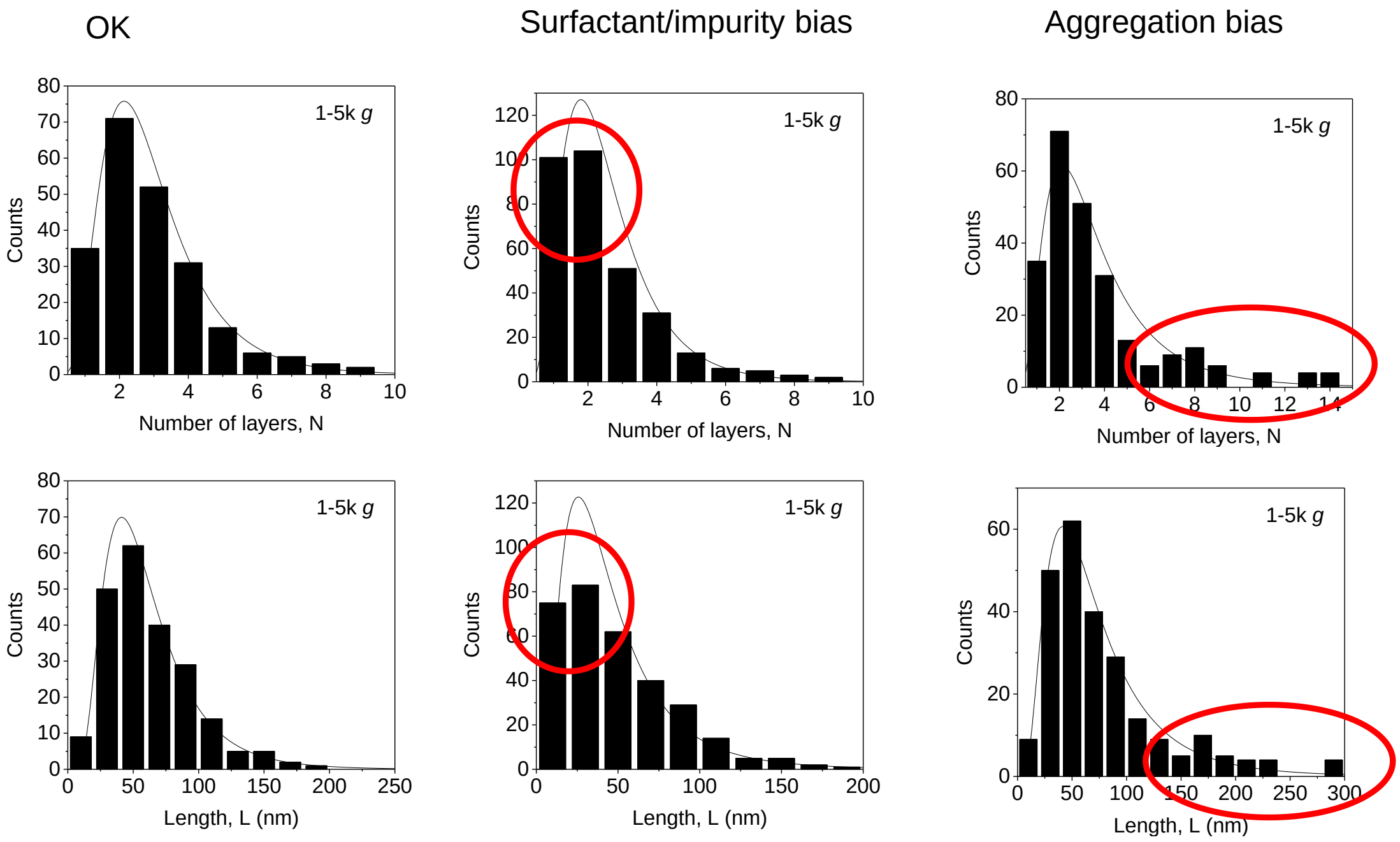

- Extract <L>, <W>, <N> (and monolayer content if applicable) from the columns

## 6. <u>Quality control</u>

A number of master curves have been gathered for variety of LPE nanosheets. These serve as quality control. Check whether your samples behave accordingly. If not, there is something wrong. Examples are shown throughout the SI and briefly summarised here.

i) Plots of log <L> and log <N> vs central *g* (if you do a size selection cascade) always scale linearly with log *g*
ii) Arithmetic and volume fraction weighted mean scale linearly (master curve for all materials). If this is not the case, check whether some outliers need to be excluded from statistical evaluation.
iii) The monolayer volume fraction is related to <N> and <N>$_{Vf}$.
iv) Plots of area <L*W> vs <N>, or <L> vs <N> are powerlaws (i.e. linear on log-log scale)

# 6 Supporting references